\documentclass{SciPost}
\usepackage[utf8]{inputenc}
\usepackage[T1]{fontenc}
\usepackage{cleveref}%
\usepackage{caption}
\usepackage{subcaption}
\usepackage{placeins} %
\usepackage{tikz}					%
\usepackage{pgfplots}
\pgfplotsset{compat=1.18}
\usetikzlibrary{shapes,arrows,backgrounds,arrows.meta, positioning, calc}
\usepackage{sidecap} %
\usepackage{amsmath}
\usepackage{xcolor}
\usepackage[frozencache=true,cachedir=minted-cache]{minted}
\usepackage{listings}
\usepackage{mathtools}
\usepackage{tcolorbox} %
\usepackage{wrapfig}
\usepackage{xstring}
\usepackage{mathtools}
\usepackage{siunitx}
\usepackage{braket}
\usepackage{array}
\usepackage{booktabs}
\usepackage{pifont}
\usepackage{adjustbox}
\hypersetup{
    colorlinks,
    linkcolor={red!50!black},
    citecolor={blue!50!black},
    urlcolor={blue!80!black}
}

\usepackage[bitstream-charter]{mathdesign}
\DeclareSymbolFont{usualmathcal}{OMS}{cmsy}{m}{n}
\DeclareSymbolFontAlphabet{\mathcal}{usualmathcal}

\fancypagestyle{SPstyle}{
\fancyhf{}
\lhead{\colorbox{scipostblue}{\bf \color{white} ~SciPost Physics Codebases }}
\rhead{{\bf \color{scipostdeepblue} ~Submission }}

\fancyfoot[C]{\textbf{\thepage}}
}

\newenvironment{codeblock}{%
  \VerbatimEnvironment
  \begin{minted}[%
  frame=single,
  linenos,
  breaklines,
  fontsize=\ttfamily\footnotesize,
  mathescape,
  escapeinside={ßß},
  autogobble
  ]{python}%
}{%
  \end{minted}
}

\tcbuselibrary{skins}
\newtcolorbox[auto counter, number within=section]{brainstorming}{
  colback=gray!10, %
  colframe=gray!80, %
  boxrule=0.5mm, %
  arc=4mm, %
  auto outer arc,
  width=\textwidth,
  boxsep=5mm, %
  fonttitle=\footnotesize, %
  fontupper=\footnotesize, %
  skin=enhanced, %
  before upper={%
    \setlength{\itemsep}{-0.5em} %
  },
  top=1mm, %
  bottom=1mm %
}

\newcommand{\posL}[0]{x}
\newcommand{\posIdx}[0]{n}
\newcommand{\posD}[0]{\posL_\posIdx}
\newcommand{\posMin}[0]{\posL_\mathrm{min}}

\newcommand{\freqL}[0]{f}
\newcommand{\freqIdx}[0]{m}
\newcommand{\freqD}[0]{\freqL_\freqIdx}
\newcommand{\freqMin}[0]{\freqL_\mathrm{min}}

\newcommand{\samplesPos}[0]{g_n}
\newcommand{\samplesPosFFT}[0]{g_n^\mathrm{fft}}
\newcommand{\samplesPosInt}[2]{#1_n^\mathrm{int}(#2)}
\newcommand{\samplesFreq}[0]{G_m}
\newcommand{\samplesFreqFFT}[0]{G_m^\mathrm{fft}}

\newcommand{\freqCan}{-\mathrm{floor}(0.5N) \, \Delta \freqL}
\newcommand{\freqCanSym}{\freqMin^\text{sym}}
\newcommand{\posCan}{-\mathrm{floor}(0.5N) \, \Delta \posL}
\newcommand{\posCanSym}{\posMin^\text{sym}}

\newcommand{\fou}[0]{\mathcal{F}}
\newcommand{\ifou}[0]{\widehat{\mathcal{F}}}
\newcommand{\fft}[0]{\texttt{fft}_m}
\newcommand{\ifft}[0]{\texttt{ifft}_m}

\newcommand{\dftNoNExp}[1]{\textcolor{red}{e^{#1 2\pi i \ m \Delta \freqL \ n \Delta \posL}}}
\newcommand{\fftExp}[1]{\textcolor{red}{e^{#1 2\pi i \ \frac{m n}{N}}}}

\newcommand{\freqZeroShift}[1]{\textcolor{orange}{e^{\textcolor{orange}{#1} 2\pi i \ \freqMin \ n \Delta x}}}
\newcommand{\xZeroShift}[1]{\textcolor{brown}{e^{\textcolor{brown}{#1} 2\pi i \ \posMin \  m \Delta \freqL}}}
\newcommand{\phaseCorrection}[1]{\textcolor{blue}{e^{\textcolor{blue}{#1} 2\pi i \ \posMin \ \freqMin}}}

\newcommand{\first}[1]{#1}

\newcommand{\code}[1]{\texorpdfstring{\mintinline[breaklines,breakanywhere]{py}{#1}}{#1}}
\newcommand{\codeNoBreak}[1]{\texorpdfstring{\mintinline[]{py}{#1}}{#1}}

\begin{document}

\pagestyle{SPstyle}

\begin{center}{\Large \textbf{\color{scipostdeepblue}{
FFTArray: A Python Library for the Implementation of Discretized Multi-Dimensional Fourier Transforms\\
}}}\end{center}

\begin{center}\textbf{
Stefan J. Seckmeyer\textsuperscript{1$\star$},
Christian Struckmann\textsuperscript{1},
Gabriel Müller\textsuperscript{1},
Jan-Niclas Kirsten-Siem\ss\textsuperscript{1} and
Naceur Gaaloul\textsuperscript{1$\dagger$}
}\end{center}

\begin{center}
{\bf 1} Institut für Quantenoptik, Leibniz Universität Hannover, Welfengarten 1, D-30167, Hannover, Germany
\\[\baselineskip]
$\star$ \href{mailto:seckmeyer@iqo.uni-hannover.de}{\small seckmeyer@iqo.uni-hannover.de}\,,\quad
$\dagger$ \href{mailto:gaaloul@iqo.uni-hannover.de}{\small gaaloul@iqo.uni-hannover.de}
\end{center}

\section*{\color{scipostdeepblue}{Abstract}}
\textbf{\boldmath{%
Partial differential equations describing the dynamics of physical systems rarely have closed-form solutions.
Fourier spectral methods, which use Fast Fourier Transforms (FFTs) to approximate solutions, are a common approach to solving these equations.
However, mapping Fourier integrals to discrete FFTs is not straightforward, as the selection of the grid as well as the coordinate-dependent phase and scaling factors require special care.
Moreover, most software packages that deal with this step integrate it tightly into their full-stack implementations.
Such an integrated design sacrifices generality, making it difficult to adapt to new coordinate systems, boundary conditions, or problem-specific requirements.
To address these challenges, we present FFTArray, a Python library that automates the general discretization of Fourier transforms.
Its purpose is to reduce the barriers to developing high-performance, maintainable code for pseudo-spectral Fourier methods.
Its interface enables the direct translation of textbook equations and complex research problems into code, and its modular design scales naturally to multiple dimensions.
This makes the definition of valid coordinate grids straightforward, while  coordinate grid specific corrections are applied with minimal impact on computational performance.
Built on the Python Array API Standard, FFTArray integrates seamlessly with array backends like NumPy, JAX and PyTorch and supports Graphics Processing Unit acceleration.
The code is openly available at \url{https://github.com/QSTheory/fftarray} under Apache-2.0 license.
}}
\vspace{\baselineskip}

\noindent\textcolor{white!90!black}{%
\fbox{\parbox{0.975\linewidth}{%
\textcolor{white!40!black}{\begin{tabular}{lr}%
  \begin{minipage}{0.6\textwidth}%
    {\small Copyright attribution to authors. \newline
    This work is a submission to SciPost Physics Codebases. \newline
    License information to appear upon publication. \newline
    Publication information to appear upon publication.}
  \end{minipage} & \begin{minipage}{0.4\textwidth}
    {\small Received Date \newline Accepted Date \newline Published Date}%
  \end{minipage}
\end{tabular}}
}}
}
\vspace{10pt}
\noindent\rule{\textwidth}{1pt}
\tableofcontents
\noindent\rule{\textwidth}{1pt}
\vspace{10pt}

\section{Introduction}

\FloatBarrier
Many interesting and important physical systems are modeled by differential equations that often lack closed-form solutions.
Fourier integrals often allow to formulate approximate solutions for these systems.
Although performing these integrals analytically may still remain intractable, they can often be evaluated numerically by discretizing them and using Fast Fourier Transform (FFT) algorithms.

A prominent example is the Schrödinger equation, which governs the evolution of quantum mechanical systems ranging from single-particle dynamics to many-body phenomena.
While exact solutions exist for idealized cases like box-shaped potentials and the quantum harmonic oscillator, most practical applications require numerical methods such as the split-step Fourier method~\cite{Feit1982,Hairer2002}.
These numerical simulations allow the validation and improvement of analytical solutions as well as to go beyond our current understanding of complex physical systems.

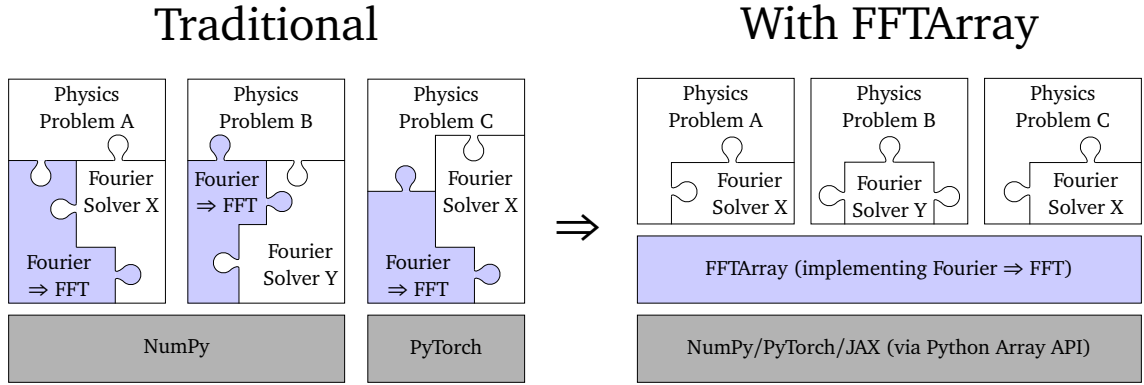
\begin{figure}
\centering
\begin{adjustbox}{width=\textwidth}
\begin{tikzpicture}
\tikzset{
  puzzle/.pic={
    \fill[scale=0.3,draw=black]
      (0,0) to[out=0,in=-45] (-0.15,0.5)
      arc(225:-45:0.6 and 0.6)
      to[out=-135,in=180] (0.6,0);
  }
}

\newcommand{\puzzle}[3]{
  pic[rotate=#1,yscale=#2,fill=#3]{puzzle}
  ++(#1:0.18)
}

  \node [
    draw,
    rectangle,
    minimum width=6cm,
    minimum height=1.2cm,
    align=center,
    fill=black!30,
    anchor=south west
  ]
  (numpy1) at (0,0) {NumPy};

  \node [
    draw,
    rectangle,
    minimum width=2.8cm,
    minimum height=4cm,
    align=center,
    anchor=south west
  ]
  (probA1) at ($ (numpy1.north west) + (0, 0.2) $) {};

  \node[
    align=center,
    anchor=north,
  ]
  (iprobA1) at ($ (probA1.north) + (0, 0) $) {Physics\\Problem A};

  \node[
    align=center,
    anchor=south west,
  ]
  (ifouA1) at ($ (probA1.south west) + (0.2, 0) $) {Fourier\\$\Rightarrow$ FFT};

  \node[
    align=center,
    anchor=east,
  ]
  (isolvA1) at ($ (probA1.east) + (0, 0) $) {Fourier\\Solver X};

  \begin{scope}[on background layer]
    \fill[color=blue!20]
      ($(ifouA1.south east) + (0.3,0)$)
      |- (isolvA1.west |- ifouA1.north)
      |- ($(probA1.west |- iprobA1.south) + (0,-0.45)$)
      |- ($(ifouA1.south east) + (0.3,0)$);
  \end{scope}

  \draw (isolvA1.west |- ifouA1.north) -| ($(ifouA1.south east) + (0.3,0.6)$) \puzzle{270}{1}{blue!20} -- ($(ifouA1.south east) + (0.3,0.0)$);
  \draw  (isolvA1.west |- ifouA1.north) -- ++(0, 0.6) \puzzle{90}{1}{white} -- ($(isolvA1.west |- iprobA1.south) + (0,-0.45)$);
  \draw
    ($(probA1.west |- iprobA1.south) + (0,-0.45)$)
    -- ++(0.5, 0) \puzzle{0}{-1}{white}
    -- ++(1.2,0)  \puzzle{0}{1}{white}
    -- ($(probA1.east |- iprobA1.south) + (0,-0.45)$)
  ;

  \node [
    draw,
    rectangle,
    minimum width=2.8cm,
    minimum height=4cm,
    align=center,
    anchor=south east
  ]
  (probB1) at ($ (numpy1.north east) + (0, 0.2) $) {};

  \node[
    align=center,
    anchor=north,
  ]
  (iprobB1) at ($ (probB1.north) + (0, 0) $) {Physics\\Problem B};

  \node[
    align=center,
    anchor=north west,
  ]
  (ifouB1) at ($ (probB1.west |- iprobB1.south) + (0, -0.5) $) {Fourier\\$\Rightarrow$ FFT};

  \node[
    align=center,
    anchor=south east,
  ]
  (isolvB1) at ($ (probB1.south east) + (0, 0.2) $) {Fourier\\Solver Y};

  \draw  ($(probB1.west |- iprobB1.south) + (0,-0.45)$) -- ++(0.5, 0) \puzzle{0}{1}{blue!20} -- ++(1.2,0)  \puzzle{0}{-1}{white} -- ($(probB1.east |- iprobB1.south)  + (0,-0.45)$);
  \draw  ($(ifouB1.east |- iprobB1.south) + (0,-0.45)$)
  -- ++(0, -0.65) \puzzle{270}{1}{blue!20}
    -- ($ (ifouB1.east |- isolvB1.north) + (0,0.2)$)
    -- ($(isolvB1.west |- isolvB1.north) + (-0.3,0.2)$)
    -- ++(0, -0.6) \puzzle{270}{-1}{white}
    -- ($(isolvB1.west |- probB1.south) + (-0.3,0)$)
    ;

  \begin{scope}[on background layer]
    \fill[color=blue!20]
      (probB1.south west)
      |- ($(ifouB1.east |- iprobB1.south)+ (0,-0.45)$)
      |- ($ (isolvB1.west |- isolvB1.north) + (-0.3,0.2)$)
      |- (probB1.south west)
      ;
  \end{scope}

  \node [
    draw,
    rectangle,
    minimum width=2.8cm,
    minimum height=1.2cm,
    fill=black!30,
    align=center,
    anchor=south west,
  ] (pytorch1) at ($ (numpy1.south east) + (0.4, 0) $) {PyTorch};

  \node [
    draw,
    rectangle,
    minimum width=2.8cm,
    minimum height=4cm,
    align=center,
    anchor=south west
  ]
  (probC1) at ($ (pytorch1.north west) + (0, 0.2) $)  {};

  \node[
    align=center,
    anchor=north,
  ]
  (iprobC1) at ($ (probC1.north) + (0, 0) $) {Physics\\Problem C};

  \node[
    align=center,
    anchor=south west,
  ]
  (ifouC1) at ($ (probC1.south west) + (0.2, 0) $) {Fourier\\$\Rightarrow$ FFT};

  \node[
    align=center,
    anchor=east,
  ]
  (isolvC1) at ($ (probC1.east) + (0, 0) $) {Fourier\\Solver X};

  \draw  ($ (probC1.west |- iprobC1.south) + (0, -1.)$)
    -- ++(0.6, 0) \puzzle{0}{1}{blue!20}
    -- ($ (isolvC1.west |- iprobC1.south) + (0, -1.) $)
    -- ($ (isolvC1.west |- iprobC1.south) + (0, 0) $)
    -- ++(0.6, 0) \puzzle{0}{-1}{white}
    -- (probC1.east |- iprobC1.south)
    ;
  \draw
    (isolvC1.west |- ifouC1.north)
    -- ($(ifouC1.north east) + (0.3,0)$)
    -- ++(0, -0.4) \puzzle{270}{1}{blue!20}
    -- ($(ifouC1.south east) + (0.3,0)$)
    ;
  \draw  (isolvC1.west |- ifouC1.north) -- (isolvC1.west |- iprobC1.south);

  \begin{scope}[on background layer]
    \fill[color=blue!20]
      (probC1.south west)
      |- ($ (isolvC1.west |- iprobC1.south) + (0, -1.)$)
      |- ($(ifouC1.north east) + (0.3,0)$)
      |- (probC1.south west)
      ;
  \end{scope}

  \node [
    draw,
    rectangle,
    minimum width=9cm,
    minimum height=1.2cm,
    align=center,
    fill=black!30,
    anchor=south west,
  ] (arraylibs) at ($ (pytorch1.south east) + (2, 0) $) {NumPy/PyTorch/JAX (via Python Array API)};

  \node [
    draw,
    rectangle,
    fill=blue!20,
    minimum width=9cm,
    minimum height=1.2cm,
    align=center,
    anchor=south east,
  ]
  (fftarray) at ($ (arraylibs.north east) + (0, 0.2) $) {FFTArray (implementing Fourier $\Rightarrow$ FFT)};

  \node[
    align=center,
    anchor=east,
  ]
  at ($ (fftarray.north west) + (-0.5, 0.15) $) {\Huge $\Rightarrow$};

  \node [
    draw,
    rectangle,
    minimum width=2.8cm,
    minimum height=2.6cm,
    align=center,
    anchor=south west,
  ]
  (probA2) at ($ (fftarray.north west) + (0, 0.2) $) {};

  \node[
    align=center,
    anchor=north,
  ]
  (iprobA2) at ($ (probA2.north) + (0, 0) $) {Physics\\Problem A};

  \node[
    align=center,
    anchor=east,
  ]
  (isolvA2) at ($ (probA2.east) + (0, -0.75) $) {Fourier\\Solver X};

  \draw  ($ (isolvA2.west |- probA2.south) + (-0.6, 0)$)
    -- ++(0, 0.5) \puzzle{90}{-1}{white}
    -- ($ (isolvA2.west |- iprobA2.south) + (-0.6, -0.5)$)
    -- ++(1,0) \puzzle{0}{1}{white}
    -- ($ (probA2.east |- iprobA2.south) + (0, -0.5)$);

  \node [
    draw,
    rectangle,
    minimum width=2.8cm,
    minimum height=2.6cm,
    align=center,
    anchor=south,
  ]
  (probB2) at ($ (fftarray.north) + (0, 0.2) $) {};

  \node[
    align=center,
    anchor=north,
  ]
  (iprobB2) at ($ (probB2.north) + (0, 0) $) {Physics\\Problem B};

  \node[
    align=center,
    anchor=south,
  ]
  (isolvB2) at ($ (probB2.south) + (0., 0) $) {Fourier\\Solver Y};

  \draw ($(isolvB2.east |- probB2.south) + (0., 0)$)
    -- ($(isolvB2.east) + (0,-0.09)$) \puzzle{90}{-1}{white}
    -- ($ (isolvB2.east |- isolvB2.north) + (0., 0.1)$)
    -- ($(isolvB2.north) + (0.09,0.1)$) \puzzle{180}{-1}{white}
    -- ($ (isolvB2.west |- isolvB2.north) + (0, 0.1)$)
    -- ($(isolvB2.west) + (0,0.09)$) \puzzle{270}{-1}{white}
    -- ($ (isolvB2.west |- isolvB2.south) + (0, 0.)$)
  ;

  \node [
    draw,
    rectangle,
    minimum width=2.8cm,
    minimum height=2.6cm,
    align=center,
    anchor=south east,
  ]
  (probC2) at ($ (fftarray.north east) + (0, 0.2) $) {};

  \node[
    align=center,
    anchor=north,
  ]
  (iprobC2) at ($ (probC2.north) + (0, 0) $) {Physics\\Problem C};

  \node[
    align=center,
    anchor=east,
  ]
  (isolvC2) at ($ (probC2.east) + (-0.2, -0.75) $) {Fourier\\Solver X};

  \draw  ($ (isolvC2.west |- probC2.south) + (-0.2, 0)$)
    -- ++(0, 0.5) \puzzle{90}{1}{white}
    -- ($ (isolvC2.west |- iprobC2.south) + (-0.2, -0.5)$)
    -- ++(1., 0) \puzzle{0}{1}{white}
    -- ($ (probC2.east |- iprobC2.south) + (0, -0.5)$)
    ;

  \node[
    anchor=base,
  ]
  (tradLabel) at ($ (probB1.north) + (0, 0.7) $) {\Huge Traditional};

  \node[
    anchor=base,
  ]
  (newLabel) at ($ (probB2.north) + (0, 0.7) $) {\Huge With FFTArray};

\end{tikzpicture}
\end{adjustbox}
\caption{Conceptual overview of FFTArray's role in scientific software architecture.
Traditional software couples the definition of the physics problem, the Fourier-based differential equation solver and the discretization of Fourier transforms with Fast Fourier Transforms (FFTs) within a monolithic framework.
These components interact with each other via problem-specific internal APIs, creating complex interdependent codebases.
FFTArray decouples the implementation of discretized Fourier transforms from the solver and physics problem.
This architectural simplification enables researchers to focus on core physics and solver logic without managing low-level FFT implementation details, resulting in more maintainable and reusable scientific code.}
\label{fig:FFTArrayPosition}
\end{figure}

Implementing a spectral Fourier solver for a particular problem can be split into three parts (\cref{fig:FFTArrayPosition}).
At the input level, the user defines their system, representative of a concrete set of potentially time-dependent differential equations.
In the case of the Schrödinger equation, this could be the initial state and the Hamiltonian of the system.
The solver algorithm transforms these into a list of discrete steps to compute the current state of the system from its previous state.
In the case of a spectral Fourier method, like a split-step Fourier solver, each of these steps contains multiple analytical Fourier integrals.
These analytical expressions are discretized over a finite spatial volume with a finite sample spacing in order to evaluate them numerically.
This involves the translation of the analytical Fourier integrals into Fast Fourier Transforms and must take into account the concrete system, the requirements of the solver as well as the coupling between the sample spacing in position and frequency space, described by the Sampling Theorem~\cite{Shannon1949}.
These three blocks are built on top of a general numerical array library, which in the case of Python are libraries like NumPy, JAX and PyTorch~\cite{harris2020array,jax2018github,Ansel2024a}.

Existing implementations often entangle these logically distinct blocks - physical system definition, solver algorithms, and FFT-based discretization (\cref{fig:FFTArrayPosition}) - into application-specific frameworks~\cite{Antoine2014,Antoine2015,Schloss2018,Smith2022,Fioroni2024}.
This monolithic design creates three major limitations:
\begin{enumerate}
    \item \textbf{Lack of generality}: Each implementation needs to focus on a specific subset of systems in order to keep complexity in check. Examples for such limitations are only simulating a single wave function or supporting only exactly two dimensions.
    \item \textbf{Code duplication}: Due to the limited scope, one needs a new implementation for each significant change in the simulated system or used solver. Since each implementation might use special properties like symmetric grids to simplify their use of FFTs, the discretization of the analytic Fourier transforms into FFTs is re-derived and implemented.
    \item \textbf{Code complexity}: Tight coupling between implementation shortcuts and special-case logic makes the code diverge significantly from its original mathematical formulation. Any modification to one part of the system risks breaking interconnected components, requiring coordinated updates and understanding across the entire codebase.
\end{enumerate}

We have generally observed that managing these limitations presents a substantial challenge in the development of scientific simulations as systems grow in complexity.
To conclude, integrating system definitions, solvers, and Fourier transforms into a monolithic program with narrow scope makes these programs challenging to maintain, reuse or adapt to new problems.

Here, we present the Python library FFTArray, which implements the numerical approximation of a Fourier transform with an FFT on arbitrarily placed multi-dimensional grids.
Encapsulating this level cleanly allows numerical simulations to focus on the other two blocks and therefore be significantly simpler (as illustrated in \cref{fig:FFTArrayPosition}).
The design and implementation of FFTArray focuses on three goals:
\begin{enumerate}
    \item \textbf{From formulas to code}: Physicists can directly map analytical equations involving Fourier transforms to code without mixing discretization details with physics. This enables rapid prototyping of diverse physical models and solver strategies.
    \item \textbf{State-of-the-art performance}: We achieve state-of-the-art performance on Graphics Processing Units (GPUs) when solving the Schrödinger equation via split-step algorithms for large quantum systems with more than $10^{9}$ samples and more than $10^4$ time steps.
    \item \textbf{Seamless multidimensionality}: Dimensions are broadcast by name which enables a uniform API to seamlessly transition from single- to multi-dimensional systems.
\end{enumerate}

FFTArray is a pure Python library built on top of the Python Array API standard in order to be accessible and maintainable in the future.
This design ensures compatibility with the ubiquitous NumPy for smaller calculations in any environment, while also enabling GPU acceleration with JAX and PyTorch.

Despite its broad applicability, we note that FFTArray was developed in the context of simulating the atom-light interaction, dynamics and interferometry of ultracold atomic ensembles, particularly Bose-Einstein condensates (BECs)~\cite{Pethick2008}.
Thus, our examples and benchmarking focus on problems within this domain.
The time evolution of these systems is determined by a nonlinear Schrödinger equation, referred to as the Gross–Pitaevskii equation (GPE)~\cite{Gross1963,pitaevskii1961vortex,Pichery2023d}.
A concrete system can be simulated by solving this equation using the split-step method~\cite{Fitzek2020}.

This paper is organized as follows:
\Cref{sec:DiscretizeFourier} introduces the mathematical framework for approximating the continuous Fourier transform on arbitrary coordinate grids using FFTs.
It also highlights special cases of coordinate grid choices which are often chosen to simplify the use of the FFT.
\Cref{chap:TheLibrary} outlines the rationale behind the design of the FFTArray library that allows near native performance and facilitate seamless interaction with various array libraries.
In \cref{chap:Examples}, we present multiple application examples, ranging from performing a derivative to computing the quantum mechanical ground state of a coupled two-species mixture of Bose-Einstein condensates confined in a harmonic potential.
We analyze the computational precision of FFTArray by comparing it with the analytical solution of the single-species isotropic quantum harmonic oscillator.
\Cref{chap:performance} showcases the performance of FFTArray using NumPy and JAX on various central processing units (CPUs) and GPUs.
This chapter also serves as a starting guide for developing efficient implementations of different solver algorithms based on FFTArray.

\section{Discretization of the Fourier Transform}\label{sec:DiscretizeFourier}

\first{The Fourier transform converts an analytical function from its original domain $x$ into its representation in the domain of the conjugate variable $f$, which we will refer to as the frequency domain~\cite{Stein2003}.}
The unit of $f$ is given by the inverse of the unit of the variable $x$ in the original domain $\left[f\right] = \left[ x \right] ^{-1}$.
For a function $g(x): \mathbb{R} \to \mathbb{C}$ (including $\mathbb{R} \to \mathbb{R}$), defined in the original domain, the Fourier transform $\fou$ returns a complex-valued function $G(f): \mathbb{R} \to \mathbb{C}$ in the frequency domain.
This function gives the amplitudes and phases for all frequencies $f$ which make up the original function.
The inverse Fourier transform $\ifou$ converts a function from the frequency domain back to the original domain:
\begin{align}
    \fou&: \ G(\freqL) = \int_{-\infty}^{\infty}dx \ g(x)\ e^{- 2 \pi i \freqL\posL},\quad \forall\ \freqL\in \mathbb R \label{eq:Fourier},\\
    \ifou&: \ g(\posL) = \int_{-\infty}^{\infty}d\freqL\ G(\freqL)\ e^{2 \pi i \freqL\posL},\quad \forall\ x \in \mathbb R \label{eq:InverseFourier}.
\end{align}

The integrals in \cref{eq:Fourier,eq:InverseFourier} must exist in order for the aforementioned definitions to be valid.
For Schwartz functions $\mathcal{S}(\mathbb{R})$, the Fourier transformation defines a pointwise automorphism~\cite{Stein2003}.
Its domain can be extended to distributions~\cite{Hormander2003} and to $L^2(\mathbb{R})$ functions by defining \cref{eq:Fourier} as the limit of the image of $\mathcal{S}(\mathbb{R})$ functions that converge to a function in $L^2(\mathbb{R})$~\cite{Stein2005}.

Note, that the above definition of the Fourier transform uses the linear frequency $f$ as opposed to the circular frequency $\omega=2 \pi f$.
The FFTArray library chooses the name "position space" for the original domain, which we will use from here on.
This does not affect its applicability to Fourier transforms in time, where the position space is typically called "time domain".

\first{For many functions it is not feasible to evaluate their Fourier transform analytically.}
Beyond trivial or well-known analytical functions it is often impractical to calculate the integral in \cref{eq:Fourier} analytically.
Moreover, the sampled values of any measurement of a physical quantity or the output of a numerical algorithm do not have an analytical expression to perform the Fourier transform on.

\subsection{A Discretized Fourier Transform}
\first{For the cases where analytically evaluating the Fourier transform is not feasible, one can construct a discretized analog of the Fourier transform.}
FFTArray focuses on a regular N-point grid with $N$ equidistant samples $x_n$ in position space and $f_m$ in frequency space:
\begin{align}
    \posD &\coloneqq \posMin + \posIdx  \Delta \posL, \quad \posIdx = 0, \ldots, N-1 , \label{eq:xn} \\
    \quad \freqD &\coloneqq \freqMin + \freqIdx \Delta \freqL, \quad \freqIdx = 0, \ldots, N-1 \label{eq:km}.
\end{align}
The sample spacings $\Delta \posL, \Delta \freqL > 0$ describe the distance between two samples and $\posMin, \freqMin$ are the smallest samples in position and frequency space, respectively.

One approach to discretize the function is to sample it in either position or frequency space without any prefiltering or postfiltering:
\begin{align}
    \samplesPos &\coloneqq g(\posD) , \\
    \samplesFreq &\coloneqq G(\freqD).
\end{align}

Using these definitions, we approximate the integrals from \cref{eq:Fourier} and \cref{eq:InverseFourier} as Riemann sums leading to a general discretized Fourier transform (gdFT and gdIFT):
\begin{align}
    \samplesFreq  &= \Delta \posL \sum_{n=0}^{N-1} \samplesPos \ e^{-2 \pi i \ \freqD \posD } \quad \text{(gdFT)}, \label{eq:DiscreteFourier}\\
    \samplesPos  &= \Delta \freqL \sum_{m=0}^{N-1} \samplesFreq \ e^{2 \pi i \ \freqD \posD} \quad \text{(gdIFT)}  \label{eq:DiscreteInverseFourier}.
\end{align}

Due to the discretized position space we can only distinguish frequencies within the range $f_\text{period}=1/(\Delta x)$~\cite{Trefethen2000}.
The same holds for the discrete frequency space which then results in the following coupling between position and frequency space\footnote{\Cref{eq:dft-reciprocity} is for $\freqL$ being in the unit of cycles.
Without loss of generality it can be substituted by a variable in any unit. For example for an angular frequency $k = 2 \pi \freqL$ \cref{eq:dft-reciprocity} would be $2 \pi = N \Delta k \Delta \posL$.}:
\begin{align}
    \posL_\text{period} &\coloneqq N \Delta \posL = \frac{1}{\Delta \freqL},\\
    \freqL_\text{period} &\coloneqq N \Delta \freqL = \frac{1}{\Delta \posL},\\
    1 &= N \Delta \freqL \Delta \posL. \label{eq:dft-reciprocity}
\end{align}
With \cref{eq:dft-reciprocity} the two transforms \cref{eq:DiscreteFourier} and \cref{eq:DiscreteInverseFourier} are exact inverses to each other.

\first{We emphasize that discretizing a continuous function and the Fourier transform has several non-trivial implications.}
We highlight the essentials in the context of this work and refer the reader to refs.~\cite{Pharr2023, Trefethen2000, Oppenheim2013, Unser2000} for extended discussions.
The most central aspect of sampling is the Nyquist Shannon Sampling Theorem:
If a function has finite bandwidth $B$ (i.e. no spectral content beyond $B$), sampling with $\Delta \posL \leq 1/(2B)$ allows exact reconstruction~\cite{Shannon1949}.
Uniform sampling with $\Delta \posL$ (and infinitely many samples) replicates the continuous spectrum with period $1/(\Delta \posL)$.
If the bandwidth of the sampled function is larger than $1/(2 \Delta \posL)$, these replicas of the function's frequency spectrum overlap.
The contributions of any overlapping frequencies will appear as an additional amplitude and phase under a different "alias" frequency.
Therefore, this process is also called aliasing~\cite{Pharr2023}.
Since most functions in practice are not strictly band-limited (e.g. a Gaussian wave packet has infinite extent in both position and frequency) care has to be taken to reduce aliasing.
Depending on the application, this can be done for example by prefiltering or using a higher sampling rate.
For a more exhaustive treatment of the different trade-offs in sampling continuous functions on an N-point grid, see for example refs.~\cite{Jerri1977, Unser2000}.

\first{The grid on which the g(I)DFT in \cref{eq:DiscreteFourier,eq:DiscreteInverseFourier} operate is periodic.}
In physics simulations this property can often be utilized as periodic boundary conditions~\cite{Trefethen2000}.
On the other hand, an open domain can be approximated by zero-padding and/or using absorbing layers at the domain boundaries~\cite{Muga2004,Oppenheim2013}.

\first{Note, that these observations do not constrain the choice of the offsets $\posMin$ and $\freqMin$.}
In most band-limited cases the energy content of the function is centered at $f=0$ which makes centering the frequency space around zero the best choice.
In other cases an asymmetric window can be beneficial since it is known that only specific frequencies appear in the sampled function.
Examples for this are single-sideband modulated signals~\cite{Shannon1949} or a quantum mechanical wavefunction describing a Gaussian wave packet with a non-zero velocity.

\subsection{Implementation}
\first{The general discretized Fourier transform, \cref{eq:DiscreteFourier,eq:DiscreteInverseFourier}, could be implemented directly and then used as a discrete analog to the continuous Fourier transform.}
But in practice, there are multiple optimizations possible to ensure the best computational performance.
The most important optimization is to replace the naive sums by using Fast Fourier Transform (FFT) algorithms.
The resulting expression can then often be simplified even further which is described in \cref{sec:specialCases}.

\first{From this point on, we extend the mathematical notation by functions which take an array as the input and return an array.}
The appearance of an index like $n$ inside the argument of the function $\texttt{dft}_m \left( g_n \right)$ marks $\texttt{dft}$ as a function which takes an array in position space as an argument and returns an array in frequency space.
The subscript $n$ or $m$ denotes the space of their result.
\begin{align}
    \texttt{dft}_m \left( g_n \right)
    &\equiv
    \texttt{dft}_m \left(\left\{ g_n \right\}_{n=0}^{N-1}\right)
    \coloneqq
    \sum_{n=0}^{N-1} g_n \ \fftExp{-} ,\label{eq:DFT}
    \\
    \texttt{idft}_n \left( G_m \right)
    &\equiv
    \texttt{idft}_n \left(\left\{ G_m \right\}_{n=0}^{N-1}\right)
    \coloneqq
    \frac{1}{N} \sum_{m=0}^{N-1} G_m \ \fftExp{+}\label{eq:iDFT}.
\end{align}

The DFT and inverse DFT have multiple conventions for phase and scaling factors.
The above definitions follow NumPy~\cite{harris2020array}, the standard library of scientific computing in Python.
Evaluating these sums naively has a computational time complexity of $\mathcal{O}\left(N^2\right)$.
They can be computed more efficiently with Fast Fourier transform (FFT) algorithms in $\mathcal{O}\left(N \log N\right)$ time.

\first{To be able to calculate \cref{eq:DiscreteFourier,eq:DiscreteInverseFourier} in $\mathcal{O}\left(N \log N\right)$ steps we decompose them into separate phase factors and express them in terms of the \code{fft} and \code{ifft}:}
\begin{align}
    \text{(gdFT)} \quad \samplesFreq
    &= \Delta \posL \ \sum_{n=0}^{N-1} \samplesPos \ e^{-2 \pi i \ \left( \freqMin + \freqIdx \Delta \freqL \right) \left( \posMin + \posIdx \Delta \posL \right) } \\
    &= \Delta \posL
        \ \xZeroShift{-}
        \ \phaseCorrection{-}
        \ \sum_{n=0}^{N-1} \samplesPos\
        \ \underbrace{\dftNoNExp{-}}_{\fftExp{-}}
        \ \freqZeroShift{-}
        \\
    &= \Delta \posL
        \ \xZeroShift{-}
        \ \phaseCorrection{-}
        \ \fft \left(
            \samplesPos \ \freqZeroShift{-}
        \right), \label{eq:gdFT_with_fft}
\end{align}
\begin{align}
    \text{(gdIFT)} \quad \samplesPos
    &= \Delta \freqL \ \sum_{m=0}^{N-1} \samplesFreq \ e^{2 \pi i \ \left( \freqMin + \freqIdx \Delta \freqL \right) \left( \posMin + \posIdx \Delta \posL \right) } \\
    &= \Delta \freqL
        \ \freqZeroShift{+}
        \ \sum_{m=0}^{N-1} \samplesFreq \
        \ \underbrace{\dftNoNExp{+}}_{\fftExp{+}}
        \ \xZeroShift{+}
        \ \phaseCorrection{+} \\
    &=
        \freqZeroShift{+}
        \ \Delta \freqL \ N
        \ \ifft \left(
            \samplesFreq \ \xZeroShift{+} \ \phaseCorrection{+}
        \right) \nonumber \\
    &= \freqZeroShift{+}
        \ \ifft \left(
            \samplesFreq \ \xZeroShift{+} \ \phaseCorrection{+} / \Delta \posL
        \right). \label{eq:gdIFT_with_fft}
\end{align}
Identical colors highlight exponentials with opposite signs for the forward and backward transforms.
The grouping of the factors and the use of \cref{eq:dft-reciprocity} in the last step makes the forward and backward transform symmetric which becomes important in \cref{sec:fftarray-lazy-phase}.
All remaining phase factors only depend on at most one of the indices and can therefore be applied to the values before or after the transform in linear time $\mathcal{O}(N)$.
Therefore the complete algorithm has a runtime complexity of $\mathcal{O}\left(N \log N\right)$.

\subsection{Special Cases}\label{sec:specialCases}
\first{In many common applications it is possible to remove some of the additional phase terms while still getting correct results for that use case. }
In this section we discuss some of these special cases and the simplifications they allow.
This then also shows the use-cases for the common helper functions for FFTs \code{fftshift}, \code{ifftshift} and \code{fftfreq} as defined in NumPy~\cite{NumpyFFTModule}.
These functions implement special cases of the general phase factors which are used by FFTArray.
Therefore when using the gd(I)FT via FFTArray these functions are not needed.
To properly make the connection between FFTArray and many existing tutorials and implementations, we show in this section how the gd(I)FT can be rewritten in terms of these functions in some special cases of coordinate grids.

\first{The factors $\textcolor{brown}{\exp(\pm 2\pi i \ \posMin \ m \Delta \freqL)}$ and $\textcolor{orange}{\exp(\pm 2\pi i \ \freqMin \ n \Delta x)}$ are special cases of shifting a function.}
Multiplying a function with the factor $\exp(+2\pi i \allowbreak \ \freqMin \posL)$ in position space or $\exp(-2\pi i \allowbreak \ \posMin \freqL)$ in frequency space shifts that function in the other space by $\freqMin$ or $\posMin$ respectively.
Since \cref{eq:gdFT_with_fft,eq:gdIFT_with_fft} operate on periodic functions, these shifts are cyclic.
Any part of the function that is shifted out of the position or frequency window on one side directly reappears on the other side.
\first{If $\posMin$ or $\freqMin$ are integer multiples of $\Delta\posL$ or $\Delta\freqL$, their shifts can be replaced by cyclically shifting the values in the array.}
Any values which move beyond the end of the array are moved back to the beginning.
The functions \code{fftshift}, \code{ifftshift} implement such cyclic shifts for half the length of the domain.

\subsubsection{Symmetric Frequency Space and \texorpdfstring{$\posMin=0$}{x0=0}}
Sampling a real-valued function is a common special case that can be found in many tutorials.
Starting the sampling at $\posMin=0$ is often a natural choice like for example in a time series.
With $\posMin=0$, the gd(I)FT simplifies to:
\begin{align*}
    \text{(gdFT with $\posMin=0$)} \quad \samplesFreq &= \Delta \posL
        \ \fft \left(
            \samplesPos \ \freqZeroShift{-}
        \right),\\
    \text{(gdIFT with $\posMin=0$)} \quad \samplesPos
        &= \freqZeroShift{+}
        \ \ifft \left(
            \samplesFreq / \Delta \posL
        \right).
\end{align*}

The frequency space representation of such a function is conjugate symmetric ($G(f)=\overline{G(-f)}$) and therefore frequency space must be chosen symmetrically:
\begin{align}
    \freqCanSym = \freqCan.
\end{align}
Since $\freqCanSym$ is an integer multiple of $\Delta\freqL$, the exponential $\freqZeroShift{\pm}$ can be replaced by a simple shift of the values.
These shifts by $\freqCanSym$ are implemented in \code{fftshift} and its inverse \code{ifftshift}.
Replacing the remaining phase factors with these functions reduces the gd(I)FT to the more commonly known form:
\begin{align*}
    \text{(gdFT with $\posMin=0$, $\freqMin=\freqCanSym$)} \quad \samplesFreq &= \Delta \posL
        \ \texttt{fftshift}_m \left( \fft \left(
            \samplesPos
        \right) \right),\\
    \text{(gdIFT with $\posMin=0$, $\freqMin=\freqCanSym$)} \quad \samplesPos
        &= \ifft \left(
        \texttt{ifftshift}_m \left(\samplesFreq / \Delta \posL
        \right) \right).
\end{align*}

\subsubsection{Symmetric Position and Frequency Space}
The special case of a position and frequency space, both symmetric around zero with zero being explicitly sampled allows to replace all phase factors with \code{fftshift} and \code{ifftshift} as well.
A subtlety of this case is that $x=0$ and $\freqL=0$ need to be sampled explicitly regardless of whether $N$ is even or odd, analogous to the symmetric frequency space case:
\begin{align}
    \posCanSym = \posCan, \label{eq:symPosCan}\\
    \freqCanSym = \freqCan \label{eq:symFreqCan}.
\end{align}
Note that the samples in both position and frequency space with these choices for $\posMin$ and $\freqMin$ are not actually symmetric for even $N$.
Recalling \cref{eq:xn} and \cref{eq:km} shows that there is one more negative than positive coordinate value.
With the coordinates properly chosen as in \cref{eq:symPosCan,eq:symFreqCan} the gd(I)FT can be written as:
\begin{align*}
    \text{(gdFT with $\posMin=\posCanSym$, $\freqMin=\freqCanSym$)} \quad \samplesFreq &= \Delta \posL
        \ \texttt{fftshift}_m \left( \fft \left(
            \texttt{ifftshift}_n \left(
            \samplesPos
            \right)
        \right) \right),\\
    \text{(gdIFT with $\posMin=\posCanSym$, $\freqMin=\freqCanSym$)} \quad \samplesPos
        &=\ \texttt{fftshift}_n \left(  \ifft \left(
        \texttt{ifftshift}_m \left(\samplesFreq / \Delta \posL
        \right) \right) \right).
\end{align*}

\subsubsection{Convolution}\label{sec:chap2exConv}
\first{Another common case where parts of the gd(I)FT can be simplified is the convolution.}
The convolution has many different applications ranging from image processing over statistics to physics~\cite{Press2007,Proakis2007,Pharr2023}.
It can be expressed with (inverse) Fourier transforms via the convolution theorem:
\begin{align}
    g(x) * h(x) &= \int_{-\infty}^{\infty} g(\tau) h(x-\tau)\, d\tau\\
    &= \ifou \left\{ \fou[g(x)] \ \fou[h(x)] \right\}.
\end{align}
This enables its computation on discretized data in $\mathcal{O}\left(N \log N\right)$ via the FFT.
In the general case of arbitrary $\posMin$ and $\freqMin$, one set of the phase and scale factors in frequency space cancels out.
However, another set remains because there are two gdFTs and only one gdIFT:
\begin{align*}
g_n * h_n &= \text{gdIFT}_n\left(\text{gdFT}_m(g_n) \ \text{gdFT}_m(h_n)\right)\\
&= \freqZeroShift{+}
        \ \ifft [\\
            &\phantom{= \quad }
                \Delta \posL
                \ \xZeroShift{-}
                \ \phaseCorrection{-}
                \ \fft \left(
                    h_n \ \freqZeroShift{-}
                \right) \\
            &\phantom{= \quad}
                \Delta \posL
                \ \xZeroShift{-}
                \ \phaseCorrection{-}
                \ \fft \left(
                    g_n \ \freqZeroShift{-}
                \right)\\
            &\phantom{= \quad}
                \xZeroShift{+} \ \phaseCorrection{+} \frac{1}{\Delta \posL}\\
            &\phantom{= \ }] \\
&= \freqZeroShift{+}
    \ \ifft (\\
    &\phantom{=\quad \quad}
        \Delta \posL
        \ \xZeroShift{-}
        \ \phaseCorrection{-}
        \ \fft \left(
        g_n \ \freqZeroShift{-}
        \right)
        \ \fft \left(
            h_n \ \freqZeroShift{-}
        \right)\\
    &\phantom{= \quad}).
\end{align*}
In this case it is important to actually apply these remaining phase and scale factors correctly to get the expected result.
For special cases of $\posMin$ and $\freqMin$ these shifts can again be replaced by \code{fftshift}, \code{ifftshift} or the identity as shown in the other examples.

\subsubsection{Derivative}\label{sec:specialCaseDerivative}
\first{Computing a derivative via the Fourier transform can be viewed as a special case of the convolution.}
In this case one of the convolved functions can be constructed directly in frequency space (for details cf. \cref{sec:ex-derivative}):
\begin{align}
    \frac{d}{dx} g(x)
    = \ifou\left\{ (2\pi i \freqL) \ \fou\left\{ g(x) \right\} \right\}.
\end{align}
Discretizing this using the gd(I)FT with $\posMin=0$ removes most phase factors:
\begin{align}
    \frac{d}{dx} g_n
    &= \text{gdIFT}_n\left(2\pi i \left(\freqMin + m \Delta\freqL \right) \ \text{gdFT}_m(g_n)\right)\\
&= \freqZeroShift{+}
    \ \ifft \left(
        2\pi i \left(\freqMin + m \Delta\freqL \right)
        \ \fft \left(
            \samplesPos \ \freqZeroShift{-}
        \right)
    \right).
\end{align}
Choosing a frequency space symmetric around zero with $\freqMin=\freqCanSym$ allows to replace the remaining phase factors with \code{fftshift}.
The result can be simplified to only require \code{fftshift} once.
Since it is then only needed in the construction of the frequency space coordinates, many FFT libraries provide a helper function called \code{fftfreq} to construct them directly:
\newcommand{\ba}[1]{\left( #1 \right)}
\newcommand{\bb}[1]{\left[ #1 \right]}
\newcommand{\bc}[1]{\left\{ #1 \right\}}
\newcommand{\bd}[1]{\left( #1 \right)}

\renewcommand{\bb}[1]{\left( #1 \right)}
\renewcommand{\bc}[1]{\left( #1 \right)}
\renewcommand{\bd}[1]{\left( #1 \right)}

\newcommand{\beginFFTFreq}{ {\textcolor{orange}{e^{+2\pi i \ \textcolor{black}{\freqCanSym} \ n \Delta x}}}
    }
\begin{align*}
\text{Set}\quad \freqMin &:= \freqCanSym\\
\Rightarrow \frac{d}{dx} g_n &= \beginFFTFreq \ \ifft
    (
    \\&\phantom{=\qquad}
    2\pi i \ba{\freqCanSym +m \Delta\freqL }
    \
    \fft \ba{\samplesPos \ {\textcolor{orange}{e^{- 2\pi i \ \textcolor{black}{\freqCanSym} \ n \Delta x}}}}
    \\&\phantom{=\quad}
    )\\
&= \ifft ( \texttt{ifftshift}_m (\\
        & \phantom{=\qquad} 2\pi i \bb{\freqCanSym +m \Delta\freqL}
        \
        \texttt{fftshift}_m \bb{
            \fft \ba{\samplesPos}
        }\\
    &\phantom{=\quad}) )\\
&= \ifft ( \\
    & \phantom{= \qquad} \texttt{ifftshift}_m \bc {
        2\pi i \bb{ \freqCanSym +m \Delta\freqL }
        } \\
    & \phantom{= \qquad} \texttt{ifftshift}_m \bc {
        \texttt{fftshift}_m \bb{
            \fft \ba{\samplesPos}
        }
    } \nonumber\\
    & \phantom{= \quad} ) \nonumber\\
&= \ifft ( \\
    & \phantom{= \qquad} 2\pi i \ \texttt{ifftshift}_m \ba{
        \freqCanSym +m \Delta\freqL
        }
            \ \fft \ba{\samplesPos}
        \nonumber\\
    & \phantom{= \quad} ) \nonumber\\
&= \ifft ( \\
    & \phantom{= \qquad} 2\pi i \ \texttt{fftfreq}_m(N, \Delta \posL)
            \ \fft \ba{\samplesPos}
        \nonumber\\
    & \phantom{= \quad} ) \nonumber.
\end{align*}

\first{As shown in the examples above, the specific possible optimizations differ a lot for different use cases.}

\section{The FFTArray Library}\label{chap:TheLibrary}

\first{FFTArray enables an easy to use general discretized Fourier transform while comprehensively addressing all special cases outlined in \cref{sec:specialCases} by implementing the general discretized Fourier transforms (gd(I)FT) in \cref{eq:gdFT_with_fft,eq:gdIFT_with_fft}.}
By providing a modular toolkit for seamlessly and efficiently manipulating discretized functions in their position and frequency space representations it replaces the handling of grid-specific complexities with the common FFT library helpers \code{fftshift}, \code{ifftshift} and \code{fftfreq}.
In contrast to the few discrete shifts supported by these methods FFTArray supports arbitrarily shifted coordinate grids.

\first{Two core classes, \code{Dimension} and \code{Array}, automatically track coordinate grids and apply FFTs and phase factors where necessary.}
The \mbox{\code{Dimension}} class (\cref{sec:dim}) encapsulates the position and frequency grids of a single dimension.
The parameters of both domains can be initialized in the way most convenient for the current problem via a constraint solver.
It ensures that the constraint in \cref{eq:dft-reciprocity} is always fulfilled and $N$ is even or a power of two.
The \mbox{\code{Array}} class (\cref{sec:array}) manages multi-dimensional sampled functions, their Fourier transforms and general mathematical operations.
It stores the sample values and \code{Dimension} objects, while tracking for each axis whether it is in position space ($\samplesPos$) or frequency space ($\samplesFreq$, as formalized in \cref{sec:DiscretizeFourier}).
Transformations between domains are executed implicitly by setting the desired space for each dimension and automatically handling the required parts of the gd(I)FT.
During arithmetic operations involving \code{Array} instances dimensions are broadcast based on their names.
Both classes are immutable and all operations on an \code{Array} create a new array which reuses the values of one of its inputs if possible.

\first{To minimize computational overhead, unnecessary scale and phase factors are automatically omitted in a user-controllable and predictable manner as detailed in \cref{sec:fftarray-lazy-phase}.}
By leveraging the Python Array API standard~\cite{Meurer2023} (cf. \cref{sec:arrayApi}), FFTArray ensures portability and allows for higher performance by utilizing hardware accelerators like GPUs.

\first{Smaller code examples are given in a Read-eval-print-loop (REPL) style.}
Python expressions and code lines are preceded with a \code{>>>} and the result of that line is printed below\footnote{Whether an output is printed as well as the shown output are not necessarily identical to what one would see in the real CPython interpreter REPL. We sometimes added the output of the assigned value which is normally not printed, switched the \code{__repr__} with the \code{__str__} representation and shortened floats in order to aid understanding.}.
Variables are prefixed with \code{np_} for NumPy Arrays, \code{dim_} for \code{fa.Dimension} objects and \code{arr_} for \code{fa.Array} objects.
The import below is used in all examples in this chapter.
\begin{codeblock}
>>> import fftarray as fa
\end{codeblock}
Additionally variables defined in earlier snippets are still available for reuse in later snippets.

\subsection{The \code{Dimension} class: Defining Coordinate Grids}\label{sec:dim}
\first{FFTArray automatically handles arbitrary coordinate grids and ensures their validity such that the user can simply choose the grid best suited to the problem.}
This design avoids mistakes when defining grids which must fulfill the dependencies given in \cref{tab:varTable}.
Additionally it standardizes which variables are used to uniquely define the grid coordinates.

\first{The \code{Dimension} class represents the coordinate grids for one dimension in position and frequency space.}
It stores the name of the dimension and the numerical parameters $N$, $\Delta\posL$, $\posMin$, and $\freqMin$. $\Delta\freqL$ can be obtained via \cref{eq:dft-reciprocity} from a given $N$ and $\Delta\posL$.
\first{To initialize a \code{Dimension} from a set of parameters which does not consist of exactly $N$, $\Delta\posL$, $\posMin$, and $\freqMin$, the system of equations in \cref{tab:varTable} must be solved.}
For example, given the grid spacings $\Delta\posL$ and $\Delta\freqL$ one would have to solve \cref{eq:dft-reciprocity} for $N$.
However, this solution for $N$ might not be an integer, which means, that either $\Delta\freqL$ or $\Delta\posL$ or both parameters need to be adapted.
Moreover, specific applications may have their own particular set of constraints on the definition on the grids, e.g., the position grid may need to be symmetric around a center point $\posL_\mathrm{middle}$ or the frequency grid may need to accommodate a function with an upper band limit of $\freqL_\mathrm{max}$.

\begin{table}
    \centering
    \begin{tabular}{|p{5.7cm}|p{2.4cm}|p{5.2cm}|}
        \hline
        \textbf{Math} & \textbf{Name in Code} & \textbf{Description} \\ \hline
        $1 = N \Delta \freqL \Delta \posL, \quad N \in \mathbb{N}_+$ & \code{n} & Number of grid points \\ \hline
        $\Delta\posL$ & \code{d_pos} & Spacing between two grid points in position space. \\ \hline
        $\Delta\freqL$ & \code{d_freq} & Spacing between two grid points in frequency space. If the unit of position space is $\left[x\right]$, the unit in frequency space is its inverse $\left[f\right] = \left[ x \right] ^{-1}$. This is a rotational frequency in cycles as opposed to an angular frequency. \\ \hline
        $\posMin$ & \code{pos_min} & The smallest position grid point. \\ \hline
        $\posL_\text{max} = \posMin + (N-1) \, \Delta\posL$ & \code{pos_max} & The largest position grid point. \\ \hline
        \begin{tabular}{@{}c@{}}
        \(
            \posL_\text{middle} =
            \begin{cases}
                0.5 \, (\posMin + \posL_\text{max} + \Delta\posL),
                \\ \text{$N$ even}  \\
                0.5 \, (\posMin + \posL_\text{max}), \\
                \text{$N$ odd}
            \end{cases}
        \)
        \end{tabular}
        & \code{pos_middle} & The middle of the position grid. \\ \hline
        $\posL_\text{extent} = x_\text{max} - \posMin$ & \code{pos_extent} & The length of the position grid. Note that this is one $\Delta \posL$ smaller than the period $\posL_\mathrm{period}$.\\ \hline
        $\freqMin$ & \code{freq_min} & The smallest frequency grid point. \\ \hline
        $\freqL_\text{max} = \freqMin + (N-1) \, \Delta\freqL$ & \code{freq_max} & The largest frequency grid point. \\ \hline
        \begin{tabular}{@{}c@{}}
        \(
            \freqL_\text{middle} =
            \begin{cases}
                0.5 \, (
                 \freqMin + \freqL_\text{max} + \Delta\freqL), \\
                 \text{$N$ even}  \\
                0.5 \, (\freqMin + \freqL_\text{max}), \\
                \text{$N$ odd}
            \end{cases}
        \)
        \end{tabular}
        & \code{freq_middle} & The middle of the frequency grid. \\ \hline
        $\freqL_\text{extent} = \freqL_\text{max} - \freqMin$ & \code{freq_extent} & The length of the frequency grid. Note that this is one $\Delta \freqL$ smaller than the period $\freqL_\mathrm{period}$.\\ \hline

    \end{tabular}
    \caption{All parameters of the constraint system between position and frequency space. Math is the naming in \cref{sec:DiscretizeFourier} while Code lists the naming adopted in the actual source code. Those names where chosen such that they are independent of the name of the used dimension to properly support multi-dimensional use-cases.
    }
    \label{tab:varTable}
\end{table}

\first{FFTArray allows to use any combination of grid parameters to initialize a \code{Dimension}}.
Internally, FFTArray uses the z3 solver~\cite{deMoura2008} for solving the constraints in \cref{tab:varTable}.
This way, the user is not required to do so by hand.
For example, a grid with $N=2048$, $\posMin=-100$, $\posL_{max}=50$ that is centered in frequency space can be initialized with\footnote{The REPL outputs in this section omit the \code{dynamically_traced_coords=False} member of the \code{Dimension} class and round floats to more readable lengths for pedagogical reasons.}:
\begin{codeblock}
>>> fa.dim_from_constraints("x", n=2048, pos_min=-100., pos_max=50., freq_middle=0.)
Dimension(name='x', n=2048, d_pos=0.0733, pos_min=-100.0, freq_min=-6.823)
\end{codeblock}

\first{The user defined parameter set may not correspond to a uniquely solvable set of equations. In such cases, FFTArray guides the user to a working solution via its error messages.}
If the given constraints allow many different solutions, a \code{NoUniqueSolutionError} is thrown, suggesting additional parameters which could complete the set of constraints leading to a unique solution.
If, on the other hand, the given constraints have no solution at all, a \code{NoSolutionFoundError} suggests parameters to be removed. \first{The solver also supports cases where an exact solution would require a non-integer number of grid points.}
In this case the error message suggests which parameters could be marked as \code{loose_params} to be automatically adapted.
These parameters are then adapted such that $N$ is even or a power of two.
Rounding up means that the extent of a space is always increased and the spacing of samples decreased.
Below we give an example, where $\Delta\posL$ has been decreased such that it fits the constraints of $\Delta\posL\leq0.1$ and $N$ being a power of two:
\begin{codeblock}
>>> fa.dim_from_constraints("x", d_pos=0.1, d_freq=0.05, pos_min=-9., freq_min=-6.4, n="power_of_two", loose_params=["d_pos"])
Dimension(name='x', n=256, d_pos=0.078125, pos_min=-9.0, freq_min=-6.4)
\end{codeblock}

Finally, the grid coordinates can be output as NumPy arrays with \code{dim.values(space)} or directly packed into a new \code{Array} via \code{fa.coords_from_dim(dim, space)} (cf. \cref{sec:initalization}).

\subsection{The \code{Array} class: Managing Values in Position and Frequency Space}\label{sec:array}
\first{The \code{Array} class streamlines the handling of the gd(I)FT by automatically managing dimension-wise scale and phase factors in both, position and frequency space.}
This eliminates the need for the user to manually track coordinate grid compatibility and apply factors when combining multidimensional arrays with each other.

\first{In contrast to a naive implementation like functions acting directly on NumPy arrays, which would redundantly apply scale and phase factors even when unnecessary, the Array class avoids such inefficiencies by tracking the current space of each Dimension and enabling transformations between spaces (see \cref{sec:transforms})}.
It automatically applies the correct factors only when required and elides them during arithmetic operations where possible (see \cref{sec:fftarray-lazy-phase,sec:fftarray-math}).
It does so by encapsulating the samples of a multidimensional function together with each  \code{Dimension}.
Correct dimension broadcasting and tracking of \code{Dimension} objects is enabled by associating each dimension with a unique name, similar to Unidata’s self-describing Common Data Model, netCDF and xarray~\cite{Brown1993, Rew1990, Hoyer2017}.

\first{As outlined in \cref{sec:arrayApi}, the operations of the Array class are built upon the Python Array API~\cite{Meurer2023} to enable portability with different array libraries and support computations on GPUs.}

\subsubsection{Initialization} \label{sec:initalization}
\first{The \code{Array} class handles the gd(I)FT by storing the function values, the \code{Dimension} object and the current space for each dimension.}
It can be initialized in various ways depending on the use case; here we highlight the two most important ones.
For a complete list of initialization functions we refer to the API reference in the documentation~\cite{fftarrayDocsCreationFunctions}.
The most common way to initialize an \code{Array} is to directly fill it with the coordinate values of a \code{Dimension}:
\begin{codeblock}
>>> dim_x: fa.Dimension = fa.dim("x", pos_min=-0.1, freq_min=0., d_pos=0.2, n=4)
Dimension(name='x', n=4, d_pos=0.2, pos_min=-0.1, freq_min=0.0)
# convert the Dimension into an Array with values given by the coordinate grid: g(x) = x
>>> arr_x: fa.Array = fa.coords_from_dim(dim_x, "pos")
<fftarray.Array (x: 2^2)> Size: 32 bytes
|dimension | space |    d     |   min    |  middle  |   max    |  extent  |
+----------+-------+----------+----------+----------+----------+----------+
|    x     |  pos  |   0.20   |  -0.10   |   0.30   |   0.50   |   0.60   |
Values<array_api_compat.numpy>:
[-0.1  0.1  0.3  0.5]
\end{codeblock}
After its initialization, the \code{Array} can be used in any arithmetic expression like for example \code{x**2}.
Alternatively, one can wrap a preexisting bare array via \code{fa.array} with \code{Dimension} objects and define its current space:
\begin{codeblock}
>>> import numpy as np
>>> dim_x = fa.dim("x", pos_min=-0.1, freq_min=0., d_pos=0.2, n=4)
Dimension(name='x', n=4, d_pos=0.2, pos_min=-0.1, freq_min=0.0)
>>> np_values = np.array([5.,6.,7.,8.]) # correspond to the coordinate defined by dims
>>> arr_pos = fa.array(np_values, [dim_x], "pos") # 1d Array in position space
<fftarray.Array (x: 2^2)> Size: 32 bytes
|dimension | space |    d     |   min    |  middle  |   max    |  extent  |
+----------+-------+----------+----------+----------+----------+----------+
|    x     |  pos  |   0.20   |  -0.10   |   0.30   |   0.50   |   0.60   |
Values<array_api_compat.numpy>:
[5. 6. 7. 8.]
\end{codeblock}
\first{Instances of \code{Array} can hold all data types of the Python Array API Standard 2024.12 \cite{Meurer2023}}.
\subsubsection{Fourier Transforms}\label{sec:transforms}
\first{A key design goal of FFTArray is to make the user deliberately select the required position or frequency space representation rather than explicitly executing transforms.}
The space for each dimension can be set individually and will trigger a gd(I)FT on the internal values.
Thereby, the code automatically documents clearly which representation is used for each operation while the actual execution of the gd(I)FT becomes implicit and can be skipped if it is unnecessary.
\begin{codeblock}
arr_freq = arr_pos.into_space("freq") # change space: "pos" -> "freq"
arr_pos = arr_freq.into_space("pos") # change space: "freq" -> "pos"
arr_pos = arr_pos.into_space("pos") # No operation done because unnecessary.
\end{codeblock}
Changing the space is only possible on a floating point \code{Array}.
When performing space-changing operations, a real valued \code{Array} is automatically upcast to a complex floating point format of the same precision.
It is also possible to set different spaces per dimension.
This can be useful when mixing dimensions like time and space within the same array, which are usually not transformed simultaneously.

\subsubsection{Arithmetic Operations and Broadcasting}\label{sec:fftarray-math}
\first{FFTArray enables writing down most computations very similarly to their analytic counterparts.}
The \code{fftarray} namespace contains all element-wise functions of the Python Array API standard~\cite{Meurer2023} like \code{sin}, \code{multiply}, etc.
These implementations of arithmetic operations are also used to support all common element-wise unary and binary Python operators between \code{Array} instances as well as with scalars.
Statistical functions like \code{sum} or \code{max} work with dimension names instead of axes indices and \code{integrate} additionally uses the spacings $\Delta \posL$ and $\Delta \freqL$ stored in \code{Dimension} as integration elements.

\first{When combining multiple arrays in an arithmetic operation their values and metadata are automatically aligned and broadcast by their \code{Dimension} name.}
The coordinate grids and space of equally named dimensions must exactly match between all operands.
Because the results of an arithmetic operation differ when done in a different space there is no automatic conversion between position and frequency space.

\first{The following example showcases a combination of the aforementioned features to define a two-dimensional Gaussian function:}
\begin{codeblock}
>>> dim_x = fa.dim_from_constraints("x", pos_min=-1., pos_max=0., n=2, freq_middle=0.)
>>> dim_y = fa.dim_from_constraints("y", pos_min=-2., pos_max=1., n=4, freq_middle=0.)
>>> arr_x = fa.coords_from_dim(dim_x, "pos")
>>> arr_y = fa.coords_from_dim(dim_y, "pos")
>>> arr_gauss_2d = fa.exp(-(arr_x**2 + arr_y**2)/0.2) # same width along x and y, centered around (x,y)=(0,0)
<fftarray.Array (x: 2^1, y: 2^2)> Size: 64 bytes
|dimension | space |    d     |   min    |  middle  |   max    |  extent  |
+----------+-------+----------+----------+----------+----------+----------+
|    x     |  pos  |   1.00   |  -1.00   | 0.00e+00 | 0.00e+00 |   1.00   |
|    y     |  pos  |   1.00   |  -2.00   | 0.00e+00 |   1.00   |   3.00   |
Values<array_api_compat.numpy>:
[[1.389e-11 4.540e-05 6.738e-03 4.540e-05]
 [2.061e-09 6.738e-03 1.000e+00 6.738e-03]]
\end{codeblock}
Note that for all operations, the \code{Dimension} objects are properly stored in the resulting \code{Array} so that the Gaussian array can be easily transformed from position into frequency space with \code{arr_gauss_2d.into_space("freq")}.

\subsubsection{Indexing} \label{sec:indexing}
\first{Indexing is supported via index or coordinate in any of the dimensions, similar to xarray.}
\begin{codeblock}
# Select the point in the middle of both x and y direction.
>>> arr_gauss_2d.sel({"x": dim_x.pos_middle, "y": dim_y.pos_middle}, method="nearest")
<fftarray.Array (x: 2^0, y: 2^0)> Size: 8 bytes
|dimension | space |    d     |   min    |  middle  |   max    |  extent  |
+----------+-------+----------+----------+----------+----------+----------+
|    x     |  pos  |   1.00   | 0.00e+00 | 0.00e+00 | 0.00e+00 | 0.00e+00 |
|    y     |  pos  |   1.00   | 0.00e+00 | 0.00e+00 | 0.00e+00 | 0.00e+00 |
Values<array_api_compat.numpy>:
[[1.]]
\end{codeblock}

When slicing, the \code{Dimension}s are sliced as well and automatically used in the resulting \codeNoBreak{Array}.
Slicing off a part of position space increases the value of \code{d_freq}, and conversely, slicing off a part of frequency space increases the value of \code{d_pos} due to their reciprocal relationship in \cref{eq:dft-reciprocity}.

\begin{codeblock}
# Select the first 3 points in y dimension.
# The Dimension object is automatically adjusted to correctly cover the selected points.
>>> arr_gauss_2d.isel({"y": slice(0,3)})
<fftarray.Array (x: 2^1, y: 3)> Size: 48 bytes
|dimension | space |    d     |   min    |  middle  |   max    |  extent  |
+----------+-------+----------+----------+----------+----------+----------+
|    x     |  pos  |   1.00   |  -1.00   | 0.00e+00 | 0.00e+00 |   1.00   |
|    y     |  pos  |   1.00   |  -2.00   |  -1.00   | 0.00e+00 |   2.00   |
Values<array_api_compat.numpy>:
[[1.389e-11 4.540e-05 6.738e-03]
 [2.061e-09 6.738e-03 1.000e+00]]
\end{codeblock}
\first{Indexing with sub steps is not supported because the domain of the other space could be adjusted in multiple ways.}
Increasing the step size in one space reduces the extent of the other space.
This extent reduction could be done on either end of the space or on both ends in different ways.
Since there is no sensible default for this reduction, sub steps are not supported.

\subsection{Lazy Phase Factor Application}\label{sec:fftarray-lazy-phase}

\first{One of the major design goals for FFTArray is to achieve high computational performance by avoiding unnecessary computations, particularly the application of scale and phase factors during gd(I)FTs.}
FFTArray can skip applying these factors if they would cancel out during back-and-forth transformations.
The user-accessible values are always given in the representation with all phase and scale factors applied. 
Skipping certain factors in-between operations avoids floating-point inaccuracies which could otherwise accumulate if the factors were actually applied and reversed on each transform.
Users retain control over this behavior as part of the programming model: if desired, they can explicitly influence or disable it.
This explicit control ensures predictable outcomes, as opposed to relying on automatic optimizations, which may vary unpredictably across program versions or configurations.

\first{In order to automate the scale and phase factor application while still ensuring user control, we introduce the concept of ``lazy application''.}
After performing an (i)FFT using \code{into_space}, the scale and phase factors are not directly applied but instead, the resulting \code{Array} tracks that they are missing via the flag \code{factors_applied=False}.
This flag is separate for each dimension.
All operations on \code{Array} objects in FFTArray take \code{factors_applied} into account to ensure the correct result and avoid applying these factors if possible.

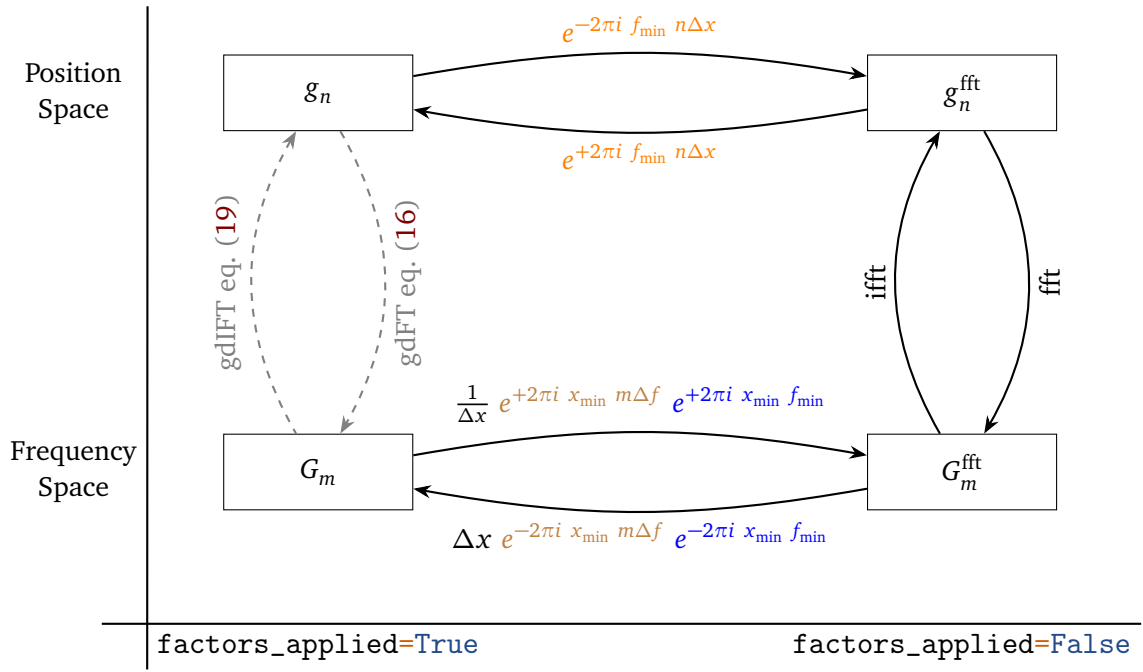
\begin{figure}[htp]
    \centering
    \begin{tikzpicture}[
          node distance=3cm and 5cm,
          box/.style={draw, rectangle, minimum width=2.5cm, minimum height=1cm},
          ->, >=Stealth
        ]

        \node[box] (A) {$\samplesPos$};
        \node[box, right=of A, xshift=1cm] (B) {$\samplesPosFFT$};
        \node[box, below=of A, yshift=-1cm] (D) {$\samplesFreq$};
        \node[box, right=of D, xshift=1cm] (C) {$\samplesFreqFFT$};

        \node[draw=none, below=of D, yshift=1.5cm] (E) {\code{factors_applied=True}};
        \node[draw=none, below=of C, yshift=1.5cm] (F) {\code{factors_applied=False}};
        \draw[-, thick] ($(E.north west) + (-0.6,0)$) -- (F.north east);

        \node[draw=none, left=of A, align=center, text width=1.7cm, xshift=4cm] (G) {Position \\ Space};
        \node[draw=none, left=of D, align=center, text width=1.7cm, xshift=4cm] (H) {Frequency \\ Space};
        \draw[-, thick] (E.south west) -- ($(G.north east) + (0,0.6)$);

        \draw[->, thick, bend left=10] (A) to node[above] {$\freqZeroShift{-}$} (B);
        \draw[<-, thick, bend right=10] (A) to node[below] {$\freqZeroShift{+}$} (B);

        \draw[->, thick, bend left=30] (B) to node[midway, left, rotate=90, transform shape, anchor=north] {fft} (C); %
        \draw[<-, thick, bend right=30] (B) to node[midway, left, rotate=90, transform shape, anchor=south] {ifft} (C); %

        \draw[gray, dashed, thick, ->, bend left=30] (A) to node[midway, left, rotate=90, transform shape, anchor=north] {\textcolor{gray}{gdFT \cref{eq:gdFT_with_fft}}} (D);
        \draw[gray, dashed, thick, <-, bend right=30] (A) to node[midway, left, rotate=90, transform shape, anchor=south] {\textcolor{gray}{gdIFT \cref{eq:gdIFT_with_fft}}} (D);

        \draw[->, thick, bend left=10] (C) to node[below] {$\Delta \posL \ \xZeroShift{-} \ \phaseCorrection{-}$} (D);
        \draw[<-, thick, bend right=10] (C) to node[above] {$\frac{1}{\Delta \posL} \ \xZeroShift{+} \ \phaseCorrection{+}$} (D);

    \end{tikzpicture}
    \caption{
        The four different internal states for the values of an FFTArray.
        By default (\code{eager=False}) each operation on the FFTArray minimizes the number of state transitions.
    }
    \label{fig:StateDiag}
\end{figure}

\first{The flag \code{factors_applied} implements a new set of states of \code{Array}s internal values on top of the current \code{space} flag, namely, $\samplesPosFFT$ and $\samplesFreqFFT$.}
The new states are required directly before computing the (I)FFT and are derived from \cref{eq:gdFT_with_fft,eq:gdIFT_with_fft} with the colored phase factors being the same.
They are called $\samplesPosFFT$ and $\samplesFreqFFT$:
\begin{align}
    \samplesPosFFT &\coloneqq\samplesPos \, \freqZeroShift{-}, \label{eq:samplesPosInt}\\
    \samplesFreqFFT &\coloneqq \samplesFreq \, \xZeroShift{+} \, \phaseCorrection{+} \, / \Delta \posL. \label{eq:samplesFreqInt}
\end{align}
With these definitions, the gd(I)FT from $\samplesPos$ to $\samplesFreq$ or vice versa can be split into three separate steps as depicted in \cref{fig:StateDiag}.
For the gdFT, the first step is multiplying the position space values $\samplesPos$ with $\freqZeroShift{-}$, which results in $\samplesPosFFT$.
The second step is performing the FFT on $\samplesPosFFT$ to compute $\samplesFreqFFT$.
In the third step, we multiply $\samplesFreqFFT$ with $\xZeroShift{+} \, \phaseCorrection{+}\, / \Delta \posL$ to get the actual frequency space values $\samplesFreq$.
The advantage of splitting this process into three individual steps is that the final transitions $\samplesPosFFT$ to $\samplesPos$ ($\samplesFreqFFT$ to $\samplesFreq$) to the states with all phase factors applied can be skipped or simplified for many operations as shown in \cref{subsec:lazyAdd,subsec:lazyMul,subsec:lazyDiv,subsec:lazyAbs}.

By default, the lazy evaluation minimizes state transitions in \code{factors_applied} in \cref{fig:StateDiag} for all \code{Array} operations.
The user can deactivate this behavior by setting the attribute \code{eager: Tuple[bool]} to \code{True} for each dimension.
In this case, the phase and scale factors are applied directly after each space change.
Combining two \code{Array} instances raises an error if their \code{eager} attributes do not match.

\first{All operations behave, up to numerical accuracy, as if they had been executed with the phase and scale factors applied.}
An \code{Array} object (\code{arr}) only allows public access to $\samplesPos$ (\code{arr.values("pos")}) and $\samplesFreq$ (\code{arr.values("freq")}).
In cases in which multiple calculations require the factors to be applied to the same \code{Array} it can improve performance to explicitly change the value of \code{factors_applied} with \code{arr = arr.into_factors_applied(True)}.

\first{In order to take advantage of the lazy phase factor application, the functions \code{abs}, \codeNoBreak{add}, \codeNoBreak{subtract}, \codeNoBreak{multiply} and \codeNoBreak{divide} use optimized code paths.}
These optimized implementations apply to both the free-standing functions in the \code{fftarray} name space as well as their unary and binary counterparts on the \code{Array} class.
\first{Apart from \code{abs}, the above functions require two \code{Array} objects $g_n$ and $h_n$ as arguments and thus need separate rules for each combination of \code{factors_applied}.}
This chapter only focuses on a single dimension since phase and scale factor application is independent for each dimension.
Additionally it only focuses on position space, since except for \code{abs} the same rules apply to frequency space just with the other phase and scale factors from \cref{eq:samplesPosInt,eq:samplesFreqInt}.

The values stored internally in an \code{Array} are given by
\begin{align}
    \samplesPosInt{g}{s} \coloneqq
   \begin{cases*}
        \samplesPos \quad & \text{if   \code{factors_applied=True}} \\
        \samplesPosFFT \quad & \text{if  \code{factors_applied=False}}  \\
    \end{cases*}
    &= g_n \, \left[\freqZeroShift{-}\right]^s,
\end{align}
using \ref{eq:samplesPosInt}.
In order to be able to write \code{factors_applied} in analytical expressions, we encode it into a variable $s$:
\begin{align}
    s \coloneqq
   \begin{cases*}
        0 & \text{if  \code{factors_applied=True}}  \\
        1 & \text{if  \code{factors_applied=False}}  \\
    \end{cases*}.
\end{align}

The correct full representation is obtained via moving the exponential to the other side:
\begin{align}
    g_n = \samplesPosInt{g}{s} \, \left[\freqZeroShift{+}\right]^s.
\end{align}

\subsubsection{Addition and subtraction}\label{subsec:lazyAdd}

Addition has two input arrays with user-facing values $g_n$ and $h_n$ which should be added to get the result $f_n := g_n + h_n$.
The derivation of the optimized rules for addition starts in terms of the two internal Array values $\samplesPosInt{g}{s_1}$ and $\samplesPosInt{h}{s_2}$ with their respective \code{factors_applied} states $s_1$ and $s_2$.
It is not always possible to completely avoid the application of scale and phase factors and still get a result of the form $\samplesPosInt{f}{s^\mathrm{res}}$ with $s^\mathrm{res} \in \{0,1\}$.
Therefore we introduce possible adjustments to each input as $s_1^\mathrm{op}, \, s_2^\mathrm{op} \in \{-1, 0, 1\}$.
These adjustments can be used to switch the representation of the values in the array object between $\samplesPos$ and $\samplesPosFFT$ before the operation.
For addition this results in:
\begin{align}
    \begin{split}
\samplesPosInt{g}{s_1}\left[\freqZeroShift{-}\right]^{s_1^\mathrm{op}} & + \samplesPosInt{h}{s_2}\left[\freqZeroShift{-}\right]^{s_2^\mathrm{op}} \\
&= g_n \left[\freqZeroShift{-}\right]^{s_1+s_1^\mathrm{op}} + h_n \left[\freqZeroShift{-}\right]^{s_2+s_2^\mathrm{op}}
\end{split}\\
    &\stackrel{!}{=} f_n \left[\freqZeroShift{-}\right]^{s^\mathrm{res}},
    \quad f_n \coloneqq g_n + h_n \\
    \Rightarrow s^\mathrm{res} &\stackrel{!}{=} s_1 + s_1^\mathrm{op} = s_2 + s_2^\mathrm{op}\label{eq:addEq}.
\end{align}

Now we need to solve \cref{eq:addEq} while keeping $s_1^\mathrm{op}=0$ and $s_2^\mathrm{op}=0$ as much as possible in order to avoid having to apply phase and scale factors to the input operands.
In the case of $s_1=s_2$ this is directly possible with $s_1^\mathrm{op}=s_2^\mathrm{op}=0$ leading to $s^\mathrm{res}=s_1=s_2$.
If $s_1 \neq s_2$, one of the two input \code{Array}s needs to be adjusted, so either $s_1^\mathrm{op} \neq 0$ or $s_2^\mathrm{op} \neq 0$.
Since the choice whether to adjust $s_1$ or $s_2$ has no performance implications, the \code{eager} attribute acts as a tie breaker.
If \code{eager=False}, $s_1^\mathrm{op}$ and $s_2^\mathrm{op}$ are chosen such that $s^\mathrm{res}=1$ which corresponds to \code{factors_applied=False}.
If \code{eager=True}\footnote{In this case \code{factors_applied=False} is pretty uncommon because it requires that the user manually changed either the \code{factors_applied} or \code{eager} attribute. But for example manually setting \code{factors_applied=False} on an \code{Array} with \code{eager=True} is a valid operation.}, the $s_1^\mathrm{op}$ and $s_2^\mathrm{op}$ are chosen such that $s^\mathrm{res}=0$ and therefore the resulting array will have \code{factors_applied=True}.
The logic outlined above is encoded in \cref{tab:addLUT}.
These lookup tables are also used to implement that logic for each operation in the actual library.
In each operation they are consulted for each dimension separately.
The above derivation can be done identically for subtraction.

\begin{table}
    \centering
    \begin{tabular}{cccccc}
        eager & $s_1$ & $s_2$ & $s_1^\mathrm{op}$ & $s_2^\mathrm{op}$ & $s^\mathrm{res}=s_1 + s_1^\mathrm{op} = s_2 + s_2^\mathrm{op}$ \\
        \midrule
        False & 1 & 1 & 0 & 0 & 1 \\
        False & 1 & 0 & 0 & 1 & 1 \\
        False & 0 & 1 & 1 & 0 & 1 \\
        False & 0 & 0 & 0 & 0 & 0 \\
        \midrule
        True & 1 & 1 & 0 & 0 & 1 \\
        True & 1 & 0 & -1 & 0 & 0 \\
        True & 0 & 1 & 0 & -1 & 0 \\
        True & 0 & 0 & 0 & 0 & 0 \\
        \bottomrule
    \end{tabular}
    \caption{
        Look-up-table for \code{add} and \code{subtract}.
        It encodes which inputs need phase and scale factors applied for each dimension.
        If possible the factors are factored out.
        \code{eager} acts as a tie breaker when any of the two inputs could be adjusted in order to get a correct result.
    }
    \label{tab:addLUT}
\end{table}

\subsubsection{Multiplication}\label{subsec:lazyMul}
\first{In the case of \code{multiply} the commutativity of complex multiplication can be exploited.}
The multiplication of two arrays $g_n$ and $h_n$ can be written as

\begin{align}
    \begin{split}
    \samplesPosInt{g}{s_1}\left[\freqZeroShift{-}\right]^{s_1^\mathrm{op}} &\times \samplesPosInt{h}{s_2}\left[\freqZeroShift{-}\right]^{s_2^\mathrm{op}} \\
    &= g_n \left[\freqZeroShift{-}\right]^{s_1+s_1^\mathrm{op}} \times h_n \left[\freqZeroShift{-}\right]^{s_2+s_2^\mathrm{op}}
    \end{split} \\
    &= (g_n \times h_n) \left[\freqZeroShift{-}\right]^{s_1+s_1^\mathrm{op} + s_2+s_2^\mathrm{op}} \\
    &\stackrel{!}{=} f_n \left[\freqZeroShift{-}\right]^{s^\mathrm{res}}, \quad f_n \coloneqq  g_n \times h_n \\
    \Rightarrow s^\mathrm{res} &\stackrel{!}{=} s_1+s_1^\mathrm{op}+s_2+s_2^\mathrm{op} \label{eq:mulEq}.
\end{align}

\Cref{tab:mulLUT} solves the resulting \cref{eq:mulEq} such that it minimizes the number of entries where $s_j^\mathrm{op} \neq 0$ and therefore also minimizes the amount of additional arithmetic.
Only the case of $s_1=s_2=1$ requires the application of additional phase factors, which we have arbitrarily chosen to apply to the second \code{Array}.

\begin{table}
    \centering
    \begin{tabular}{cccccc}
        eager & $s_1$ & $s_2$ & $s_1^\mathrm{op}$ & $s_2^\mathrm{op}$ & $s^\mathrm{res}=s_1 + s_1^\mathrm{op} + s_2 + s_2^\mathrm{op}$ \\
        \midrule
        False/True & 1 & 1 & 0 & -1 & 1 \\
        False/True & 1 & 0 & 0 & 0 & 1 \\
        False/True & 0 & 1 & 0 & 0 & 1 \\
        False/True & 0 & 0 & 0 & 0 & 0 \\
        \bottomrule
    \end{tabular}
    \caption{
        Look-up-table for \code{multiply}.
        It encodes which inputs need phase and scale factors applied for each dimension.
        Since the multiplication of the inputs commutes with the factors, they can be propagated through without applying them except in the case of two phase factors which require the application of one of them.
    }
    \label{tab:mulLUT}
\end{table}

\subsubsection{Division}\label{subsec:lazyDiv}
\first{Division is similar to multiplication but with the difference that the signs of $s_2$ and $s_2^\mathrm{op}$ in the equality condition for $s^\mathrm{res}$ are flipped:}

\begin{align}
    \frac{\samplesPosInt{g}{s_1}\left[\freqZeroShift{-}\right]^{s_1^\mathrm{op}}}{\samplesPosInt{h}{s_2}\left[\freqZeroShift{-}\right]^{s_2^\mathrm{op}}} &= \frac{g_n \left[\freqZeroShift{-}\right]^{s_1+s_1^\mathrm{op}}}{ h_n \left[\freqZeroShift{-}\right]^{s_2+s_2^\mathrm{op}}} \\
    &= \frac{g_n}{h_n} \left[\freqZeroShift{-}\right]^{s_1+s_1^\mathrm{op} - s_2-s_2^\mathrm{op}} \\
    &\stackrel{!}{=}
    f_n \left[\freqZeroShift{-}\right]^{s^\mathrm{res}}, \quad f_n \coloneqq \frac{g_n}{h_n} \\
    \Rightarrow s^\mathrm{res}& \coloneqq s_1 + s_1^\mathrm{op} - s_2 - s_2^\mathrm{op} \label{eq:divEq}.
\end{align}
Solving \cref{eq:divEq} in \cref{tab:divLUT} also requires an extra phase factor in only in one case.
Again the operand can be chosen arbitrarily, though one needs to take care to use the correct sign.

\begin{table}
    \centering
    \begin{tabular}{cccccc}
        \rule{0pt}{1.1em} eager & $s_1$ & $s_2$ & $s_1^\mathrm{op}$ & $s_2^\mathrm{op}$ & $s^\mathrm{res}=s_1 + s_1^\mathrm{op} - s_2 - s_2^\mathrm{op}$\\
        \midrule
        False/True & 1 & 1 & 0 & 0 & 0 \\
        False/True & 1 & 0 & 0 & 0 & 1 \\
        False/True & 0 & 1 & 0 & -1 & 0 \\
        False/True & 0 & 0 & 0 & 0 & 0 \\
        \bottomrule
    \end{tabular}
    \caption{
        Look-up-table for \code{divide}.
        It encodes which inputs need phase and scale factors applied for each dimension.
        Compared to \code{multiply} the signs of the second operand are flipped but still only one case needs an actual correction to implement the operation correctly.
    }
    \label{tab:divLUT}
\end{table}

\subsubsection{Absolute values}\label{subsec:lazyAbs}

\first{\code{abs(x)} removes the phase of a complex number.}
Therefore, any not yet-applied phase-factors can simply be dropped and the result will always be with \code{factors_applied=True}.
If the values are in frequency space and \code{factors_applied=False}, the scale factor $\left[\Delta \freqL N\right]^s$ needs to be applied before or after computing the absolute value of the internal values.
The frequency space scale factors $\left[\Delta \freqL N\right]^s$ are computed after the \code{abs} operation, because it is more efficient to apply them to a real-valued array instead of a complex-valued array.

\subsubsection{Showcase}
\first{The logic described in this section can be used to elide the application of phase and scale factors in a wide class of algorithms, which we demonstrate in \cref{chap:Examples}.} Below we showcase a compact implementation of these optimizations.
\begin{codeblock}
import fftarray as fa

# Compute the dimension properties.
dim_x = fa.dim_from_constraints("x", pos_min=-1., pos_max=1., n=1024, freq_middle=0.)

# Initialize the coordinate grids in position and frequency space.
# Those are real-valued and therefore have to have factors_applied=True.
# They default to eager=False.
arr_x = fa.coords_from_dim(dim_x, "pos") ß\label{code:initpos}ß# $\samplesPos$
arr_f = fa.coords_from_dim(dim_x, "freq") # $\samplesFreq$

# The result of the square with factors_applied=True is again factors_applied=True
arr_pos1 = arr_x**2 # $\samplesPos$
# Changing the space leaves the array with factors_applied=False. The factors have not been applied yet.
arr_freq1 = arr_pos1.into_space("freq") ß\label{code:lazy1}ß# $\samplesFreqFFT$

arr_freq2 = arr_freq1 * arr_f ß\label{code:lazy2}ß# $\samplesFreqFFT$, multiplication of factors_applied=False and factors_applied=True leads to factors_applied=False, no factors are actually applied during this operation.

# Because arr_freq2 is in the fft representation (G^fft_m) the ifft can be applied directly.
# Therefore this is only a call to ifft, no factors in frequency or position space necessary before the transformation.
arr_pos2 = arr_freq2.into_space("pos") # $\samplesPosFFT$
# eager=False acts as a tie breaker, so the result has factors_applied=False.
arr_freq3 = arr_freq2 + 5 ß\label{code:lazy4}ß# $\samplesFreqFFT$, if eager=True it would be $\samplesFreq$

arr_freq3 = fa.exp(arr_freq2) ß\label{code:lazy5}ß# $\samplesFreq$, factors applied before expontential function
np_arr_freq2 = arr_freq2.values("freq") ß\label{code:lazy6}ß# values of $\samplesFreq$ in a plain NumPy array

arr_freq4 = fa.abs(arr_freq2) ß\label{code:lazy7}ß# $\samplesFreq$, only scaling factors were applied since $|\samplesFreq|=\Delta\posL|\samplesFreqFFT|$
\end{codeblock}

\subsection{Python Array API}\label{sec:arrayApi}
\first{FFTArray is built on top of the Python Array API to leverage the speed-ups offered by modern hardware accelerators like GPUs.}
The specific needs of hardware accelerators for deep learning and scientific compute led in the last years to the creation of multiple new python libraries for array computing.
Each of these libraries has different trade-offs.
NumPy is almost universally available in the Python ecosystem and has low start-up and per operation overhead.
JAX and PyTorch both enable significant speed-ups on GPUs but have different designs and methods to translate a Python program to run on a GPU.
A library like FFTArray is in principle agnostic to these details and could be built on top of any of these libraries.
But the different trade-offs and histories of these libraries cause them to have different application programming interfaces (APIs).
To enable a library like FFTArray to take advantage of all of these libraries from a single source, the Python Array API standard was created.
It is developed by the Consortium for Python Data API Standards \cite{Meurer2023} and defines a common minimal set of functionality.
Adoption of this standard is facilitated by the \code{array-api-compat} library~\cite{ArrayAPICompat}.
It provides a wrapper over libraries like NumPy, PyTorch and JAX to fix any standard-violating behavior of the individual array libraries.

\first{All array operations in FFTArray, from basic arithmetic to (i)FFTs, are forwarded to the underlying library via \code{array-api-compat}.}
Every Array API compliant library provides a namespace which we will call \code{xp}.
This namespace exposes at least a standardised set of functionality like \code{xp.sin} or \code{xp.fft.fftn}.
Every arithmetic operation on an \code{Array} from direct additions to functions like \code{fa.sin} are automatically dispatched to the functions of the Array API namespace \code{xp} of the array.
As an example \code{fa.sin(arr).values("pos")} is equivalent to \code{xp.sin(arr.values("pos"))}.
This also means that \code{array-api-compat} and the wrapped array library define any not standardized behavior of FFTArray.
A notable example for not standardized behavior which is commonly used are the generally more relaxed type promotion rules.
For example \code{np.asarray(True)+2} results in 3 with NumPy although this upcasting behavior (\code{bool} to \code{int}) is not guaranteed by the standard.
When using NumPy as the backend for FFTArray this upcast is performed while with other backends it might not.

\first{The array library can be different for each individual \code{Array}.}
An \code{Array} can be initialized with values from any Array API compatible library, e.g., \code{np.ndarray}.
The Array API namespace is then automatically deduced.
The other array creation functions can optionally be given an Array API namespace.
If it is not possible to determine the used array library from the input like in the case of a list, FFTArray uses a user-configurable default namespace which itself defaults to NumPy.
The Array API namespace of any \code{Array} can also always be inspected via \code{arr.xp} and changed via \code{new_arr = arr.into_xp(xp)}.
This conversion behavior currently always goes through a NumPy array and only supports explicitly implemented libraries because it is not covered by the standard at the moment.

If the user attempts to mix multiple different namespaces, an error is thrown because it is unclear in which namespace the operation should be executed.
Therefore, the user needs to ensure that all arrays which are combined in an operation use the same underlying array library.
In the example below, we show how to set it to the \code{jax.numpy} namespace.
In this case, any operations on the arrays \code{arr_g_x_jax} or \code{arr_lin_jax} are executed by JAX.
\begin{codeblock}
>>> import jax.numpy as jnp
>>> import fftarray as fa
>>> dim_x = fa.dim_from_constraints("x", pos_min=-1., pos_max=1., n=4, freq_middle=0.)

>>> arr_g_x_jax = fa.coords_from_dim(dim_x, "pos", xp=jnp) # set namespace explicitly to jax.numpy
<fftarray.Array (x: 2^2)> Size: 16 bytes
|dimension | space |    d     |   min    |  middle  |   max    |  extent  |
+----------+-------+----------+----------+----------+----------+----------+
|    x     |  pos  |   0.67   |  -1.00   |   0.33   |   1.00   |   2.00   |
Values<jax.numpy>:
[-1.         -0.3333333   0.33333337  1.        ]

>>> arr_lin_jax = fa.array(jnp.linspace(0., 1.5, 4), dim_x, "pos")
<fftarray.Array (x: 2^2)> Size: 16 bytes
|dimension | space |    d     |   min    |  middle  |   max    |  extent  |
+----------+-------+----------+----------+----------+----------+----------+
|    x     |  pos  |   0.67   |  -1.00   |   0.33   |   1.00   |   2.00   |
Values<jax.numpy>:
[0.  0.5 1.  1.5]
\end{codeblock}

\subsubsection{JAX Tracing}
\first{The tracing feature of JAX is often required to reach high computational performance when using JAX as the \code{xp}.}
Tracing extracts the computation graph of a Python function by executing it with placeholder values that retain only the shape and data type of arrays, not their actual values.
This graph is then compiled for efficient execution, particularly on GPUs.
It can also be modified to enable features like gradient computation.
For tracing to work, JAX requires custom types (e.g. \code{Array} and \code{Dimension}) to mark which members are dynamic (replaced by placeholders) and which are static during computation.

\first{FFTArray supports JAX tracing by providing the necessary implementations.}
Users must register \code{Array} and \code{Dimension} as JAX-compatible data structures by calling \code{fa.jax_register_pytree_nodes()} before use.

\first{By default, all members of the \code{Dimension} class are marked as static during tracing.}
This inserts all member values directly into the generated computation graph, enabling efficient reuse of compiled code (e.g., in loops).
However, this prevents dynamic updates to grids during execution, as changes would require re-tracing the function.
To enable dynamic grids (e.g., for moving domains) the creation functions of \code{Dimension} have the parameter \code{dynamically_traced_coords}.
Setting it to \code{True} makes $\posMin$, $\freqMin$ and $\Delta \posL$ as well as all derived parameters except for $N$ (since JAX requires fixed shapes) dynamic at trace time.
This allows to reuse the same function for different \code{Dimension}s but comes with a restriction in usability.
Since most properties of \code{Dimension} are dynamic in this case, it cannot be checked at trace time whether one \code{Dimension} is equal to another.
Therefore if two arrays each contain a \code{Dimension} with the same name but different tracers, they cannot be combined with each other as shown below:
\begin{codeblock}
    import pytest
    import fftarray as fa
    import jax
    fa.jax_register_pytree_nodes()
    fa.set_default_xp(jax.numpy)

    dim_x = fa.Dimension("x", 4, 0.5, 0., 0., dynamically_traced_coords=True)
    
    @jax.jit
    def my_fun(dim1: fa.Dimension) -> fa.Array:
        arr1 = fa.coords_from_dim(dim1, "pos")
        arr2 = fa.coords_from_dim(dim1, "pos")

        # Works, because both arrays use the same dimension with the same tracers.
        return arr1+arr2

    my_fun(dim_x)
    
    @jax.jit
    def my_fun_not_dynamic(dim1: fa.Dimension, dim2: fa.Dimension) -> fa.Array:
        arr1 = fa.coords_from_dim(dim1, "pos")
        arr2 = fa.coords_from_dim(dim2, "pos")

        # Addition requires all dimensions with the same name to be equal, this is explicitly checked before the operation.
        # The check for equality fails with a `jax.errors.TracerBoolConversionError` because the coordinate grids' values of the `Dimension`s are only known at runtime.
        # If `dynamically_traced_coords` above were set to False, the exact values of `dim1` and `dim2` were available at trace time and therefore this addition would succeed.
        return arr1+arr2


    
    with pytest.raises(jax.errors.TracerBoolConversionError):
        my_fun_not_dynamic(dim_x, dim_x)
\end{codeblock}
This can be solved by passing each \code{Dimension} instance exactly once into the jitted function.
Note that when passing the same \code{Dimension} object as part of two different \code{FFTArray} objects, each \code{Dimension} instance gets its own distinct tracer.
For example two \code{FFTArray} objects which contain a \code{Dimension} named \code{"x"} could not be combined inside a jitted function if they were passed in as parameters.
Using \code{dynamically_traced_coords=True} requires very careful engineering of the code. Therefore, it defaults to \code{False} to cover the more common cases of static coordinate grids.

\section{Examples}\label{chap:Examples}

In this section, we demonstrate applications of FFTArray using various examples where gdFTs come into play.
\Cref{sec:ex-derivative} demonstrates how to numerically compute a derivative.
\Cref{sec:ex-gpe} describes how to use the split-step Fourier method to solve the Schrödinger equation.
This method is then used in \cref{sec:ex-bragg} to simulate a matter-wave beam splitter using Bragg diffraction.
\Cref{sec:ex-harmSingle} and \Cref{sec:ex-dualspecies} use a variation of Fourier split-step called imaginary time evolution to find the ground state of matter waves in a harmonic trap.
\Cref{sec:ex-harmSingle} implements a single species wave function without self-interaction in a two-dimensional isotropic harmonic oscillator and evaluates the precision of the solution against the precise analytic solution.
\Cref{sec:ex-dualspecies} extends that to two interacting Bose-Einstein condensates in a harmonic trap.

\subsection{Derivative}\label{sec:ex-derivative}

\first{The Fourier transform can be used to compute the n-th order derivative of a function $g(x): \allowbreak \mathbb{R} \to \mathbb{C}$ with:}
\begin{align}
    \frac{\partial^n}{\partial x^n} g(x)
    = \ifou\left\{ (2\pi i \freqL)^n \fou\left\{ g(x) \right\} \right\}.
    \label{eq:derivative}
\end{align}

Note that directly discretizing this relation as shown in this chapter is only one way to numerically compute a derivative and roughly equivalent to a highest-order difference formula.
If the signal showcases strong discontinuities including at the periodic boundaries of the sampled domain, other approaches like a lower order differencing formula can lead to better results.
Such approaches can also be implemented with an FFT by a convolution with a different kernel~\cite{Moin2010,Sunaina2018}.
We showcase the implementation of \cref{eq:derivative} with a modulated Gaussian and both its analytic and numeric derivative:
\begin{align}
    g(x) &=
    \cos(x) \ e^{\frac{-(x-1.25)^2}{25}},
    \label{eq:derivative-func}\\
    \frac{\partial}{\partial x} g(x) &=
    \left(\frac{-2 (x-1.25)}{25} \ \cos(x) - \sin(x) \right) \ e^{\frac{-(x-1.25)^2}{25}}.
    \label{eq:derivative-func-diff}
\end{align}
This test function and the x grid in the example code are both not symmetric around zero in order to show a general case where the the phase factors cannot be simplified.
An important property of the test function is that it goes to zero on the edges of the sampled domain.
If one extends the domain on both sides to get values on the boundaries even closer to zero, the precision of the derivative increases further in this case.
The analytical function and its derivative are plotted alongside their numerical counterparts in \cref{fig:derivativeTestFun}.
\begin{codeblock}
import numpy as np
import fftarray as fa

# Test function and its derivative
g = lambda x: fa.cos(x)*fa.exp(-(x-1.25)**2/25.)
g_d1 = lambda x: ((-(2*(x-1.25))/25.)*fa.cos(x) - fa.sin(x))*fa.exp(-(x-1.25)**2/25.)

dim_x = fa.dim_from_constraints("x", # dimension name
    pos_min=-40., pos_max=50., d_pos=.5, # position space grid
    freq_middle=0., # frequency grid offset
    loose_params=["d_pos"], # The resulting d_pos in dim_x will be made smaller than the input d_pos such that N is a power of two.
)
x = fa.coords_from_dim(dim_x, "pos") # position space coordinate grid
f = fa.coords_from_dim(dim_x, "freq") # frequency space coordinate grid
sampled_fn = g(x) # sample the function in position space

# Compute the derivative
order = 1 # Order of the derivative
derivative_kernel = (2*np.pi*1.j*f)**order
g_d1_numeric = (sampled_fn.into_space("freq")*derivative_kernel).into_space("pos") ß\label{code:derivativeKernelApplication}ß

# Compute the expected result directly from the analytic derivative.
d1_analytic = g_d1(x)

# Compare the numeric and analytical result.
# In this example with these domains they are equal to at least eleven decimal digits.
np.testing.assert_array_almost_equal(g_d1_numeric.values("pos"), d1_analytic.values("pos"), decimal=11) ß\label{code:DerivativeAssert}ß
\end{codeblock}
The \code{assert} in line \ref{code:DerivativeAssert} shows that this example is precise to up to 11 decimal digits.
If the zero padding on the sides is chosen larger, the precision can also be higher.
In this example the cancellation of the phase factors (reduced to \code{fftshift} and \code{ifftshift}) in \cref{sec:specialCaseDerivative} happens automatically in line \ref{code:derivativeKernelApplication}.

\begin{figure}
    \centering
    \begin{subfigure}{0.48\textwidth}
        \includegraphics[width=\linewidth]{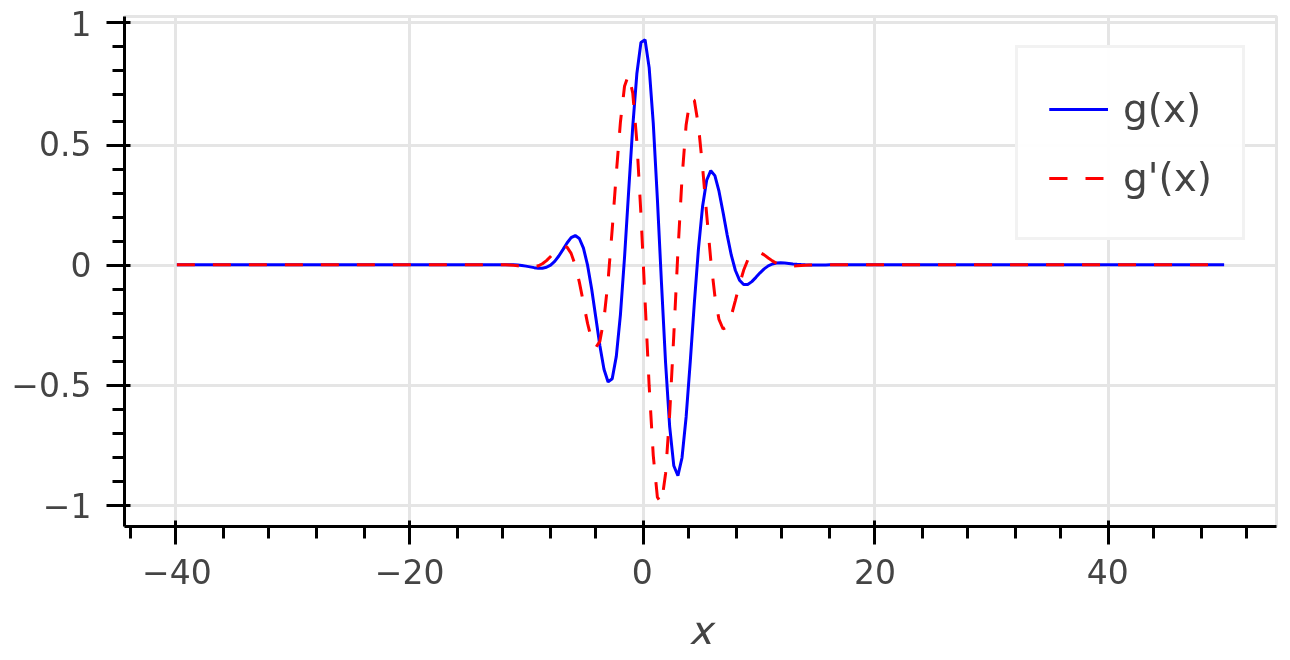}
    \end{subfigure}\hfill
    \begin{subfigure}{0.48\textwidth}
        \includegraphics[width=\linewidth]{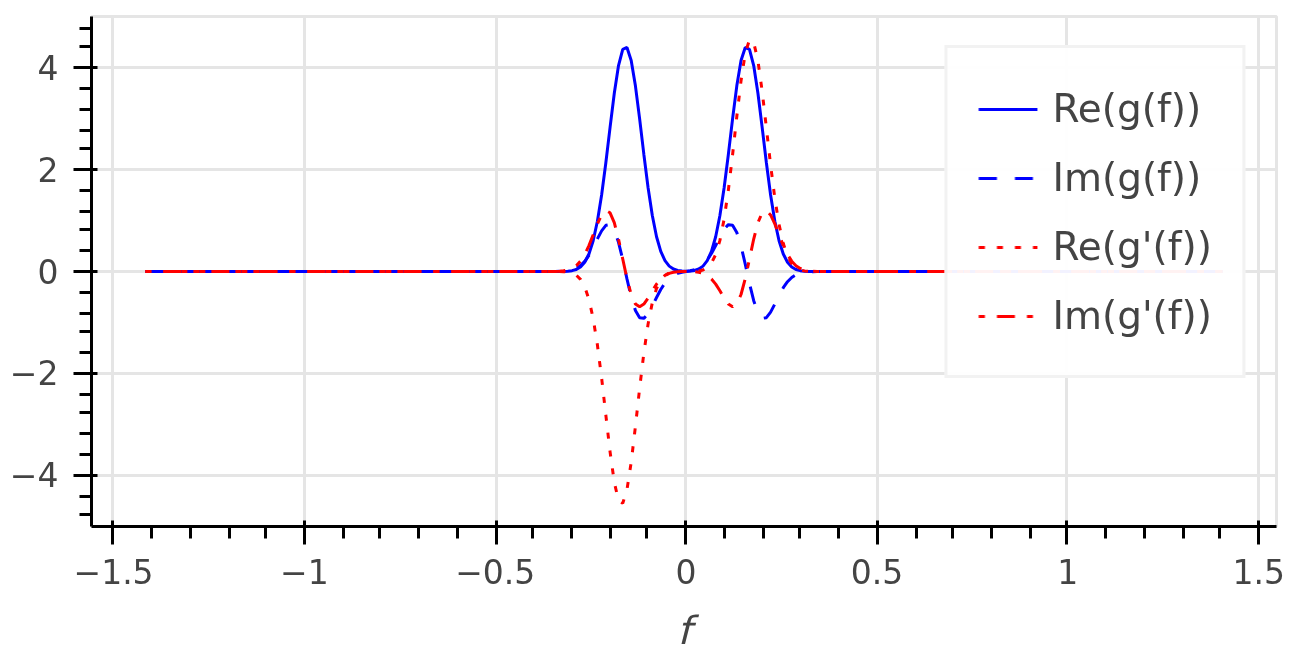}
    \end{subfigure}
    \caption{For illustration purposes a plot of the example function, \cref{eq:derivative-func}, and its first derivative, \cref{eq:derivative-func-diff}, used to demonstrate differentiation using the gdFT.
    The left plot is in position space and the right plot in frequency space.
    Both are plotted in the exact domain and sample density used in the example code.
    }
    \label{fig:derivativeTestFun}
\end{figure}

\subsection{Solving the Schrödinger Equation}\label{sec:ex-gpe}
\first{The Schrödinger equation is the central wave equation of quantum mechanics describing the time evolution of a single particle}:
\newcommand{\vecV}[1]{\mathbf{#1}}
\newcommand{\rPos}[0]{\vecV{r}}
\newcommand{\POp}[0]{T_\rPos}
\newcommand{\VOp}[0]{V(\rPos, t)}
\newcommand{\HOp}[0]{H(\rPos, t)}
\newcommand{\dimIdx}[0]{l}
\newcommand{\expc}[1]{\exp \left( #1 \right)}
\begin{align}
    \POp &\coloneqq \frac{- \hbar^2 \nabla_\rPos^2}{2m}\\
    i\hbar\frac{\partial}{\partial t} \Psi(\rPos, t) &=\HOp \Psi(\rPos,t) \\
    &= \left (\POp + \VOp \right ) \Psi(\rPos,t)
\end{align}
where $\POp$ is the kinetic energy operator and $\Psi(\rPos, t): (\mathbb{R}^n, \mathbb{R}) \to \mathbb{C}$ represents the particle's wave function in a (possibly) time-dependent potential $\VOp: (\mathbb{R}^n, \mathbb{R}) \to \mathbb{R}$. $m$ denotes the mass of the particle and $\hbar$ is the reduced Planck constant.

\first{The solution for the time propagation of the Schrödinger equation is the evolution operator $U$:}
\begin{align}
    \Psi(\rPos, t+s) &= U(t+s,t) \Psi(\rPos, t)\\
    &=\mathcal T \expc{-{\frac{i}{\hbar}} \int_t^{t+s} dt' \, s \, H(\rPos, t')} \Psi(\rPos, t) \\
    \text{with } \mathcal T \left(A(t) B(t') \right) &=
    \begin{cases}
        A(t) \, B(t'), & \text{if }  t > t'\\
        B(t') \, A(t), & \text{if }  t < t'
    \end{cases}
\end{align}

Now we discretize the time evolution and only use very small time steps $\Delta t$ under the assumption that the Hamiltonian does not change too much over that time span.
For a fixed time $t$ we get a time-independent $H(\rPos, t)$ under which we evolve for some time step $\Delta t$:
\begin{align}
    U(t+\Delta t, t) \Psi(\rPos, t) &\approx\expc{-{\frac{i}{\hbar}} \, H(\rPos, t) \Delta t} \Psi(\rPos, t). \label{eq:solSchroedinger}
\end{align}
To achieve a longer time evolution each of these time steps is repeated multiple times to approximate the target dynamics.

\first{For many problems the analytical evaluation of this solution is impractical.}
To solve it numerically, the wave function and potential can be approximated by sampling the wave function in position space at a high resolution.
Evaluating the derivative contained in $\HOp$ would then require a finite difference approximation.
Exponentiating that finite difference approximation requires representing the resulting operator as a matrix with $\mathcal{O}(N^2)$ elements for the size $N$ of each dimension.

\first{To avoid storing and multiplying matrices we can make use of \cref{eq:derivative} and use a Fourier transform to turn the position derivative into a simple multiplication with $\mathbf{f}$}:
\begin{align}
    \fou \left(\nabla_\rPos^2 \Psi(\mathbf{r}) \right) &= (2 \pi \mathbf{f})^2 \, \Psi(\mathbf{f}).
\end{align}
This process is called diagonalization and causes the matrix representation of the operator in our discrete basis to become diagonal and the exponential of a diagonal matrix is just the exponential of each of its diagonal element.
However, the exponential also contains the potential operator.
That operator is diagonal in position space and would become a non-diagonal derivative when transformed into frequency space.
The evolution operator in \cref{eq:solSchroedinger} can be split into an approximate product of three diagonal operators with a second order Trotter approximation, also called split-step or Strang-Splitting~\cite{Feit1982,Hairer2002}:
\begin{align}
    &\expc{-{\frac{i}{\hbar}} H(\rPos,t) \Delta t} \Psi(\rPos,t) \nonumber \\
    & \phantom{\exp} = \expc{-{\frac{i}{\hbar}} V(\rPos,t) \frac{\Delta t}{2}} \,
    \expc{-{\frac{i}{\hbar}} \POp \Delta t} \,
    \expc{-{\frac{i}{\hbar}} V(\rPos,t) \frac{\Delta t}{2}} \Psi(\rPos,t) + \mathcal{O}(\Delta t^3) 
    \label{eq:SplitStepDoubleV}
\end{align}
or alternatively
\begin{align}
    &\expc{-{\frac{i}{\hbar}} H(\rPos,t) \Delta t}  \Psi(\rPos, t) \nonumber \\
    &  \phantom{\exp} =\expc{-{\frac{i}{\hbar}} \POp \frac{\Delta t}{2}} \,
    \expc{-{\frac{i}{\hbar}} V(\rPos,t) \Delta t} \, 
    \expc{-{\frac{i}{\hbar}} \POp \frac{\Delta t}{2}}  \Psi(\rPos, t) + \mathcal{O}(\Delta t^3).
    \label{eq:SplitStepDoubleT}
\end{align}
The error analysis for this method has been carried out in~\cite{Jahnke2000,Thalhammer2008,Neuhauser2009,Hansen2009,Kieri2015,An2021,Burgarth2024}.
The following will use \cref{eq:SplitStepDoubleV} since it is more efficient in \cref{sec:ex-dualspecies}.
With this approximation it is possible to make the kinetic energy operator diagonal in frequency space after a Fourier transform:
\newcommand{\momPropHalf}[0]{\expc{-i \frac{\hbar}{2m} \frac{\Delta t}{2} (2 \pi \mathbf{f})^2}}
\begin{align}
\fou \left( \POp \right) &= \frac{\hbar^2}{2m} (2 \pi \ \mathbf{f})^2, \\
\Rightarrow \fou \left(\expc{-{\frac{i}{\hbar}} \POp \frac{\Delta t}{2}} \right) &= \expc{-{\frac{i}{\hbar}} \left (\frac{\hbar^2}{2m} (2 \pi \mathbf{f})^2 \right ) \frac{\Delta t}{2}}\\
&= \momPropHalf.
\end{align}
These split operators can be applied to $\Psi(\rPos, t)$ by transforming it via the gdFT between position and frequency space in $\mathcal{O}(N \log N)$ time before applying each operator.
Therefore a full time step can be implemented with $\mathcal{O}(N \log N)$ time complexity by transforming $\Psi(\rPos, t)$ between the two spaces repeatedly:
\begin{align}
    \Psi_1(\mathbf{f}) &= \Psi_0(\mathbf{f}) \, \momPropHalf, \\
    \Psi_2(\rPos) &= \ifou(\Psi_1(\mathbf{f})) \, \expc{-{i \frac{1}{\hbar}} \Delta t \, \VOp}, \\
    \Psi_3(\mathbf{f}) &= \fou(\Psi_2(\mathbf{r})) \, \momPropHalf.
\end{align}

\first{With FFTArray these formulas can be translated almost line by line into code to implement a single split-step Fourier time step of $\Delta t$:}
\begin{codeblock}
from scipy.constants import hbar
import numpy as np
import fftarray as fa

def split_step(psi0: fa.Array, *,
               dt: float,
               mass: float,
               V: fa.Array,
            ) -> fa.Array:
    k_sq = 0.
    for dim in psi0.dims:
        # Using coords_from_arr ensures that attributes
        # like eager and xp do match the ones of psi.
        k_sq = k_sq + (2*np.pi*fa.coords_from_arr(psi0, dim.name, "freq"))**2

    psi1 = psi0.into_space("freq") * fa.exp((-1.j * hbar/(2*mass) * dt/2) * k_sq)
    psi2 = psi1.into_space("pos") * fa.exp((-1.j * hbar * dt) * V)
    psi3 = psi2.into_space("freq") * fa.exp((-1.j * hbar/(2*mass) * dt/2) * k_sq)
    return psi3
\end{codeblock}
\first{The for-loop to compute \code{k_sq} and the automatic vectorization of arithmetic expressions enable this whole snippet to automatically support multiple dimensions and lazy evaluation (see \cref{sec:fftarray-lazy-phase}) automatically skips unnecessary phase factors.}
If \code{psi0} has \code{factors_applied=False}, the whole \code{split_step} routine never applies a single set of scale and phase factors, because each operator application is only a multiplication.
Therefore, calling \code{split_step} multiple times in a loop does not have any per-step overhead while still supporting arbitrarily shifted coordinate grids. 
Every space change is just the call to \code{fft} or \code{ifft}, respectively.
The \code{into_space} function allows the user to pass in \code{psi0} in any space and with any value for \code{factors_applied}. 
Any necessary transformations are done automatically.
This and all following examples implement their calculations in SI units.

\first{The split-step method can be modified to find the lowest energy eigenstate of an arbitrary potential.}
This so-called imaginary time evolution is achieved by replacing the time step $\Delta t$ with an imaginary time step $\Delta t \mapsto - i \Delta t$ such that the time evolution operator becomes $\expc{-{\frac{1}{\hbar}} \HOp \Delta t}$.
This causes each time step to dampen eigenstates with higher eigenenergies faster than the ones with lower energies.
Since the whole wave function is dampened with every step, it would quickly become too small to be representable with the used floating point numbers.
To prevent that it has to be renormalized after each time step.

\first{We collected helper functions and definitions which are specific to quantum mechanical matter wave problems in a separate package called \code{matterwave}~\cite{matterwave}.}
It contains an implementation of the split-step algorithm, often used constants and helper functions for normalizing wave functions and calculating expectation values and kinetic energies.

\subsection{Bragg Diffraction of Matter Waves}\label{sec:ex-bragg}
\first{In this example we solve the Schrödinger equation with the split-step method from \cref{sec:ex-gpe} to simulate matter wave diffraction in a Bragg grating made of light.}
Bragg diffraction is one of the central atom optical operations in atom interferometry~\cite{Berman1997,Tino2014} to transfer several photon recoils of momentum without changing the internal state of the atoms~\cite{Muller2008,Chiow2011,Ahlers2016,Gebbe2021}.
This can be used to create a superposition of momentum states and thus is also referred to as a beam splitter.
This example implements a semi-classical model of Bragg diffraction.
The atom is described by a quantum-mechanical wave function.
Due to its high enough intensity the light field is described classically since there are always enough photons~\cite{Berman1997,Meystre2001,Grynberg2010}.
Two counter-propagating laser beams form a lattice which causes elastic scattering of matter waves.
This process does not change the internal state of the atom and the whole dynamics are described with the following Hamiltonian~\cite{Fitzek2020}:
\begin{align}
    H(x, t) = -\frac{-\hbar^2}{2m}  \frac{\partial^2}{\partial x^2} + 2 \hbar \Omega(t) \cos ^2 \left( k_L x - 2 \omega_r t \right) \label{eq:BraggH}
\end{align}
where $\Omega(t)$ is the time-dependent effective Rabi frequency.
$\Omega(t)$ is determined by the laser properties and is proportional to the intensity of the laser fields.
The mass $m$ in this example is for ${}^{87}\mathrm{Rb}$ and $k_L$ is the wavenumber of the utilized atomic transition which is in this case the D2 line with a wavelength of $\SI{780}{\nano\meter}$~\cite{Steck}.
The single-photon recoil frequency of this wavelength is then $\omega_r = \hbar k_L^2 / 2m$.
The initial state is defined to be a Gaussian with a momentum width of $0.01 \hbar k_L$.
The code reuses the \code{split_step} function of \cref{sec:ex-gpe} and the actual potential is simply the potential part of the Hamiltonian in \cref{eq:BraggH}.
Depending on the passed in ramp, this code simulates all Bragg regimes from Deep-Bragg to Raman-Nath as shown in~\cite{Fitzek2020}.
\begin{codeblock}
import numpy as np
from scipy.constants import hbar

# Rb87 mass in kg
mass_rb87: float = 86.909 * 1.66053906660e-27
# Rb87 D2 transition wavelength in m
lambda_L: float = 780 * 1e-9
# Bragg beam wave vector
k_L: float = 2 * np.pi / lambda_L
hbark: float = hbar * k_L
# Single-Photon recoil frequency
w_r = hbar * k_L**2 / (2 * mass_rb87)

def simulate_bragg(t_arr, dt: float, rabi_frequency, ramp_arr, xp=np, dtype=np.float64):
    # Returns the wave function after applying the Bragg beam potential to a Gaussian input state with 0.01 hbark initial momentum width. The beam is assumed to be spatially homogeneous.
    # t_arr: NumPy array containing the $t$ of each time step.
    # dt: Size of a time step
    # rabi_frequency: Sets the magnitude of $\Omega(t)$, determined by the concretely used atom transition and laser detuning and intensity.
    # ramp_arr: Scaling factor for $\Omega(t)$ for each time step.
    # $\Omega(t)$ = rabi_frequency*ramp_arr
    

    # Dimension for full sequence based on expected matter wave size and expansion speed
    dim_x: fa.Dimension = fa.dim_from_constraints("x",
        pos_extent = 2e-3,
        pos_middle = 0,
        freq_middle = 0.,
        freq_extent = 32. * k_L/(2*np.pi),
        loose_params = ["freq_extent"]
    )
    # Initialize array with position coordinates.
    x: fa.Array = fa.coords_from_dim(dim_x, "pos", xp=xp, dtype=dtype)

    # Initialize harmonic oscillator ground state
    sigma_p=0.01*hbark
    psi: fa.Array = (2 * sigma_p**2 / (np.pi*hbar**2))**(1./4.) * fa.exp(-(sigma_p**2 / hbar**2) * x**2)
    # Numerically normalize so that the norm is `1.` even though the tails of the Gaussian are cut off.
    psi *= fa.sqrt(1./fa.integrate(fa.abs(psi)**2))

    # For each time step, compute the potential and evolve the wave function in it
    for t, ramp in zip(t_arr, ramp_arr):
        V = rabi_frequency * ramp * 2. * hbar * fa.cos(
            k_L * x - 2. * w_r * t
        )**2
        psi: fa.Array = split_step(
            psi,
            dt=dt,
            mass=mass_rb87,
            V=V,
        )

    return psi
\end{codeblock}
Depending on the specific scientific scenario, this implementation of Bragg diffraction might need different inputs and control parameters.
Please note that different array libraries can require modifications to this code for peak performance, e.g. the JAX library requires to replace the above for loop with its \code{jax.lax.scan} function, see also \cref{chap:performance}.

\subsubsection{Raman-Nath Regime}
The Raman-Nath regime is characterized by a very short and bright pulse of a spatially symmetric beam splitter with a duration $\tau \ll \frac{1}{\sqrt{2 \Omega \omega_r}}$.
In this regime an analytical solution is available~\cite{Berman1997,Meystre2001}:
\begin{align}
    \left| g_n(t) \right|^2 = J_n^2(\Omega t) \label{eq:RNAnalytical}
\end{align}
where $g_n(t)$ is the amplitude of the $n$-th momentum state $\ket{2 n \hbar k}$ and $J_n$ the Bessel functions of first kind.
Following~\cite{Fitzek2020}, their demonstration scenario for this case with $\Omega=50 \omega_r$ and $\tau=\SI{1}{\mu s}$ with a rectangular intensity profile in time can be implemented with the code below.
\begin{codeblock}
rabi_frequency = 50*w_r
n_steps = 200
t_arr, dt = np.linspace(0., 1e-6, n_steps, retstep=True, endpoint=False)
ramp_arr = np.full(n_steps, 1.)

psi = simulate_bragg(t_arr, dt, rabi_frequency, ramp_arr)
\end{codeblock}

The results of this code are visualized in \cref{fig:ramanNath} and show very good agreement with the analytical solution.

\begin{figure}
    \centering
    \begin{subfigure}{0.48\textwidth}
        \includegraphics[width=\linewidth]{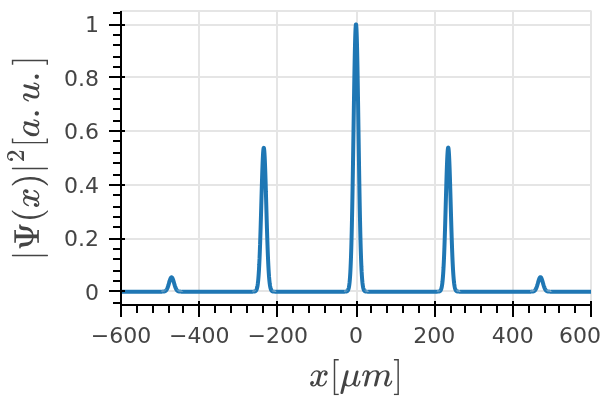}
    \end{subfigure}\hfill
    \begin{subfigure}{0.48\textwidth}
        \includegraphics[width=\linewidth]{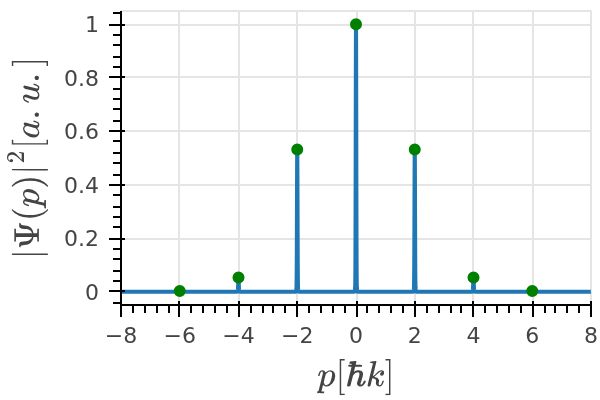}
    \end{subfigure}
    \caption{Probability density in position and momentum space after a Raman-Nath pulse with a rectangular temporal profile and $\Omega = 50 \omega_r, \tau=\SI{1}{\mu s}$ followed by $\SI{20}{ms}$ of free propagation like in~\cite{Fitzek2020}.
    The green dots mark the analytical solution from \cref{eq:RNAnalytical}.}
    \label{fig:ramanNath}
\end{figure}

\subsubsection{Bragg Regime}
The laser intensity in a two-(momentum)-level beam splitter typically has a Gaussian temporal profile to ensure broad velocity selectivity such that most atoms participate in the Rabi oscillation between the two momentum states.
The below code snippet implements a Gaussian temporal profile with $\sigma = \SI{25}{ms}$, optimized to simulate a two-level single Bragg diffraction beam splitter in 401 steps.
Special care was taken to sample the temporal profile symmetrically around $t=0$ while explicitly sampling the peak intensity.
\begin{codeblock}
# Use an odd number of steps to symmetrically sample the Gaussian
# and hit its peak at t=0 with a sample.
# Note that this snippet does not start at t=0 like above, but is symmetric around t=0.
n_steps = 401
# Rabi frequency. This specific value was found as a binary search to
# optimize a 50/50 split of the two momentum classes for this specific beam
# splitter duration and pulse form.
rabi_frequency = 25144.285917282104 # Hz
sigma_bs = 25e-6 # temporal pulse width (s)
# The Gaussian is sampled  from -4*sigma_bs to 4*sigma_bs
sampling_range_mult = 4.
t_arr, dt = np.linspace(
    start=-sampling_range_mult*sigma_bs,
    stop=sampling_range_mult*sigma_bs,
    num=n_steps,
    retstep=True,
)
# Gaussian density function
gauss = lambda t, sigma: np.exp(-0.5 * (t / sigma)**2)
# Remove the value of the Gauss at the beginning of the pulse so that
# the intensity starts and ends at zero.
gauss_offset = gauss(t = t_arr[0], sigma = sigma_bs)
ramp_arr = gauss(t = t_arr, sigma = sigma_bs) - gauss_offset

psi = simulate_bragg(t_arr, dt, rabi_frequency, ramp_arr)
\end{codeblock}

The results of this beam splitter are visualized in \cref{fig:Bragg}.
It shows that the initial atom wave function with an average momentum of $0 \hbar k$ was split cleanly into a superposition of two momentum classes of $0$ and $2 \hbar k$.
The free propagation made this momentum split also visible in position space.
\begin{figure}[htb]
    \centering
    \begin{subfigure}{0.48\textwidth}
        \includegraphics[width=\linewidth]{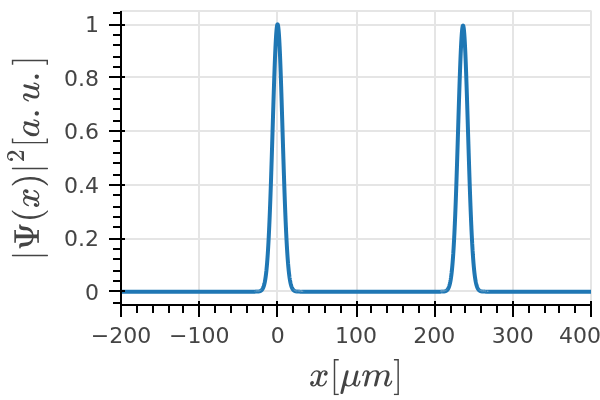}
    \end{subfigure}\hfill
    \begin{subfigure}{0.48\textwidth}
        \includegraphics[width=\linewidth]{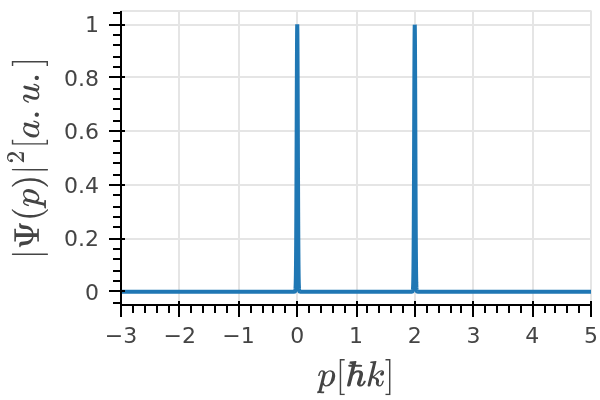}
    \end{subfigure}
    \caption{Probability density in position and momentum space of a Gaussian wave function with initial width of $\Delta p = 0.01 \hbar k$ after Bragg beam splitter with a $\sigma = \SI{25}{ms}$ Gaussian temporal profile followed by $\SI{20}{ms}$ of free propagation.
    The 50/50 split between the two momentum classes $|g,0\rangle$ and $|g,2\hbar k_L\rangle$ works very well and these ideal parameters of long smooth pulses and very sharp peaks in momentum space do not yet show visible velocity selectivity.}
    \label{fig:Bragg}
\end{figure}

\newcommand{\ndims}[0]{n}
\subsection{Finding the Ground State of the Two-Dimensional Isotropic Quantum Harmonic Oscillator}\label{sec:ex-harmSingle}
The quantum harmonic oscillator is a central model system of quantum mechanics because it can be used to approximate many other systems.
It is one of the few systems for which an exact, analytical solution for its eigenstates and eigenvectors is known.
This makes it a very good opportunity to compare the numerical precision of a simple solver based on FFTArray with its exact solution.
The isotropic quantum harmonic oscillator in $\ndims$ dimensions is defined as:
\begin{align}
    H = \frac{- \hbar^2 \nabla_\rPos^2}{2m} + \frac{1}{2} m \omega^2 \mathbf{r}^2, \quad \mathbf{r} \in {\mathbb{R}^{\ndims}}
\end{align}
In this case, the angular frequency $\omega$ of the oscillator is the same in all directions.

The solution for its ground state energy is
\begin{align}
    E_0 &= \hbar \omega \frac{\ndims}{2}.%
\end{align}

The following is a direct implementation of the imaginary time evolution described in \cref{sec:ex-gpe} for the isotropic quantum harmonic oscillator in two dimensions:
\begin{codeblock}
omega = 0.5*2.*np.pi

dim_x = fa.dim_from_constraints("x",
    pos_min=-100e-6,
    pos_max=100e-6,
    freq_middle=0.,
    n=2048,
)
y_dim = fa.dim_from_constraints("y",
    pos_min=-100e-6,
    pos_max=100e-6,
    freq_middle=0.,
    n=2048,
)

V: fa.Array = 0. # type: ignore
for dim in [dim_x, y_dim]:
    V = V + 0.5 * mass_rb87 * omega**2. * fa.coords_from_dim(dim, "pos")**2

k_sq = 0.
for dim in [dim_x, y_dim]:
    k_sq = k_sq + (2*np.pi*fa.coords_from_dim(dim, "freq"))**2

# Initialize psi as a constant function with value 1.
psi = fa.full(dim_x, "pos", 1.) * fa.full(y_dim, "pos", 1.)
for _ in range(n_steps):

    psi = psi.into_space("pos") * fa.exp((-0.5 / hbar * dt) * V)
    psi = psi.into_space("freq") * fa.exp((-1. * dt * hbar / (2*mass_rb87)) * k_sq)
    psi = psi.into_space("pos") * fa.exp((-0.5 / hbar * dt) * V)

    state_norm = fa.integrate(fa.abs(psi)**2)
    psi = psi * fa.sqrt(1./state_norm)
\end{codeblock}

\begin{figure}[htb]
    \centering
    \includegraphics[width=0.45\textwidth]{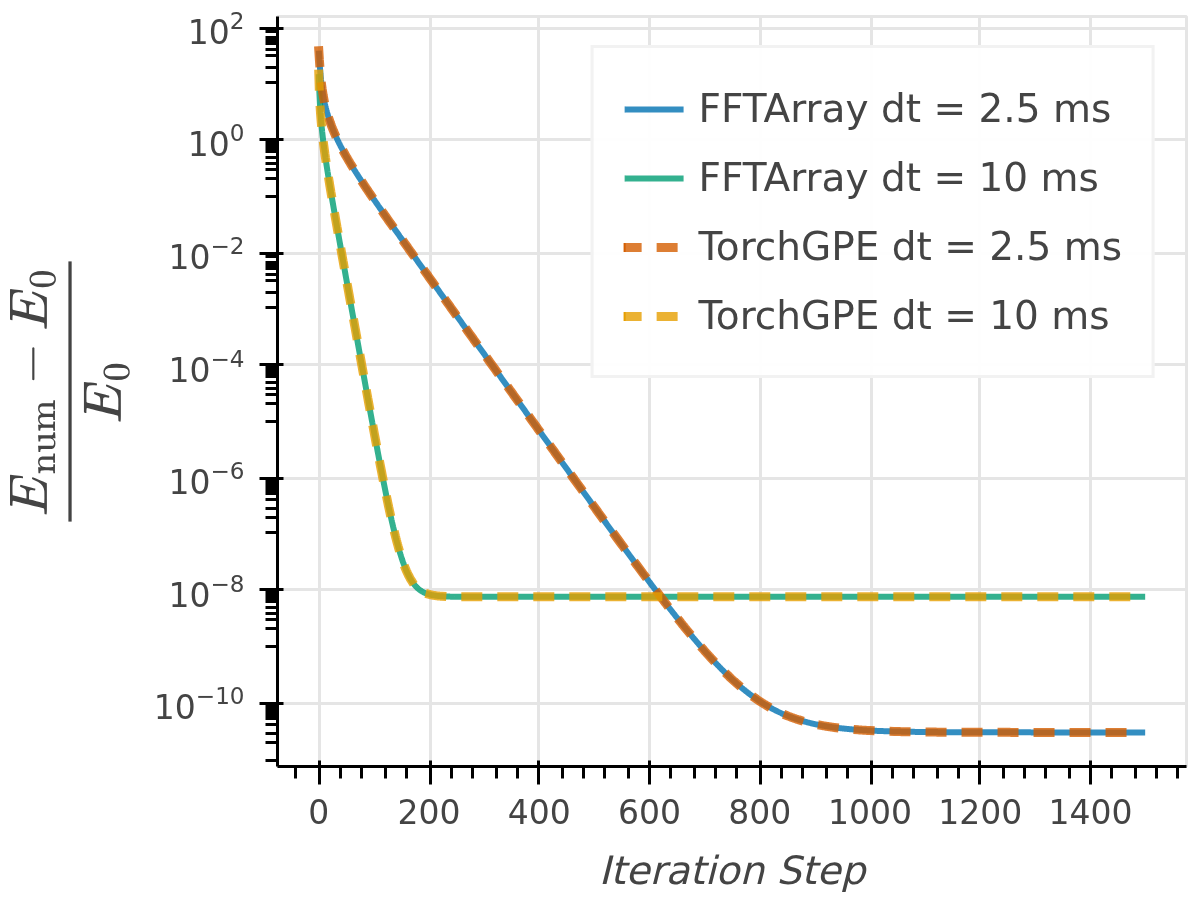}
    \caption{Absolute value of relative energy difference between the analytical and the numerical ground state of a 2D isotropic quantum harmonic oscillator for FFTArray and TorchGPE in float64.
    For both implementations, smaller time steps converge slower but are also able to more closely approach the analytic solution.
    With a time step of $dt=2.5 \, \mathrm{ms}$ both implementations reach the ground state energy with a relative error better than $10^{-9}$.}
    \label{fig:harmOsciPrecfp64}
\end{figure}

The energy of a wave function is the sum of its potential and kinetic energy:
\begin{align}
    E_\text{tot} &= E_\text{kin} + E_\text{pot},\\
    E_\text{kin} &= \frac{\hbar^2}{2 m} \int \mathrm{d^n\mathbf{f}} \, \left| \Psi(\mathbf{f}) \right|^2 (2 \pi \mathbf{f})^2,\\
    E_\text{pot} &= \frac{\hbar^2}{2 m} \int \mathrm{d^n\mathbf{r}} \, \left| \Psi(\mathbf{r}) \right|^2 V(\mathbf{r}).
\end{align}

As the metric for how well the found solution approximates the analytic solution we use the relative difference in energies between the numerical and analytical solution:
\begin{align}
    E_\mathrm{diff} &= \frac{E_\text{num} - E_0}{E_0}
\end{align}
with $E_\text{num}$ being the total energy of the numeric solution.

The resulting energies as a function of the number of time steps are shown in \cref{fig:harmOsciPrecfp64}.
Smaller time steps converge slower but are able to reach the analytical solution more precisely.
For reference we also added the results of an implementation with TorchGPE~\cite{Fioroni2024}.

\subsubsection{Single Precision Simulation (float32)}
The results in \cref{fig:harmOsciPrecfp64} are computed with double precision (float64) numbers.
However, there are only few use cases which require such high precision outside of scientific computing.
Therefore, many GPUs feature much higher single precision than double precision compute or even just single precision compute.
Examples for this are most current consumer GPUs like the NVIDIA AD102 (RTX 4090) which typically have 64 times more float32 compute than float64 compute~\cite{AdaWhitepaper}.

Each algorithm and scenario potentially require a different numerical precision.
As shown in \cref{fig:harmOsciPrecfp32}a, the precision of the result is reduced by about 4 orders of magnitude for $dt=2.5 \, \mathrm{ms}$ when doing the whole computation and evaluation in float32.
The main limit in this case is not the actual imaginary time evolution but just the evaluation of the energy of the solution.
When keeping the imaginary time evolution at float32 but computing the energy with float64 numbers, the precision is only reduced by about one order of magnitude compared to the float64 result.
Implementing such hybrid algorithms is extremely easy with FFTArray because it only requires changing the data type of the \code{psi} array before passing it to the evaluation function.

\begin{figure}
    \centering
    \begin{subfigure}{0.45\textwidth}
        \includegraphics[width=\textwidth]{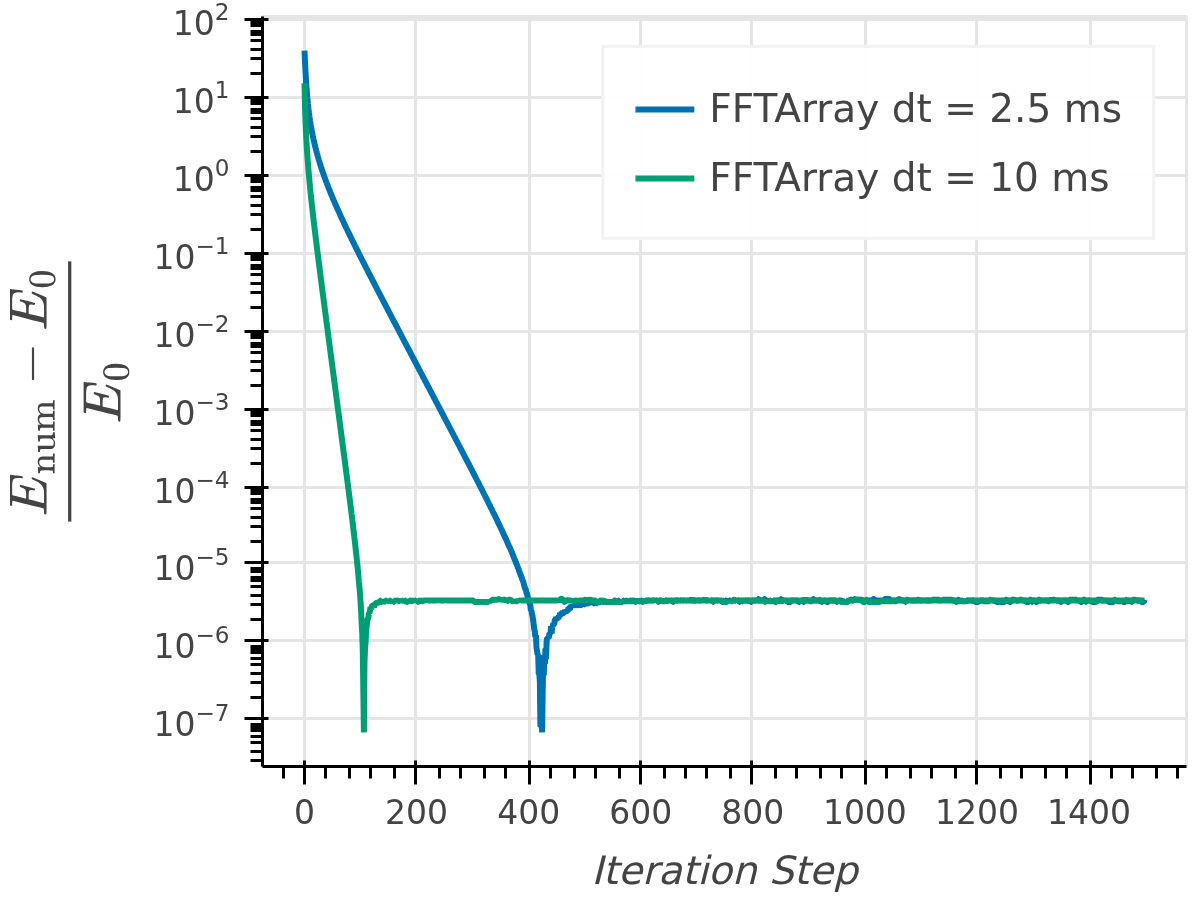}
        \caption*{(a) Imaginary time evolution and energy evaluation in float32.}
    \end{subfigure}
    \begin{subfigure}{0.45\textwidth}
        \includegraphics[width=\textwidth]{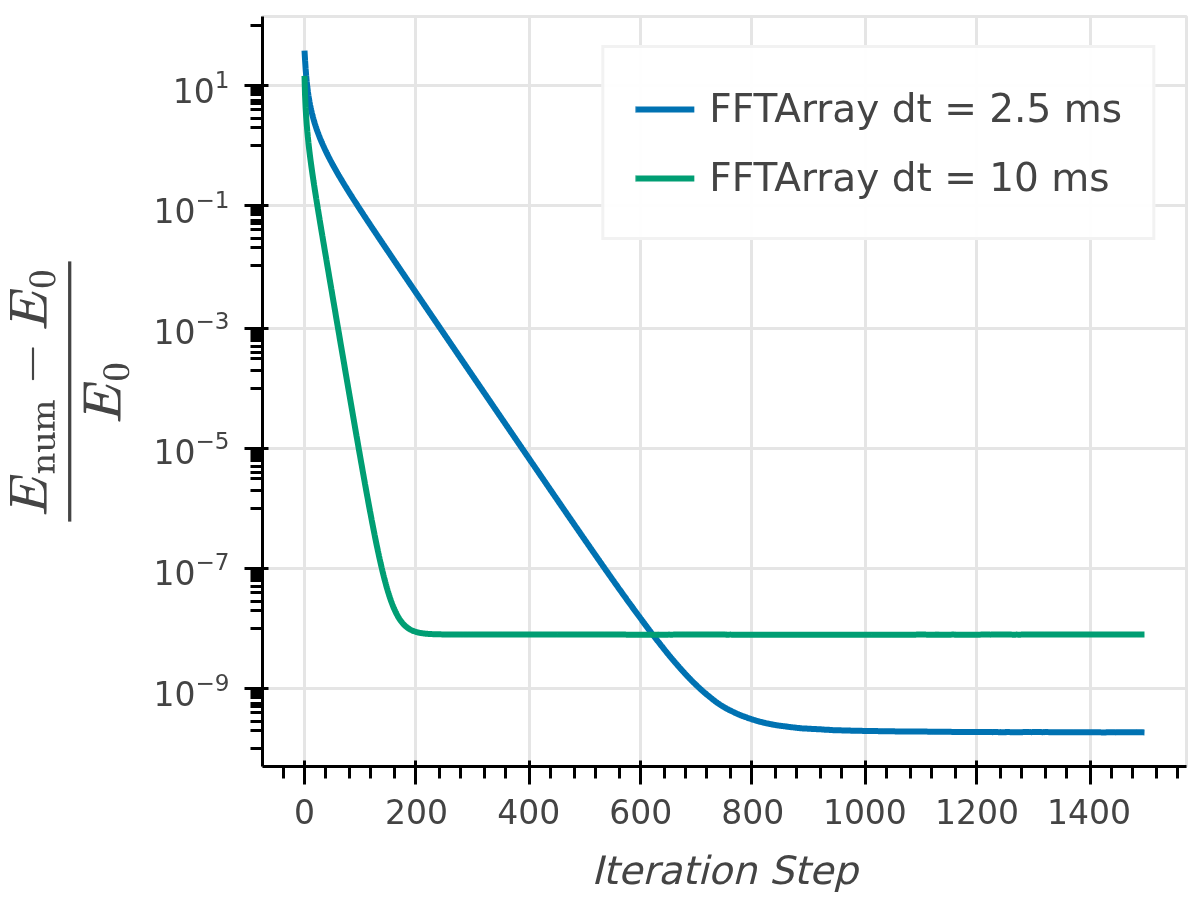}
        \caption*{(b) Imaginary time evolution in float32 and energy evaluation in float64.}
    \end{subfigure}
    \caption{Relative energy difference between the analytical and the numerical ground state of a 2D isotropic quantum mechanic oscillator in single precision (float32) with FFTArray.
    On the left, the computation of the energy of the state is also done in float32 while on the right, only the imaginary time evolution is done in float32 while the energies are evaluated in float64.
    This hybrid approach shows an improvement of three orders of magnitude.
    }
    \label{fig:harmOsciPrecfp32}
\end{figure}

\FloatBarrier

\subsection{Finding a Two-Species Ground State in a Harmonic Trap}\label{sec:ex-dualspecies}

In this example, we perform an imaginary time evolution to find the ground state of two interacting atomic species within harmonic potentials.
In particular, we describe a system of two magnetically trapped Bose Einstein condensates (BECs) of $^{87}$Rb and $^{41}$K, relevant in fundamental physics experiments~\cite{Ahlers2022, Struckmann2024}.
The traps can be described with three-dimensional anisotropic harmonic potentials with different trap frequencies for each of the two species.
Typically, these form a cigar shape with two large frequencies in two dimensions and one small frequency in the third dimension.

We describe this quantum mechanical system via a coupled Gross-Pitaevskii equation (GPE).
The GPE of a single species is a non-linear Schrödinger equation with a term that describes the self-interaction of the atoms. 
It is quantified by the scattering amplitudes $g_\text{Rb}$ and $g_\text{K}$, respectively.
The coupled GPE additionally features an additional interaction between the two different species with the scattering amplitude $g_\text{Rb,K}$.
The respective time-independent Hamiltonians for these wavefunctions are~\cite{Corgier2020,Pichery2023d}:
\begin{align}
    H_\text{Rb}(\rPos, t) &= -\frac{\hbar ^2}{2m_\text{Rb}} \nabla_\rPos^2 + V^\text{ext}_\text{Rb}\left(\rPos\right) + N_\text{Rb}\, g_\text{Rb} \left| \Psi_\text{Rb}\left(\mathbf r \right) \right|^2 + N_\text{K} \, g_\text{Rb,K} \left| \Psi_\text{K}\left(\mathbf r \right) \right|^2, \\
    H_\text{K} (\rPos, t) &= -\frac{\hbar ^2}{2m_\text{K}} \nabla_\rPos^2 + V^\text{ext}_\text{K}\left(\rPos\right) + N_\text{K}\, g_\text{K} \left| \Psi_\text{K}\left(\mathbf r \right) \right|^2 + N_\text{Rb} \, g_\text{Rb,K} \left| \Psi_\text{Rb}\left(\mathbf r \right) \right|^2,
\end{align}
\sloppy with the respective number of atoms in the BEC $N_\text{Rb,K}$, the atom mass $m_\text{Rb,K}$ and a 3-dimensional anisotropic trapping potential $V^\text{ext}_\text{Rb,K}$ like in \cref{sec:ex-harmSingle} with a frequency relation $\omega^i_\text{K} = \allowbreak \left(m_\text{Rb} / m_\text{K}\right)^{1/2} \omega^i_\text{Rb}$.

Performing the imaginary time evolution for two coupled wave functions instead of a single one can be straightforwardly implemented by adapting \ref{sec:ex-gpe}.
The total potential $V(\rPos,t)$ can be written directly as a sum of the external potential, the self-interaction and the interaction with the other species. 
This is possible as FFTArray offers full control over each time step while both, wave functions and operators, are free-standing \code{Array} objects which can be combined arbitrarily.
The following shows a single imaginary time step of the above system:

\begin{codeblock}
def imaginary_time_step_dual_species(
    psi_rb87: fa.Array,
    psi_k41: fa.Array,
    rb_potential: fa.Array,
    k_potential: fa.Array,
    dt: float,
) -> Tuple[fa.Array, fa.Array]:
    """
    Perform a single imaginary time step for the dual species GPE.
    """

    ## Calculate the potential energy operators (used for split-step and plots)
    psi_rb87 = psi_rb87.into_space("pos")
    psi_k41 = psi_k41.into_space("pos")

    psi_pos_sq_rb87 = fa.abs(psi_rb87)**2
    psi_pos_sq_k41 = fa.abs(psi_k41)**2

    self_interaction_rb87 = num_atoms_rb87 * coupling_rb87 * psi_pos_sq_rb87
    interaction_rb87_k41 = num_atoms_k41 * coupling_rb87_k41 * psi_pos_sq_k41
    V_rb87 = self_interaction_rb87 + interaction_rb87_k41 + rb_potential

    self_interaction_k41 = num_atoms_k41 * coupling_k41 * psi_pos_sq_k41
    interaction_k41_rb87 = num_atoms_rb87 * coupling_rb87_k41 * psi_pos_sq_rb87
    V_k41 = self_interaction_k41 + interaction_k41_rb87 + k_potential

    ## Imaginary time split step application

    psi_rb87 = split_step_imaginary_time(
        psi=psi_rb87,
        V=V_rb87,
        dt=dt,
        mass=m_rb87,
    )
    psi_k41 = split_step_imaginary_time(
        psi=psi_k41,
        V=V_k41,
        dt=dt,
        mass=m_k41,
    )

    return psi_rb87, psi_k41
\end{codeblock}

\begin{codeblock}
import jax.numpy as jnp
from functools import reduce

def split_step_imaginary_time(
    psi: fa.Array,
    V: fa.Array,
    dt: float,
    mass: float,
) -> fa.Array:
    """Perform an imaginary time split-step of second order in VPV configuration."""

    # Calculate half step imaginary time potential propagator
    V_prop = fa.exp((-0.5*dt / hbar) * V)
    # Calculate full step imaginary time kinetic propagator (k_sq = kx^2 + ky^2 + kz^2)
    k_sq = reduce(lambda a,b: a+b, [
        (2*np.pi * fa.coords_from_dim(dim, "freq", xp=jnp, dtype=jnp.float64))**2
        for dim in psi.dims
    ])
    T_prop = fa.exp(-dt * hbar * k_sq / (2*mass))

    # Apply half potential propagator
    psi = V_prop * psi.into_space("pos")

    # Apply full kinetic propagator
    psi = T_prop * psi.into_space("freq")

    # Apply half potential propagator
    psi = V_prop * psi.into_space("pos")

    # Normalize after step
    state_norm = fa.integrate(fa.abs(psi)**2)
    psi = psi / fa.sqrt(state_norm)

    return psi
\end{codeblock}
 
For the full code including the initialization of all functions and constants including energy tracking, we refer to the corresponding example in the repository~\cite{fftarrayDualSpeciesExample}.
\Cref{fig:DualSPeciesProbDensity} shows the final probability densities for both species.
They qualitatively match the results shown in~\cite{Pichery2023d} from where we adopted the system parameters.

Note that in this case it is beneficial to split the operator with two times the potential as in \cref{eq:SplitStepDoubleV}.
The potential has to be computed before applying the first of the split operators, in order to not degrade into an effectively first order propagation.
With the above split (\cref{eq:SplitStepDoubleV}) of the evolution operator, the wave functions are already in position space before and after each time step which allows to directly compute the self-interaction from it.
With the other split as in \cref{eq:SplitStepDoubleT}, they would be in frequency space before a time step and the potential calculation would require the execution of two additional IFFTs per time step.

\begin{figure}[htb]
    \centering
    \includegraphics[width=0.9\textwidth]{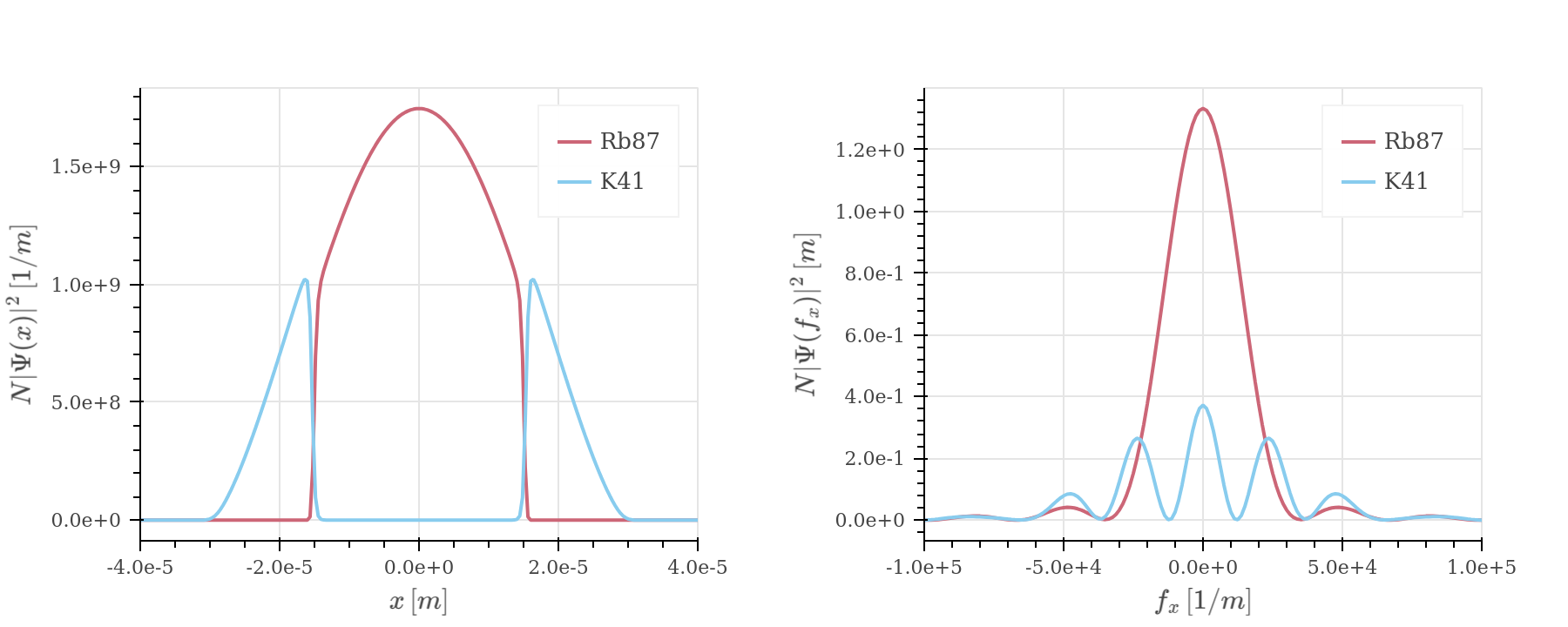}
    \caption{
        The computed ground state probability densities of the $^{87}$Rb and $^{41}$K BECs normalized to the number of atoms per species, shown in both position and frequency space.
        The repelling force between the two species splits up $^{41}$K along the weak trapping axis $x$.
        The plotted position and frequency space regions are zoomed in with respect to the actual domain sizes.
    }
    \label{fig:DualSPeciesProbDensity}
\end{figure}

\section{Computational Performance Evaluation}\label{chap:performance}
In the following, we evaluate the computational performance of FFTArray on one of our previous examples.
We compare the array libraries NumPy and JAX as well as the execution on different hardware. 
Due to the breadth of possible implementations on top of FFTArray, any performance evaluation can only look at a small subset of possible scenarios.
This section uses as an example the imaginary time evolution from \cref{sec:ex-harmSingle}.

This section evaluates whether FFTArray maintains performance comparable to directly using the underlying array library’s FFT methods. 
The comparison focuses on large domains and a split-step algorithm with many time steps since those are for us the most performance-limited scenarios.
In order to focus on those, we exclude startup and initialization costs from the time measurement as much as possible.
In order to extract the run-time per step and remove any remaining startup costs, we perform a linear fit of the results for three different numbers of time steps and take the fitted slope as our result in computational time per time step.
For more details on all these reduction steps see \cref{sec:MeasurementMethods} and \cref{sec:speedNSteps}.

We compare three different implementations of the same imaginary time evolution.
They all execute the exact same algorithm but organize the computations differently.
The first one is the code from \cref{sec:ex-harmSingle} which will be labeled as "FFTArray Direct".

The potential and time step \code{dt} are the same for each step in this scenario.
We can thus compute the propagators once before the loop and then reuse them at each step.
This insight yields the second implementation which is called "FFTArray Precomputed":
\begin{codeblock}
# <Same intialization as in the original example.>

V = get_V(psi)
k_sq = 0.
for dim in psi.dims:
    # Using coords_from_arr ensures that attributes
    # like eager and xp do match the ones of psi.
    k_sq = k_sq + (2*np.pi*fa.coords_from_arr(psi, dim.name, "freq"))**2

T_prop = fa.exp((-1. * dt * hbar / (2*mass)) * k_sq)
V_prop = fa.exp((-0.5 / hbar * dt) * V)

for _ in range(n_steps):
    psi = psi.into_space("pos") * V_prop
    psi = psi.into_space("freq") * T_prop
    psi = psi.into_space("pos") * V_prop
    
    state_norm = fa.integrate(fa.abs(psi)**2)
    psi = psi * fa.sqrt(1./state_norm)
\end{codeblock}

In order to measure how much overhead FFTArray causes inside the loop, we create a a third benchmark variant which does not use any code of FFTArray inside the loop and is called "Raw FFT":
\begin{codeblock}
# <Same intialization as in the above example.>

T_prop = fa.exp((-1. * dt * hbar / (2*mass)) * k_sq)
V_prop = fa.exp((-0.5 / hbar * dt) * V)

T_prop_arr = T_prop.values("freq")
V_prop_arr = V_prop.values("pos")
# Need the raw inner values, which are not accessible via a public API
# Using these inner values allows us to use the whole infrastructure and phase setup of FFTArray in this example.
# In the inner loop below FFTArray would not need to apply any phase and scale factors because they would cancel out.
psi = psi.into_space("pos").into_factors_applied(False)._values

for _ in range(n_steps):
    psi *= V_prop_arr
    psi = xp.fft.fftn(psi)
    psi *= T_prop_arr
    psi = xp.fft.ifftn(psi)
    psi *= V_prop_arr

    state_norm = xp.sum(xp.abs(psi)**2)*vol_elem
    psi *= xp.sqrt(1./state_norm)

# Repack the raw values correctly.
# Again there is no public API for that.
psi = fa.array(values, [x_dim, y_dim], "pos")
psi._factors_applied = (False,)*len(dims)
\end{codeblock}
This manually eliminates FFTArray completely from the inner loop in order to test whether the book-keeping of FFTArray creates measurable overhead when ensuring dimensions line up and phase factors are applied if necessary.
Doing something like this in practice also abandons the advantages FFTArray offers, which is why this code snippet needs to access private implementation details of FFTArray.

These three implementations are measured with two different array libraries.
NumPy is the base reference and only runs on CPUs.
For GPU support and potentially less overhead on CPUs, we use the JAX library.
They recommend using special structured control flow primitives~\cite{JaxStructuredControlFlow} in order to achieve best performance.
Therefore, the \code{for} loop is replaced with \code{jax.lax.scan} for these measurements.
This ensures that JAX can optimize the whole loop as one unit and operations are potentially fused which can increase performance and reduce the per-step overhead.

In order to give a broad overview of the implementations on different hardware, we chose one server and one desktop CPU (AMD Epyc 7543, AMD Ryzen 7950X3D) and GPU (NVIDIA A100, NVIDIA RTX 4090) respectively, for more details see \cref{sec:HardwareSel}.

These measurements were done at a resolution of 4096 by 4096 samples such that the values reflect the speed in the limit of large wave functions as well as possible.
With this number of samples, the complex-valued wave function has a size of 128\,MiB (float32) and 256\,MiB (float64), respectively, which does not fit in any of the caches of the tested processors.
Other domain sizes can be estimated roughly by linearly extrapolating, for details to the scaling as a function of domain size and shape, see \cref{sec:speedNSamples}.

The results in \cref{fig:benchBar} show the extremely large performance difference between CPUs and GPUs for this example.
The JAX implementation is about two orders of magnitude faster in most cases on the two GPUs compared to the two CPUs.
This shows that it is worthwhile to use the consumer GPUs over any kind of CPUs even if the computations have to happen in float64.

The tracing and batching of the \code{scan}-function of JAX is able to achieve almost the same speed in all three implementations on both CPUs and GPUs.
It therefore shows that FFTArray does not add any measurable overhead compared to using the underlying array library directly for large arrays and time steps.
This means, when using JAX, users can even take advantage of the comfort of directly writing down the formulas without a significant hit to performance on any of the tested platforms.

The NumPy implementation only runs on the CPUs and is slower than JAX on the same hardware.
"FFTArray Direct" is significantly slower than the other two implementations.
This is expected since NumPy cannot fuse multiple operations together which causes a much higher memory overhead for recomputing the propagators at each time step.
However, when pre-computing the propagators in "FFTArray Precomputed" the overhead for managing dimensions and phase factors becomes small in the tested scenario and this implementation achieves comparable speeds to the "Raw FFT" variant.
Therefore, our goal of not introducing overhead for our key use-cases is also achieved with NumPy.

The A100 is faster than the RTX 4090 in float64 while it is the other way around in float32.
This difference is less pronounced than their huge differences in peak compute performance would suggest.
Performance is most likely bound more by memory accesses than raw compute performance in many parts of the algorithm.

In order to give a comparison to an existing state-of-the-art solution, measurement results of TorchGPE are listed at the bottom of the chart for the same hardware.
The resolution in position space is the same as in the FFTArray implementation.
However, it uses twice as many samples in frequency space to reduce boundary effects~\cite{torchgpeGasClassDoc}.
As shown in \cref{fig:harmOsciPrecfp64} this does not yield a significantly better precision in this scenario. 
Since this feature cannot be deactivated at the time of writing, it is part of the overhead of TorchGPE in this scenario.
Even if its performance was a factor of two better, our solution is still multiple times faster in the measured scenario.

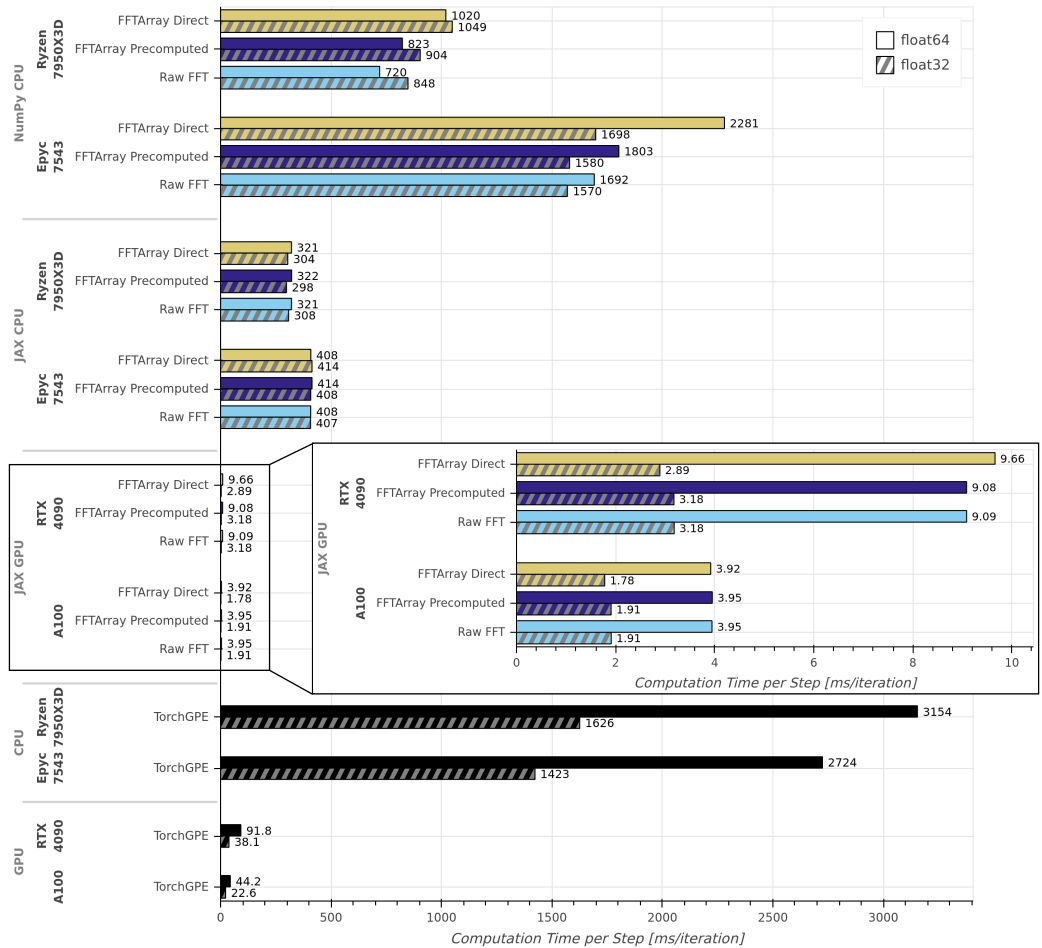
\begin{figure}
    \centering
    \begin{tikzpicture}[scale=0.8]
        \node[anchor=south west, inner sep=0] (mainplot) at (0,0) 
        {\includegraphics[width=12.8cm]{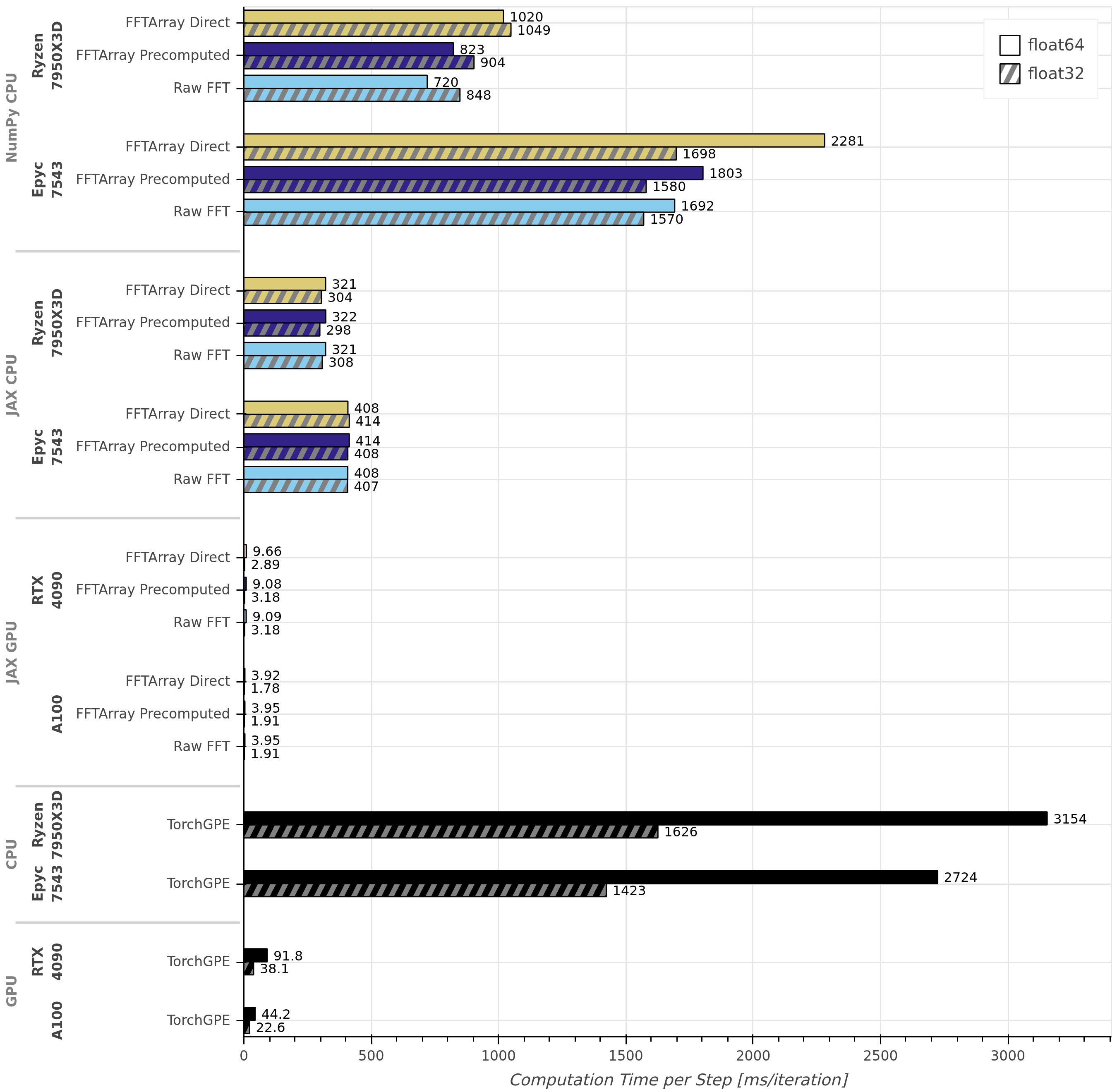}};

        \draw[black, line width=0.5pt] (0cm,4.6cm) rectangle (4.3cm, 8cm);

        \draw[line width=0.5pt, -] (4.3cm, 8cm) -- (5.01cm, 8.345cm);
        \draw[line width=0.5pt, -] (4.3cm, 4.6cm) -- (5.01cm, 4.205cm);

        \node[anchor=south west, draw, inner sep=0.5pt, line width=0.5pt] at (5cm,4.2cm)
          {\includegraphics[height=3.28cm]{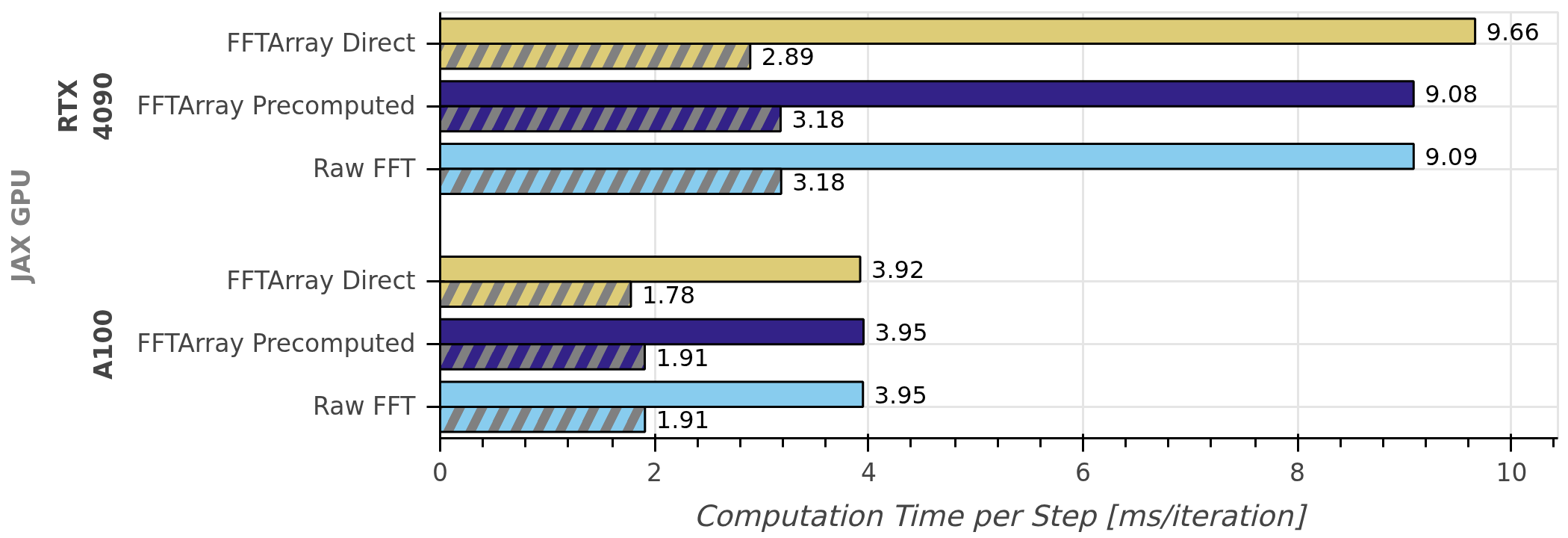}};
    \end{tikzpicture}
    \caption{Computation time per time step on the problem of finding the ground state of an isotropic 2D quantum harmonic oscillator.
    This compares different loop implementations and hardware at a resolution of 4096 by 4096 samples.
    The similar performance between "Raw FFT" and "FFTArray Precomputed" demonstrates that FFTArray introduces negligible overhead.
    GPUs are about two orders of magnitude faster than CPUs and should therefore be preferred in most cases.
    The comparison with TorchGPE shows that the performance of this implementation is better than other state-of-the-art solutions in this scenario.
    }
    \label{fig:benchBar}
\end{figure}

\section{Conclusion and Outlook}
FFTArray introduces a framework for implementing discretized Fourier transforms on arbitrarily shifted coordinate grids.
It enables researchers to translate Fourier integral formulas directly into code and easily scales from single to multiple dimensions via named dimensions.
Numerical details like choosing valid coordinate grids and implementing all necessary phase and scale factors to obtain a discretized Fourier transform are handled automatically without performance overhead for large simulations. 
By being built upon the Python Array API standard, FFTArray enables high performance on GPUs that reduces simulation run times significantly and makes large-scale 3D simulations computationally feasible.
This was already utilized successfully with in-development versions of FFTArray for multiple scientific publications~\cite{Li2024c,Lecoffre2025a,Frye-Arndt2025a,Bruneau2025}.
FFTArray allows researchers to focus on the core scientific challenges rather than the intricacies of Fourier transform implementations, therefore enabling the rapid prototyping of complex models.

Many of the library's central ideas like its constraint solver, the encapsulation of coordinates per dimension and giving the user explicit but automated control over phase and scale factor applications are language-independent and may be adapted to other programming languages.
We hope that FFTArray's modular architecture and design will empower researchers to tackle Fourier-related challenges more effectively.
Furthermore, we invite the scientific community to expand upon our existing implementations, such as our matter wave simulation package~\cite{matterwave}, as well as to contribute to this package and build new solvers on top it.

\FloatBarrier

\paragraph{Code Availability}
The code of FFTArray is openly available at \url{https://github.com/QSTheory/fftarray} and our matter wave specific library is openly available at \url{https://github.com/QSTheory/matterwave}, both under Apache-2.0 license.

\section*{Acknowledgements}
We thank our science group for feedback on earlier versions of this package and helping us charting out the variety of use cases for this package.
S.J.S. thanks Florian Fitzek for his introduction, guidance and Fortran codes for the simulation of matter wave interferometers.
We thank Eric Charron for his feedback on an intermediate version of the library.
S.J.S. thanks Alexander Hahn for discussions, guidance about Trotter product formulas and split-step algorithms and very helpful feedback to a draft of this paper.
G.M. thanks Annie Pichery for discussions about the two species ground state example.
S.J.S. thanks Michael Werner for his help and comments about the mathematical foundations of the Fourier Transform.

\paragraph{Funding information}
This work was funded by the Deutsche Forschungsgemeinschaft (German Research Foundation) under Germany’s Excellence Strategy (EXC-2123 QuantumFrontiers Grants No. 390837967), through CRC 1227 (DQ-mat) within Projects No. A05 and the German Space Agency (DLR) with funds provided by the German Federal Ministry for Economic Affairs and Climate Action (BMWK) (German Federal Ministry of Education and Research (BMBF)) due to an enactment of the German Bundestag under Grants No. 50WM2245A (CAL-II), No. 50WM2263A (CARIOQA-GE) and No. 50WM2253A (AI-Quadrat). NG acknowledges funding by the AGAPES project - grant No 530096754 within the ANR-DFG 2023 Programme. J.-N. K.-S., G.M., C.S. and S.J.S. acknowledge support from QuantumFrontiers through the QuantumFrontiers Entrepreneur Excellence Programme (QuEEP). J.-N. K.-S., G.M., S.J.S. and N.G. acknowledge funding from the EU project CARIOQA-PMP (101081775).

\begin{appendix}
\numberwithin{equation}{section}

\section{Computation Speed as a Function of Time Steps and Function Samples}\label{chap:PerformanceAppendix}

This section describes in more detail the methods, hardware and software used to create the speed measurements in \cref{chap:performance}.
In order to generate the numbers shown in \cref{fig:benchBar}, a few assumptions about the distribution of our measurement data were made, like the linearity of simulation time as a function of time steps.
These are also described and justified in this section.

The measured simulation is an imaginary time evolution to find the ground state of an n-dimensional harmonic oscillator as described in \cref{sec:ex-gpe,sec:ex-harmSingle}.
This means that among others, there are the following numerical parameters:
\begin{itemize}
    \item The number of iterations, i.e., imaginary time steps.
    \item The number of samples to discretize each of the up to three dimensions.
    \item The precision of the used floating point numbers to represent a sample. We used float32 and float64.
\end{itemize}

\subsection{Hardware Selection and System Details}\label{sec:HardwareSel}
The measurements for computation speed were done on one server and one desktop computer.

The server is a dual-socket Dell PowerEdge R7525.
The two CPUs are AMD Epyc 7543 with 32 Zen 3 cores with eight DDR4-3200 memory channels each.
The measurements were limited to one of the CPUs which was configured as a single NUMA node.
This sever also contains three NVIDIA A100 80GiB PCIe cards, with one of those used for the measurements.
The A100 is a server GPU which was specifically designed for high performance scientific computing and machine learning.
It features a peak memory bandwidth of up to 1.94 TB/s and a 1:2 ratio between float32 (19.5 TFLOPS) and float64 (9.7 TFLOPS) performance~\cite{A100ProductBrief}.

The desktop computer consists of an AMD Ryzen 7950X3D CPU and an NVIDIA RTX 4090.
The CPU is a high-end desktop CPU with 16 Zen 4 cores and two memory channels of DDR5-5200 RAM.
The 16 cores are split into two chiplets with 8 cores each.
One chiplet has an additional X3D-Cache which increases its local L3 cache to 96 MiB.
The Nvidia RTX 4090 is a high-end consumer GPU.
This means it is easier and cheaper to procure and can be used in a normal desktop whereas the A100 is only available for servers.
It features also relatively high memory bandwidth (1008 GB/s) but since its focus are graphics workloads, it only has a 1:64 ratio of float64 (1.29 TFLOPS) to float32 (82.6 TFLOPS) performance~\cite{AdaWhitepaper}.

Both systems were running Ubuntu  24.04.2 LTS with the NVIDIA driver 570.133.07 and CUDA 12.8.
The versions of the used array libraries were NumPy 2.2.6, JAX 0.6.1 and PyTorch 2.7.0.
The used version of TorchGPE was the latest public version as of 11 July 2025, \code{c02428f} on \code{https://github.com/qo-eth/TorchGPE}.

\subsection{Measurement Methodology}\label{sec:MeasurementMethods}

In order to isolate the computational speed in the limit of many time steps, the following procedure was used:
First, the simulation is run twice with two time steps in each run as a warm-up to reduce effects due to module imports, compilation caches and other initialization routines which depend on the specific system configuration and software versions.

We only time the inner loop of the simulation without any initialization or clean-up routines while ensuring that after the simulation, the computation results from the GPU have been sent back to the CPU.
To reduce run-to-run variations due to interference by other processes, each measurement was done four times and the minimum measured time was taken.
This assumes that there is a best case speed and any runtime variation is caused by interruptions which ideally do not happen.
To eliminate any startup costs, we use the fact that the simulation time scales linearly in the number of time steps (\cref{sec:speedNSteps}).
For each point in \cref{fig:benchBar,fig:nSamplesScalingJAX,fig:nSamplesScalingNumPy,fig:nSamplesScalingTorch}, we measure for three different numbers of time steps.
The base number of steps is adjusted via a calibration run such that the simulation loop takes about ten seconds, the other two time step groups are then two and four times as many time steps which results in a targeted runtime of 20 and 40 seconds, respectively.
Each of the points in the diagram \cref{fig:benchBar,fig:nSamplesScalingJAX,fig:nSamplesScalingNumPy,fig:nSamplesScalingTorch} is then the slope of a linear fit through the minimum runtime for each of the three different numbers of time steps.

\subsection{Scaling in the Number of Time Steps}\label{sec:speedNSteps}
Theoretically the run time should increase linearly as a function of the number of time steps.
As shown in \cref{fig:nStepsScaling} this is indeed the case for the "FFTArray Direct" variant and TorchGPE.
Each plot also visualizes the run-to-run variation by plotting all slower runs as grey markers.
In most cases, these are not even visible.
The points, reduced via taking the minimum, show a clear linear scaling in the number of time steps as shown by the linear fits.

\begin{figure}
    \centering
    \begin{subfigure}{0.45\textwidth}
        \includegraphics[width=\textwidth]{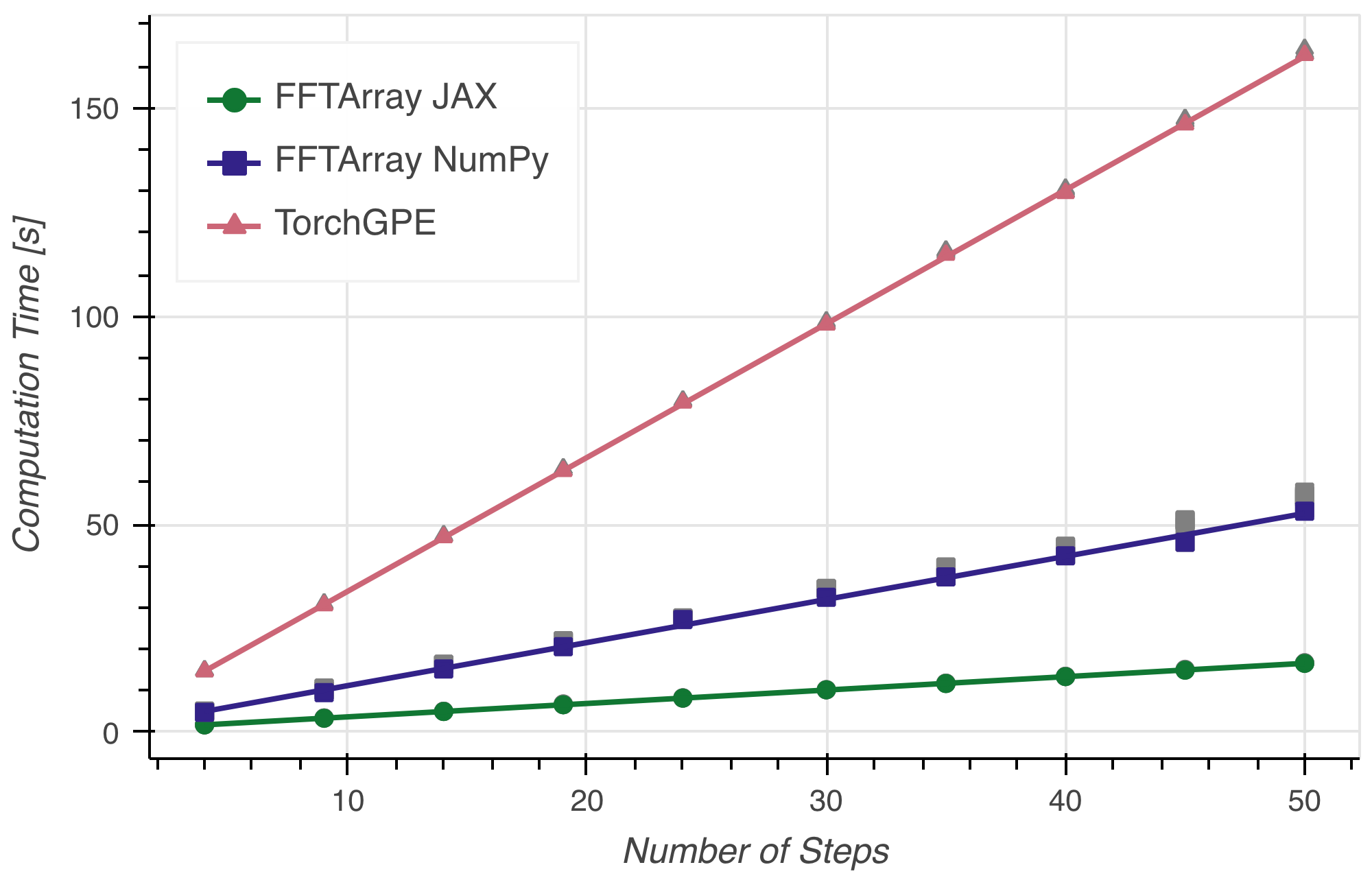}
        \caption*{(a) AMD 7950X3D}
    \end{subfigure}
    \begin{subfigure}{0.45\textwidth}
        \includegraphics[width=\textwidth]{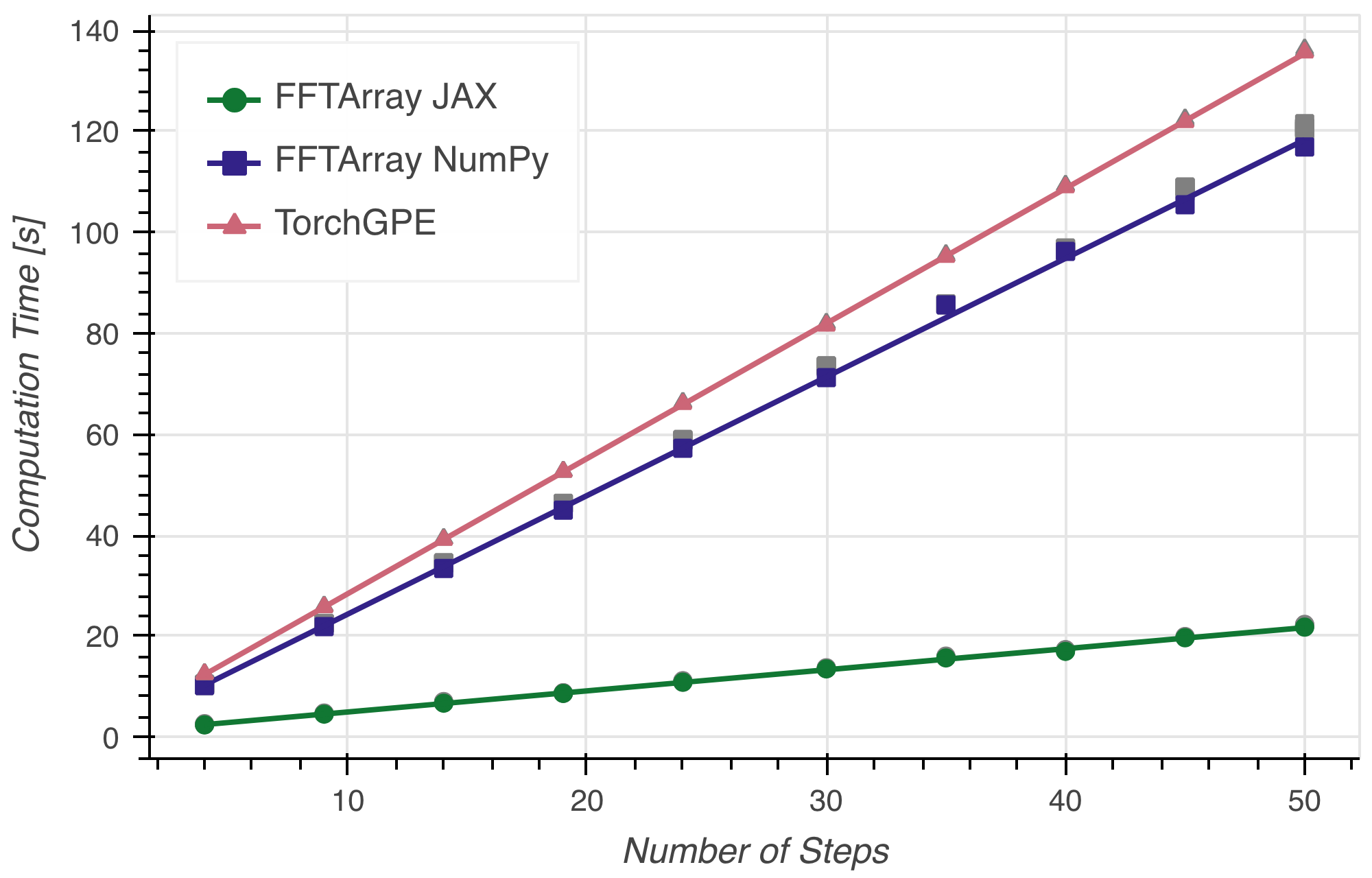}
        \caption*{(b) AMD Epyc 7543}
    \end{subfigure}
    \vfill
    \vspace{0.5cm}
    \begin{subfigure}{0.45\textwidth}
        \includegraphics[width=\textwidth]{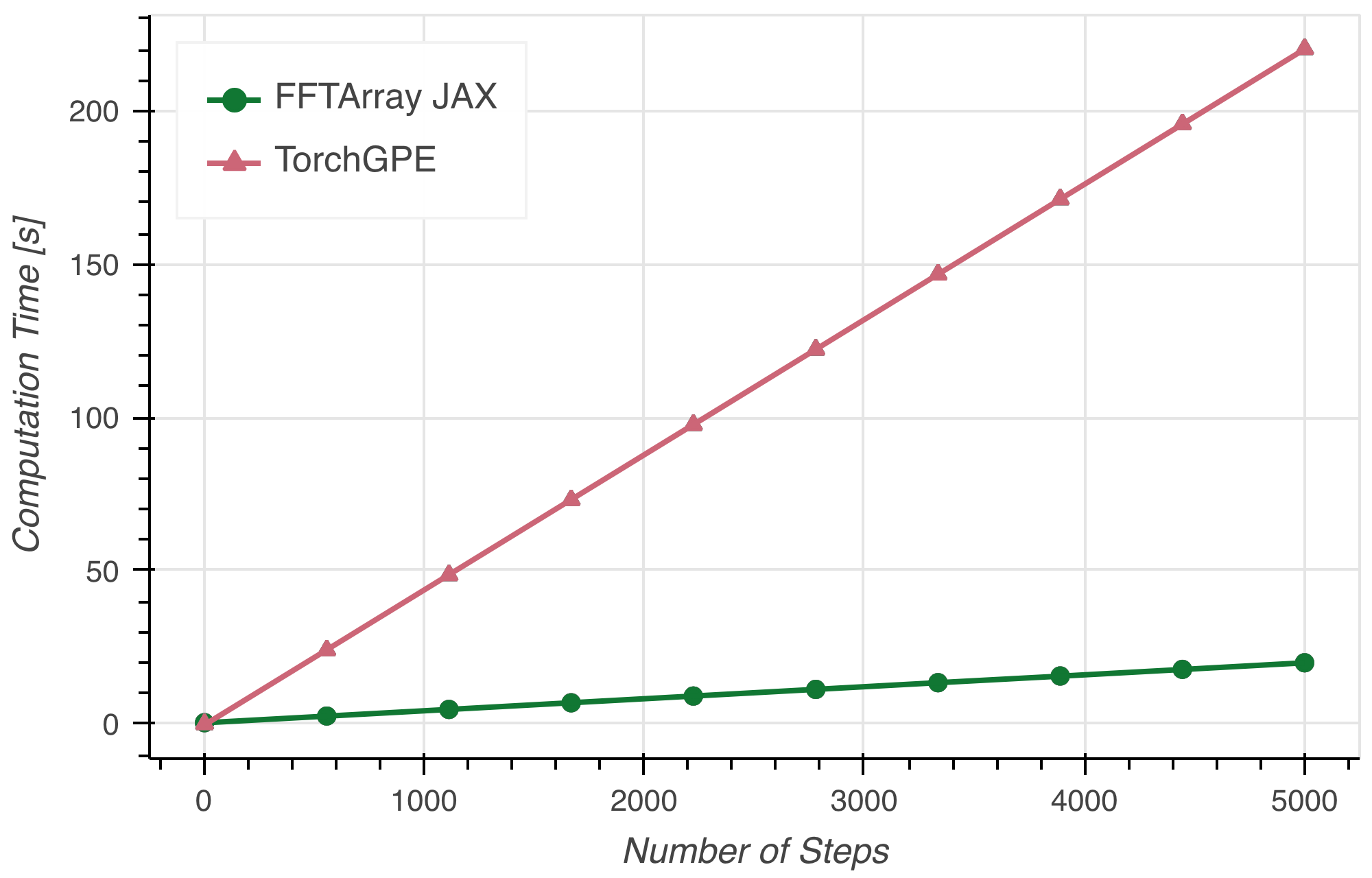}
        \caption*{(c) NVIDIA A100 80 GiB PCIe}
    \end{subfigure}
    \begin{subfigure}{0.45\textwidth}
        \includegraphics[width=\textwidth]{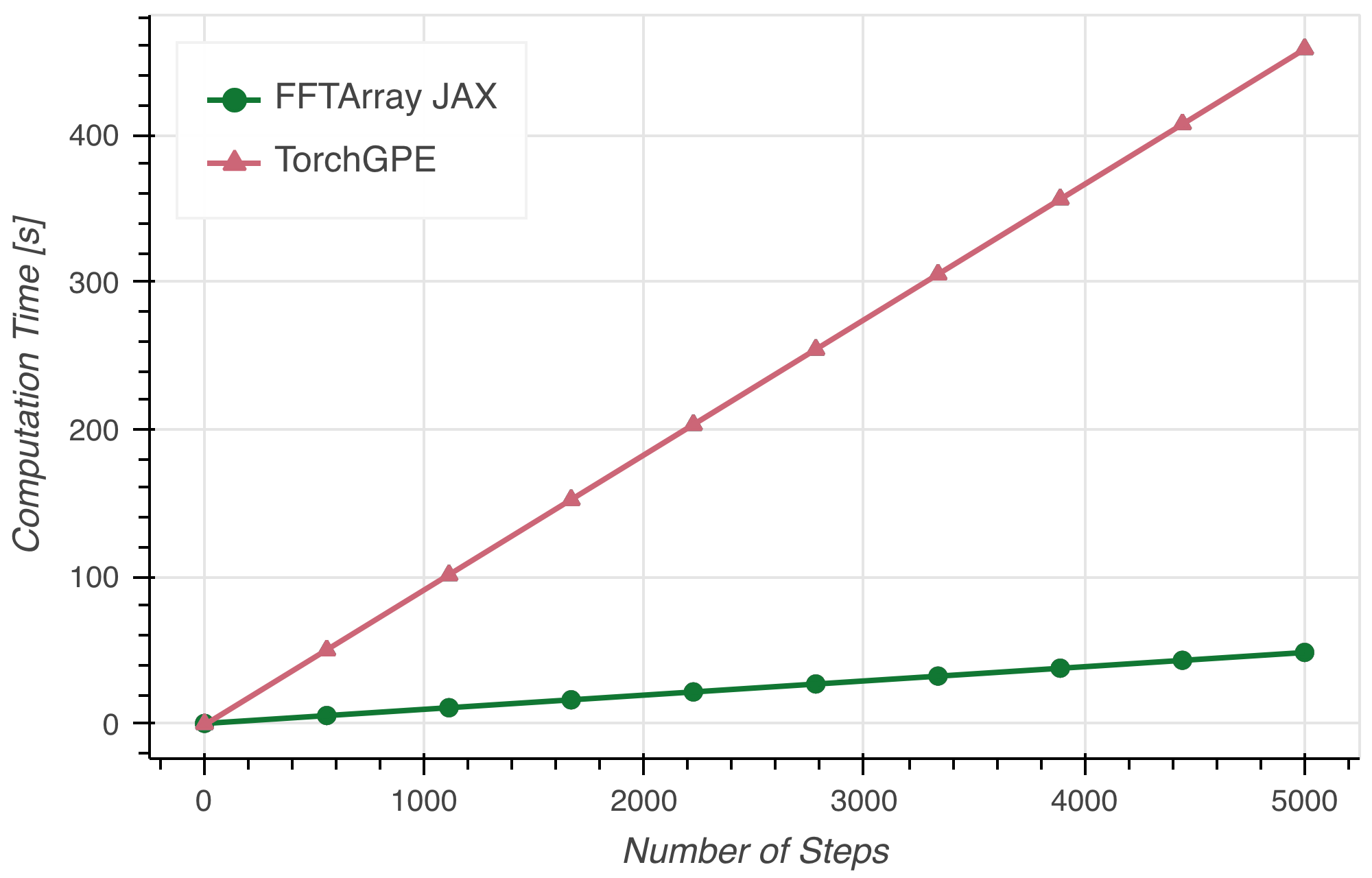}
        \caption*{(d) NVIDIA RTX 4090}
    \end{subfigure}
    \caption{The measured run time for an imaginary time evolution of 4096 by 4096 samples in an isotropic harmonic potential as a function of the number of time steps in float64 precision.
    The used implementations are "FFTArray Direct" for the FFTArray cases and TorchGPE.
    Grey markers show the durations of the runs which were not selected by the minimum reduction.
    On most points all measurements are so close together that they overlap with the selected point and are not visible.
    The lines are linear fits on the minimum run time for each duration.
    All implementations on all hardware configurations show a clear linear scaling in the number of steps which is also the expected behavior.}
    \label{fig:nStepsScaling}
\end{figure}

\FloatBarrier

\subsection{Scaling in the Number of Samples and Shape of the Domain}\label{sec:speedNSamples}

The time complexity class of the simulation algorithm as a function of samples is $\mathcal{O}(n_\text{samples} \log n_\text{samples}$) for the number of samples in each dimension.
This comes from the fact that this is the complexity class of the FFT while all other operations are in the class $\mathcal{O}(n_\text{samples})$.
However, in practice other factors can be more dominant.
There is for example a whole hierarchy of memory types with different speed and size.
The fastest and smallest are the registers, then there are two to three levels of caches and then the (V)RAM.
Additionally, not all memory accesses are equal, since most processors always have to load data at a minimum granularity.
Therefore the data layout and the memory access patterns of algorithms can have a huge impact on performance.
To get a rough overview of how important these factors are, the "FFTArray Direct" and TorchGPE implementations were tested for different grid sizes and layouts.
We always compare the same total number of samples, but these samples are distributed into different shapes:
\begin{itemize}
    \item 1D: The domain is a single dimension which contains all samples.
    \item 2D: The domain is two-dimensional with both sides having the same number of samples.
    \item 3D: The domain is three-dimensional with all three sides having the same number of samples.
    \item (64, 64, nz): The first and second dimension have 64 samples while the z dimension has a number of samples which is a power of two.
    This scenario is for example relevant for the case of a Bragg beam splitter.
    The Bragg beam splitting process requires a higher resolution in the beam direction than the other directions.
    \item (nx, 64, 64): This is in principle the same shape as the case before.
    But due to the different ordering of axes, this may change the memory access patters of the algorithm.
    This case is tested to see if that makes a significant difference.
\end{itemize}
In order to always test the exact same FFT algorithm, each axis must have a size which is a power of two.
Notably, that means that for example the 3D case can only be tested for a total number of samples of $2^{3n}$ with n being a natural number.
Additionally we only tested starting from a minimum size of 64 samples per axis and a total minimum number of samples of $2^{20}$ for FFTArray and $2^{18}$ for TorchGPE, since smaller numbers are not that interesting for our applications.
TorchGPE only officially supports 2D, so only that shape was tested.

The measurements in this section were done according to \cref{sec:MeasurementMethods} including a linear fit over multiple runs with different numbers of steps.

The results for JAX (\cref{fig:nSamplesScalingJAX}) and TorchGPE (\cref{fig:nSamplesScalingTorch}) show a mostly linear scaling for all hardware devices while NumPy (\cref{fig:nSamplesScalingNumPy}) shows a worse than linear scaling, especially in the 1D case.
The approximately linear scaling despite the theoretical complexity class of $\mathcal{O}(n_\text{samples} \log n_\text{samples}$) shows that we are apparently still in a regime of $n_\text{samples}$ where other factors dominate the run time.
With the exception of the 1D case on CPUs, all layouts show a similar performance, so a good estimation of computational speed can be done simply from the number of samples and time steps.
Notably, the 1D case is slower than the other layouts on CPU with both the NumPy and the JAX implementation but on the GPUs all layouts are comparable.

\begin{figure}
    \centering
    \begin{subfigure}{0.45\textwidth}
        \includegraphics[width=\textwidth]{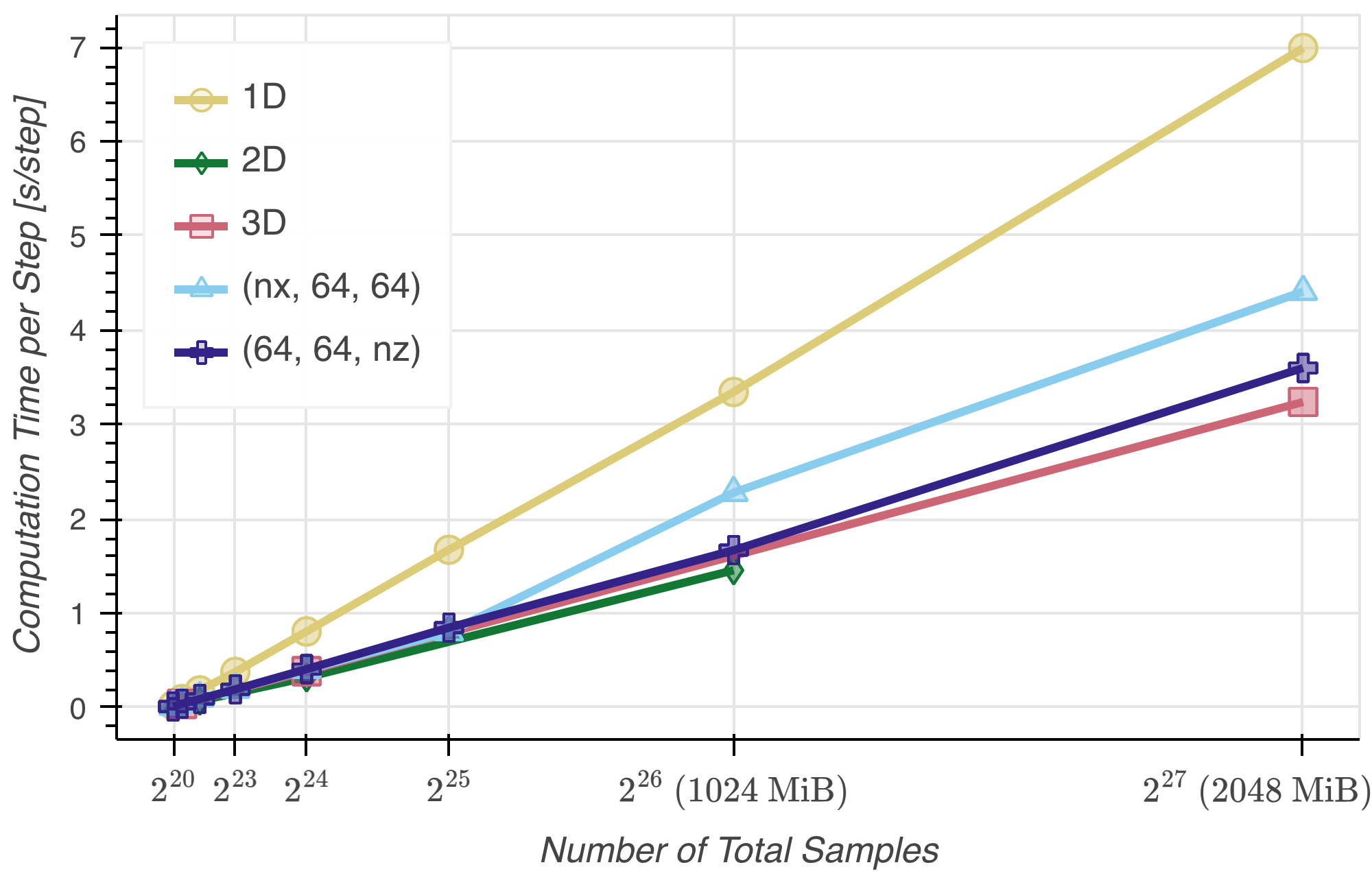}
        \caption*{(a) AMD 7950X3D}
    \end{subfigure}
    \begin{subfigure}{0.45\textwidth}
        \includegraphics[width=\textwidth]{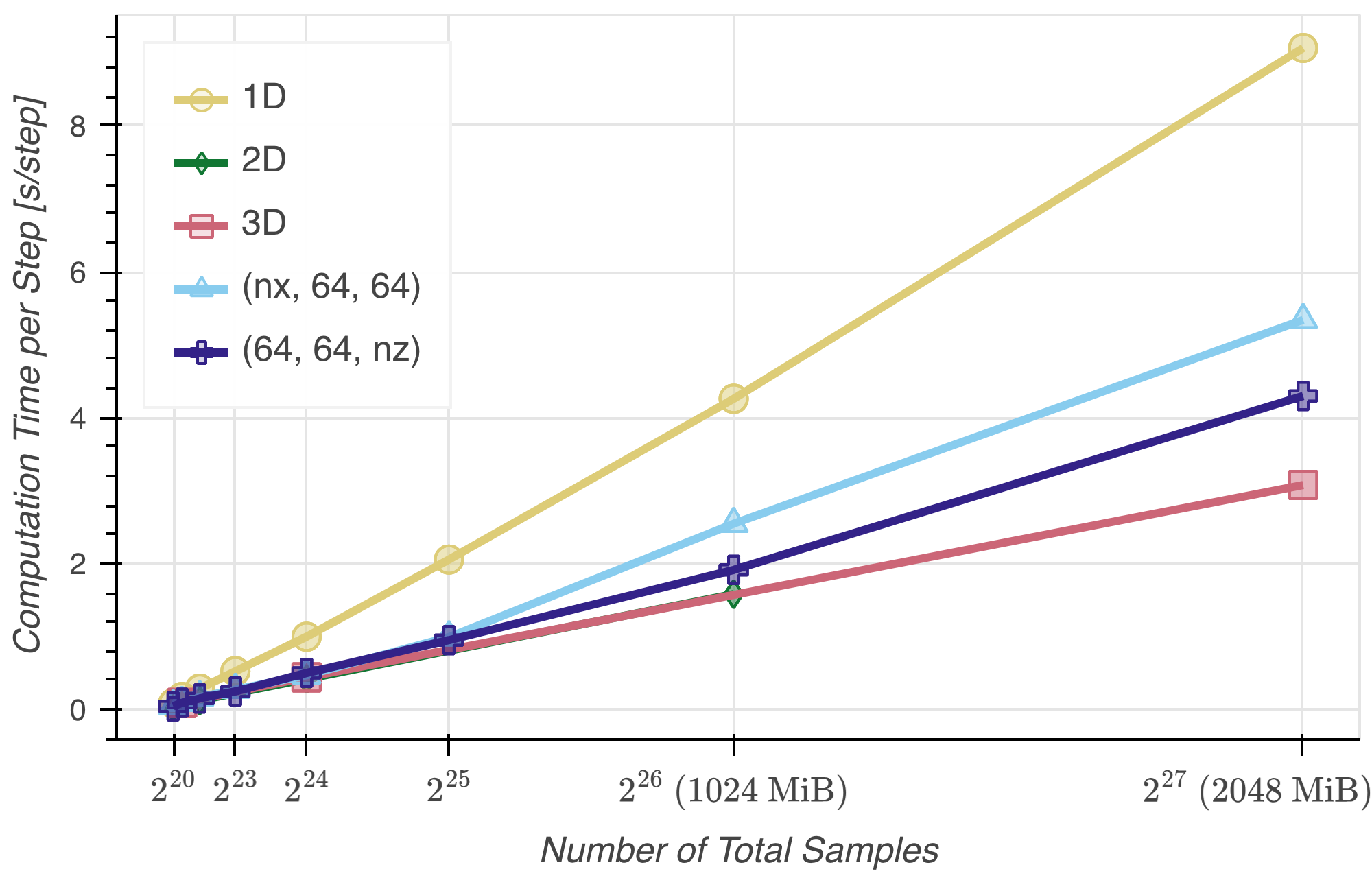}
        \caption*{(b) AMD Epyc 7543}
    \end{subfigure}
    \vfill
    \vspace{0.5cm}
    \begin{subfigure}{0.45\textwidth}
        \includegraphics[width=\textwidth]{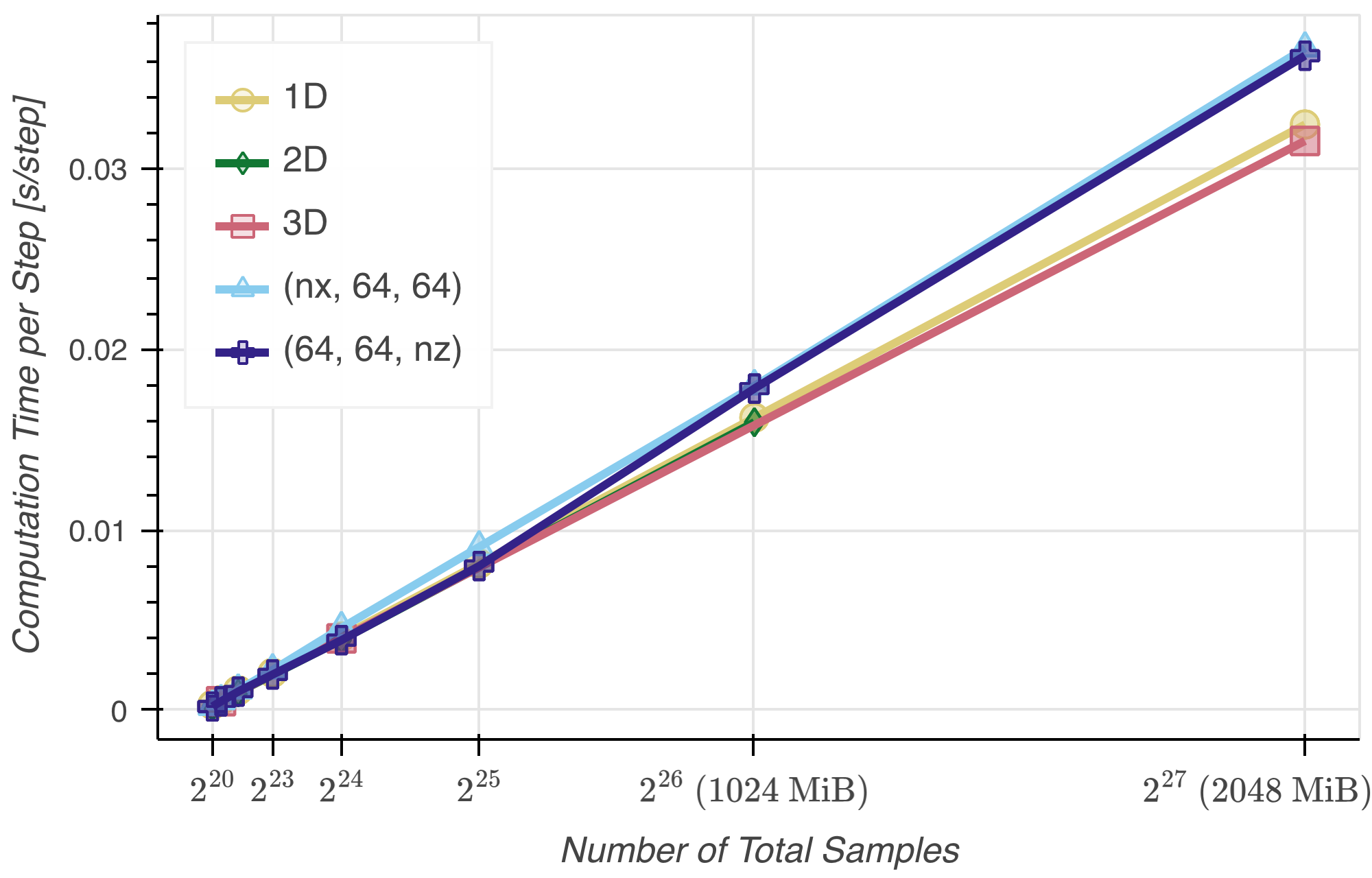}
        \caption*{(c) NVIDIA A100 80 GiB PCIe}
    \end{subfigure}
    \begin{subfigure}{0.45\textwidth}
        \includegraphics[width=\textwidth]{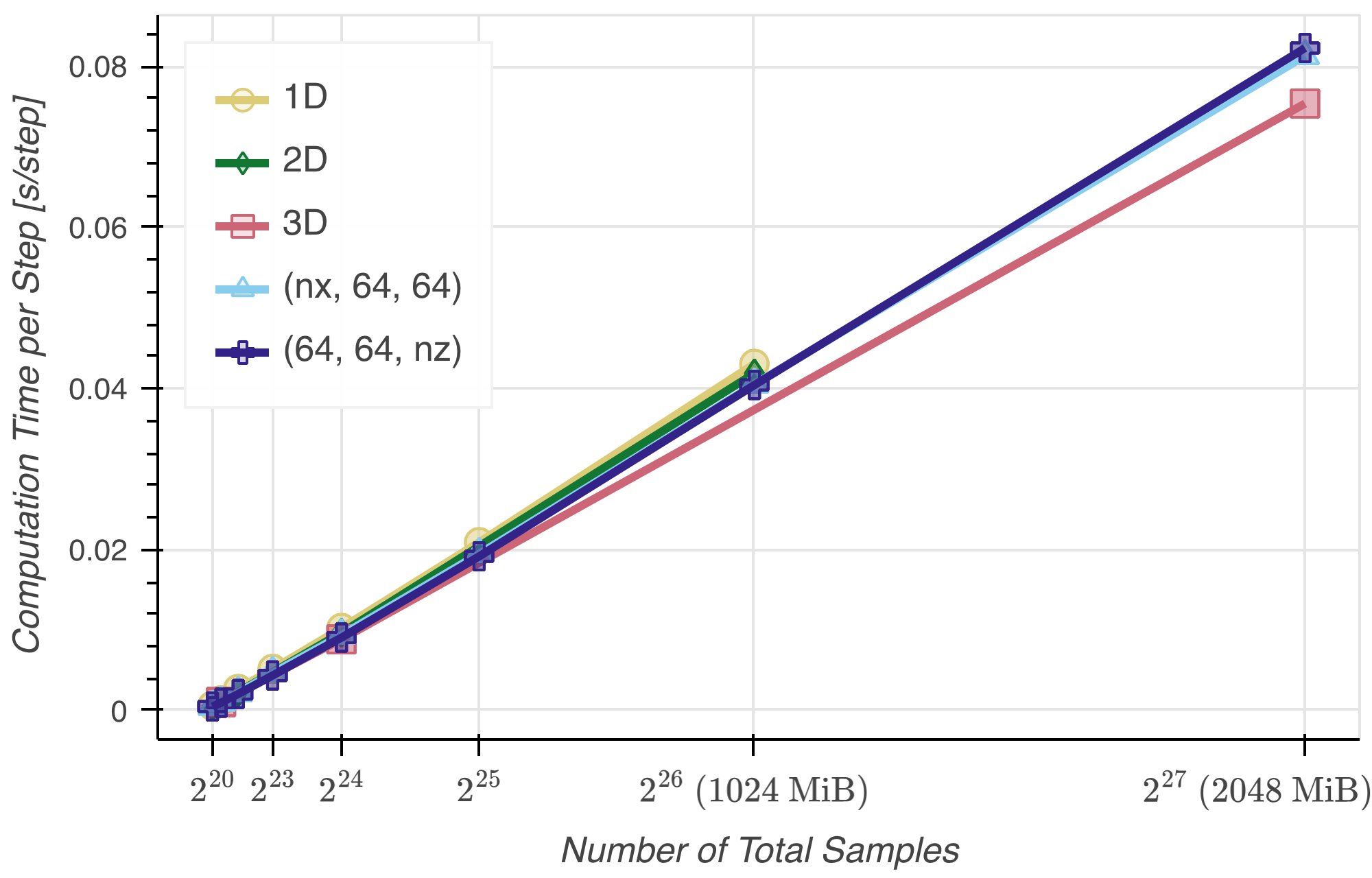}
        \caption*{(d) NVIDIA RTX 4090}
    \end{subfigure}
    \caption{
    Computation time per step as a function of the number of samples and their layout with the JAX "FFTArray Direct" implementation in float64.
    Even though the theoretical scaling is $\mathcal{O}(n_\text{samples} \log n_\text{samples})$, JAX effectively scales linearly in the number of samples.
    All layouts are reach similar speeds with the exception of the 1D case on CPUs which is significantly slower.
    On GPUs the 1D case also achieves similar speeds to the other layouts.
    All points are directly connected for better visual clarity.
    }
    \label{fig:nSamplesScalingJAX}
\end{figure}

\begin{figure}
    \centering
    \begin{minipage}{0.45\textwidth}
        \includegraphics[width=\textwidth]{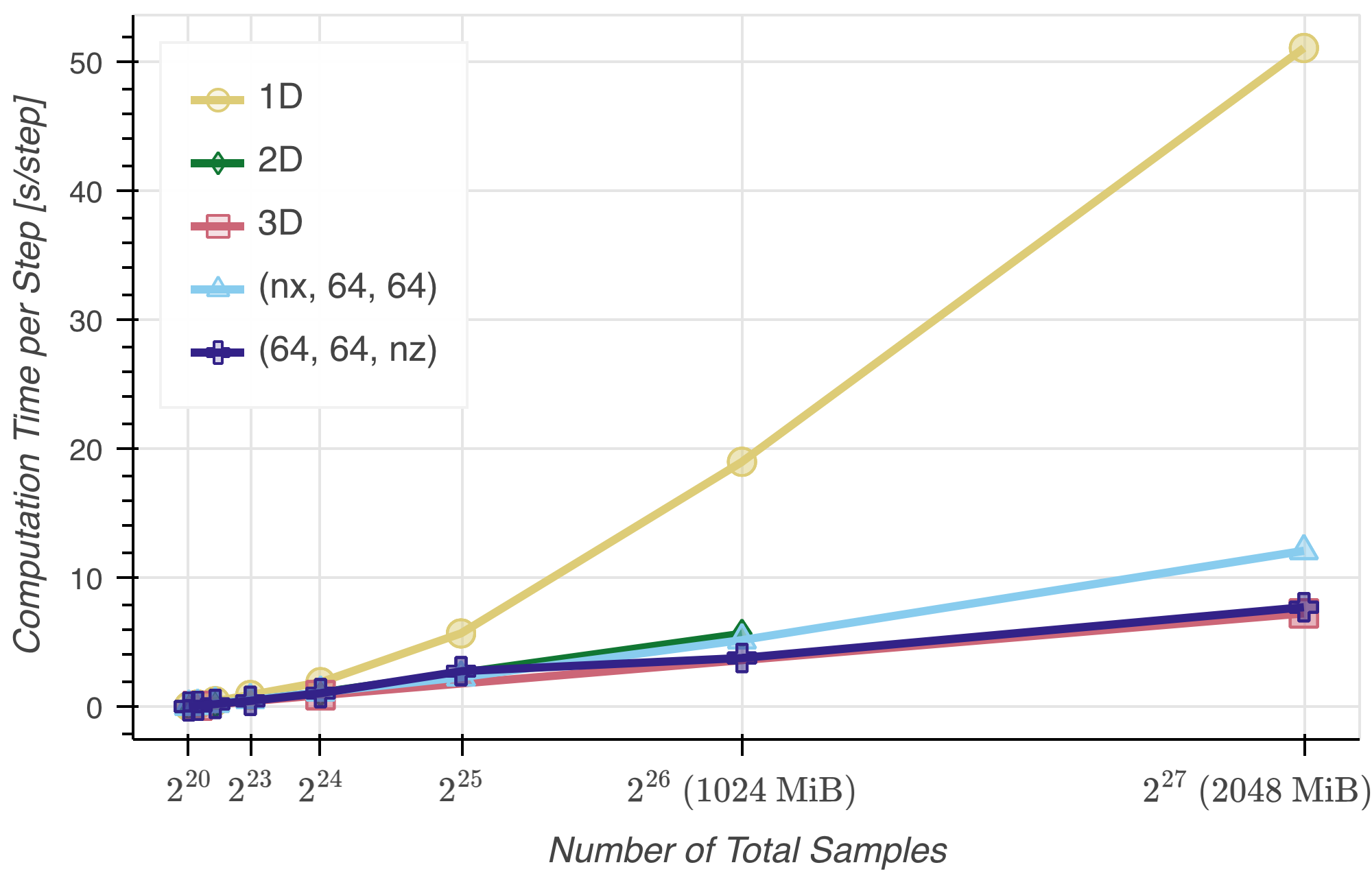}
        \caption*{(a) AMD 7950X3D}
    \end{minipage}
    \begin{minipage}{0.45\textwidth}
        \includegraphics[width=\textwidth]{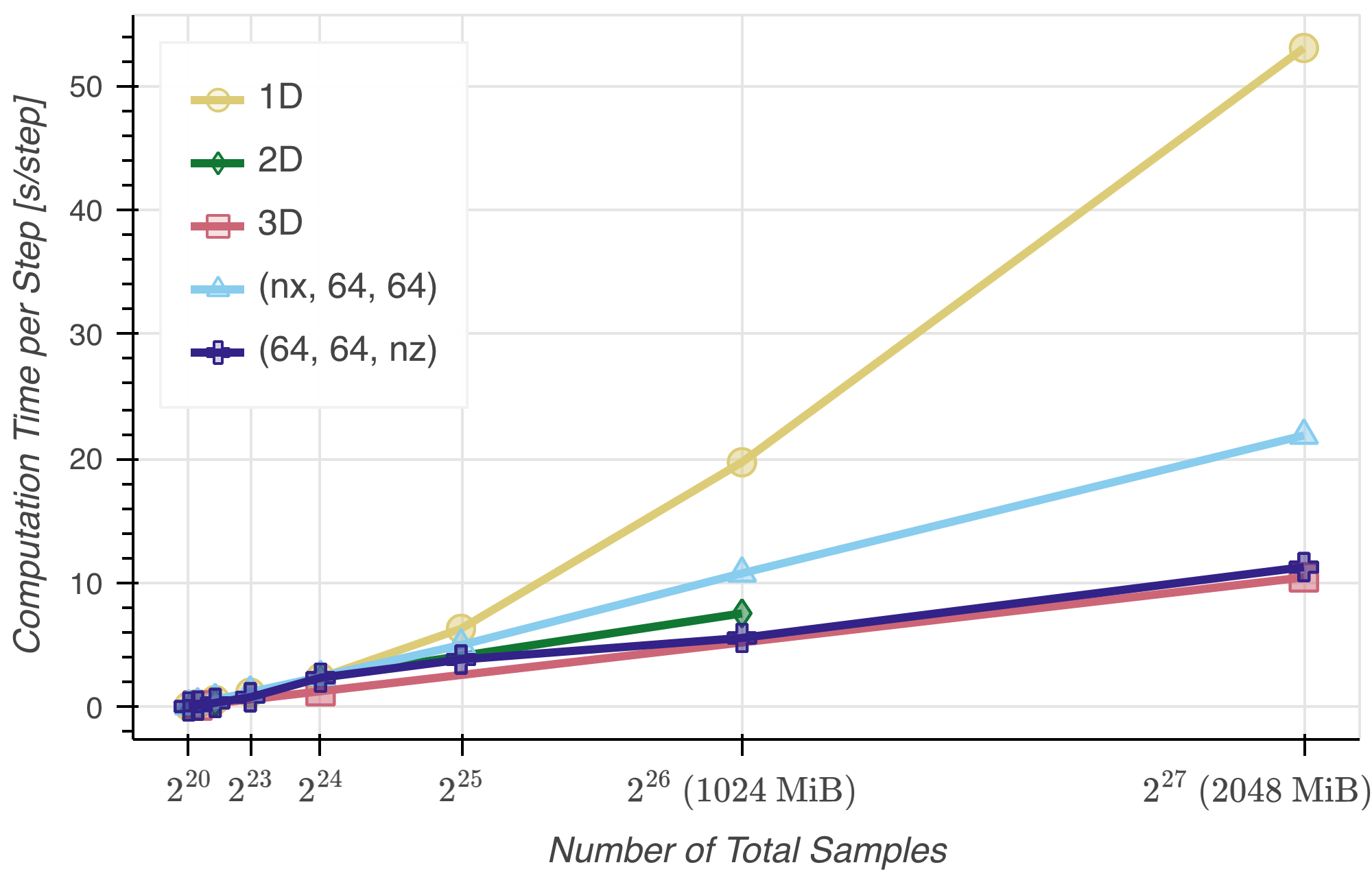}
        \caption*{(b) AMD Epyc 7543}
    \end{minipage}
    \caption{
    Computation time per step as a function of the number of samples and their layout with the NumPy "FFTArray Direct" implementation in float64.
    NumPy shows a slower than linear scaling in the number of samples.
    All layouts are simulated at similar speeds with the exception of the 1D case which is significantly slower.
    All points are directly connected for better visual clarity.}
    \label{fig:nSamplesScalingNumPy}
\end{figure}

\begin{figure}
    \centering
    \begin{minipage}{0.45\textwidth}
        \includegraphics[width=\textwidth]{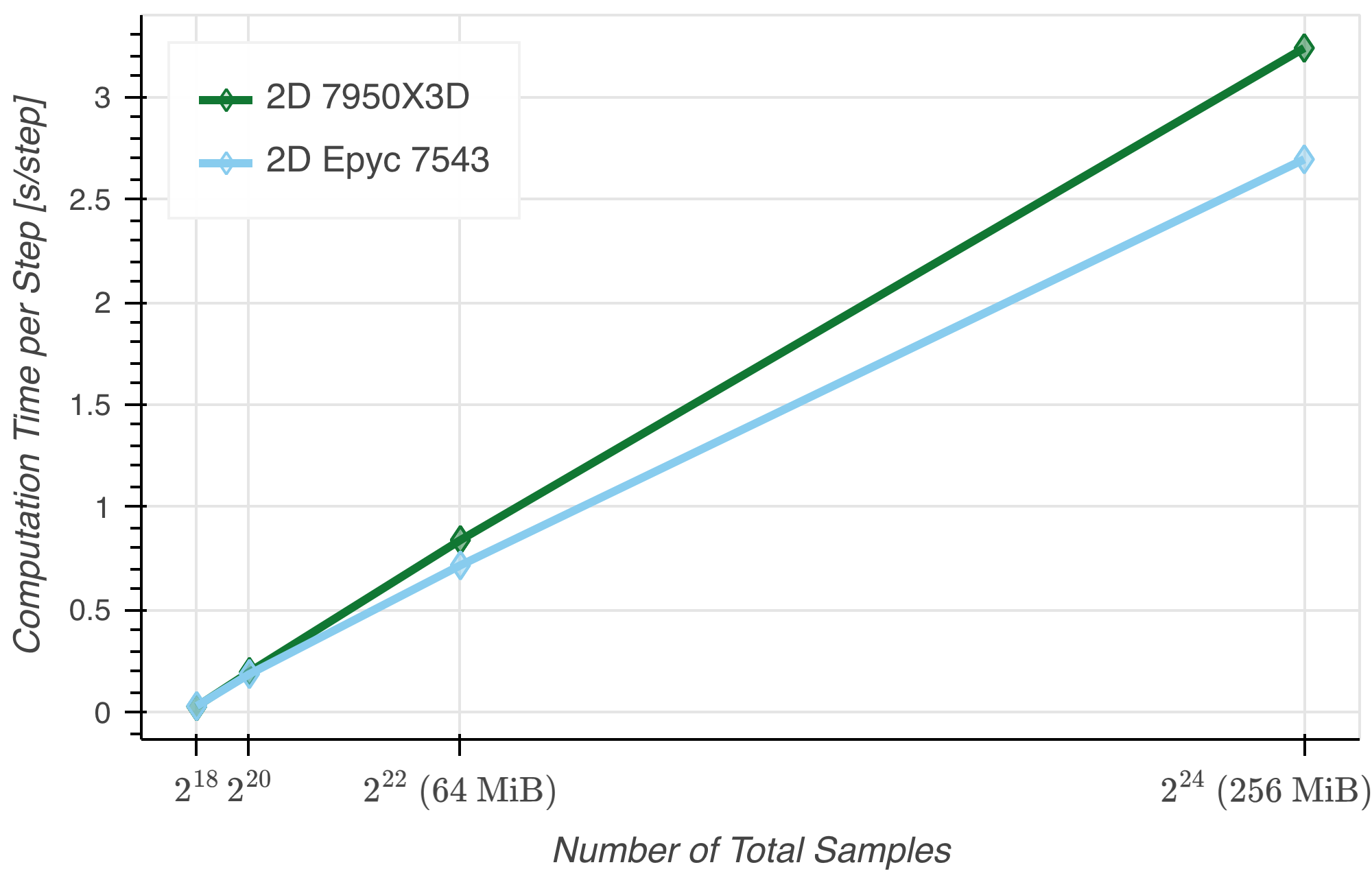}
        \caption*{(a) CPU}
    \end{minipage}
    \begin{minipage}{0.45\textwidth}
        \includegraphics[width=\textwidth]{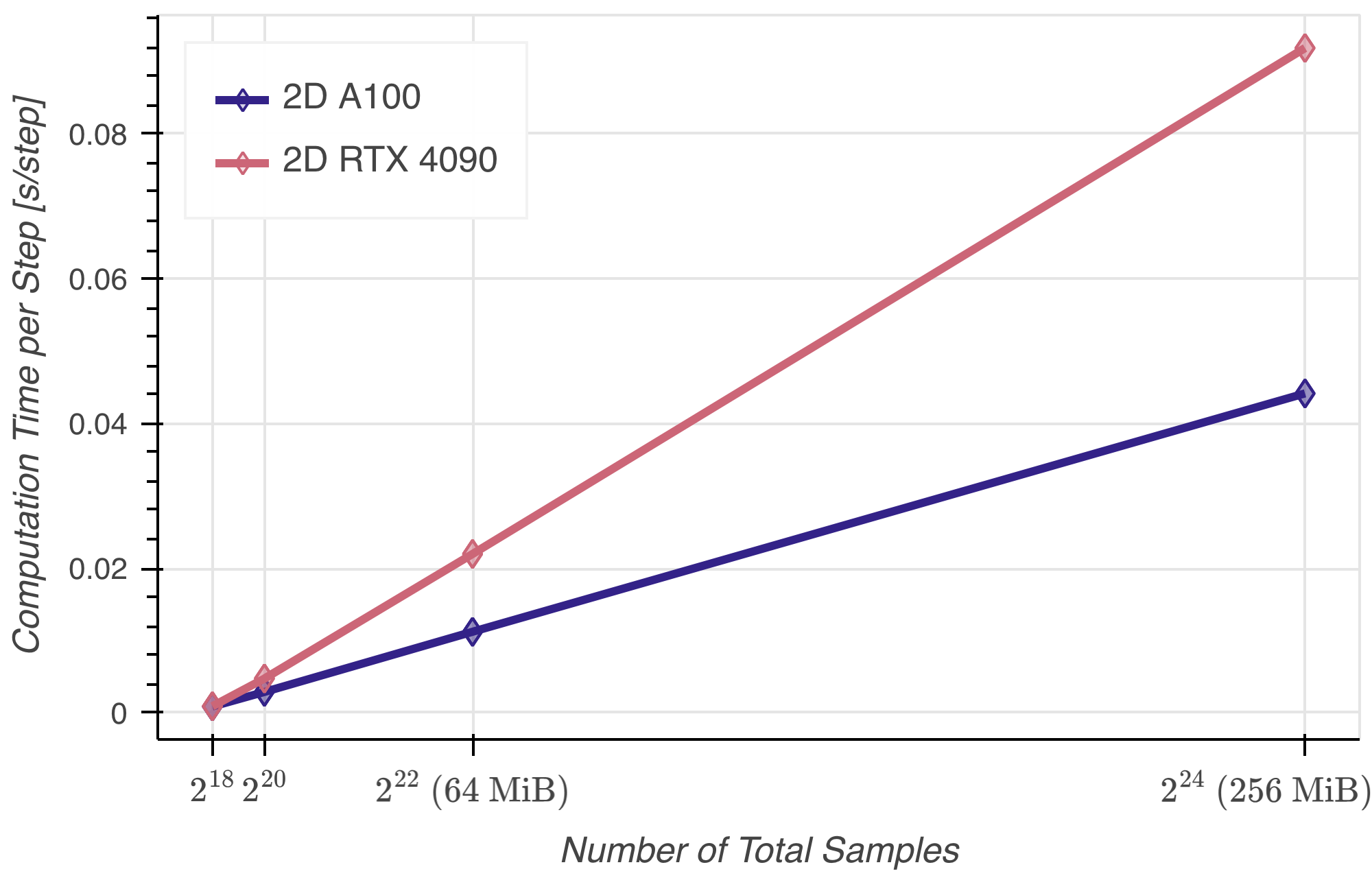}
        \caption*{(b) GPU}
    \end{minipage}
    \caption{
    Computation time per step as a function of the number of samples and their layout with the TorchGPE implementation in float64.
    Even though the theoretical scaling is $\mathcal{O}(n_\text{samples} \log n_\text{samples})$, TorchGPE effectively scales linearly in the number of samples.
    Only the 2D configuration was tested since TorchGPE does not support 1D or 3D domains.
    All points are directly connected for better visual clarity.}
    \label{fig:nSamplesScalingTorch}
\end{figure}

\FloatBarrier

\end{appendix}

\bibliography{FFTArray,software}

\begin{thebibliography}{10}
\providecommand{\url}[1]{\texttt{#1}}
\providecommand{\urlprefix}{URL }
\expandafter\ifx\csname urlstyle\endcsname\relax
  \providecommand{\doi}[1]{doi:\discretionary{}{}{}#1}\else
  \providecommand{\doi}{doi:\discretionary{}{}{}\begingroup
  \urlstyle{rm}\Url}\fi
\providecommand{\eprint}[2][]{\url{#2}}

\bibitem{Feit1982}
M.~J. Feit, J.~A. Fleck and A.~Steiger,
\newblock \emph{Solution of the {{Schr{\"o}dinger}} equation by a spectral
  method},
\newblock Journal of Computational Physics \textbf{47}, 412 (1982),
\newblock \doi{10.1016/0021-9991(82)90091-2}.

\bibitem{Hairer2002}
E.~Hairer, G.~Wanner and C.~Lubich,
\newblock \emph{Geometric {{Numerical Integration}}}, vol.~31 of \emph{Springer
  {{Series}} in {{Computational Mathematics}}},
\newblock Springer Berlin Heidelberg, Berlin, Heidelberg,
\newblock ISBN 978-3-662-05020-0 978-3-662-05018-7,
\newblock \doi{10.1007/978-3-662-05018-7} (2002).

\bibitem{Shannon1949}
C.~Shannon,
\newblock \emph{Communication in the {{Presence}} of {{Noise}}},
\newblock Proceedings of the IRE \textbf{37}(1), 10 (1949),
\newblock \doi{10.1109/JRPROC.1949.232969}.

\bibitem{harris2020array}
C.~R. Harris, K.~J. Millman, S.~J. {van der Walt}, R.~Gommers, P.~Virtanen,
  D.~Cournapeau, E.~Wieser, J.~Taylor, S.~Berg, N.~J. Smith, R.~Kern, M.~Picus
  \emph{et~al.},
\newblock \emph{Array programming with {{NumPy}}},
\newblock Nature \textbf{585}(7825), 357 (2020),
\newblock \doi{10.1038/s41586-020-2649-2}.

\bibitem{jax2018github}
J.~Bradbury, R.~Frostig, P.~Hawkins, M.~J. Johnson, C.~Leary, D.~Maclaurin,
  G.~Necula, A.~Paszke, J.~Vander{P}las, S.~Wanderman-{M}ilne and Q.~Zhang,
\newblock \emph{{JAX}: composable transformations of {P}ython+{N}um{P}y
  programs},
\newblock \url{https://github.com/google/jax},
\newblock Version 0.6.1 (2018).

\bibitem{Ansel2024a}
J.~Ansel, E.~Yang, H.~He, N.~Gimelshein, A.~Jain, M.~Voznesensky, B.~Bao,
  P.~Bell, D.~Berard, E.~Burovski, G.~Chauhan, A.~Chourdia \emph{et~al.},
\newblock \emph{{{PyTorch}} 2: {{Faster}} machine learning through dynamic
  python bytecode transformation and graph compilation},
\newblock In \emph{29th {{ACM}} International Conference on Architectural
  Support for Programming Languages and Operating Systems, Volume 2 ({{ASPLOS}}
  '24)}. ACM,
\newblock \doi{10.1145/3620665.3640366} (2024).

\bibitem{Antoine2014}
X.~Antoine and R.~Duboscq,
\newblock \emph{{{GPELab}}, a {{Matlab}} toolbox to solve
  {{Gross}}--{{Pitaevskii}} equations {{I}}: {{Computation}} of stationary
  solutions},
\newblock Computer Physics Communications \textbf{185}(11), 2969 (2014),
\newblock \doi{10.1016/j.cpc.2014.06.026}.

\bibitem{Antoine2015}
X.~Antoine and R.~Duboscq,
\newblock \emph{{{GPELab}}, a {{Matlab}} toolbox to solve
  {{Gross}}--{{Pitaevskii}} equations {{II}}: {{Dynamics}} and stochastic
  simulations},
\newblock Computer Physics Communications \textbf{193}, 95 (2015),
\newblock \doi{10.1016/j.cpc.2015.03.012}.

\bibitem{Schloss2018}
J.~Schloss and L.~O'Riordan,
\newblock \emph{{{GPUE}}: {{Graphics Processing Unit Gross--Pitaevskii
  Equation}} solver},
\newblock Journal of Open Source Software \textbf{3}(32), 1037 (2018),
\newblock \doi{10.21105/joss.01037}.

\bibitem{Smith2022}
B.~D. Smith, L.~W. Cooke and L.~J. LeBlanc,
\newblock \emph{{{GPU-accelerated}} solutions of the nonlinear
  {{Schr{\"o}dinger}} equation for simulating {{2D}} spinor {{BECs}}},
\newblock Computer Physics Communications \textbf{275}, 108314 (2022),
\newblock \doi{10.1016/j.cpc.2022.108314}.

\bibitem{Fioroni2024}
L.~Fioroni, L.~Gravina, J.~Stefaniak, A.~Baumg{\"a}rtner, F.~Finger, D.~Dreon
  and T.~Donner,
\newblock \emph{A {{Python GPU-accelerated}} solver for the
  {{Gross-Pitaevskii}} equation and applications to many-body cavity {{QED}}},
\newblock SciPost Physics Codebases p.~38 (2024),
\newblock \doi{10.21468/SciPostPhysCodeb.38}.

\bibitem{Pethick2008}
C.~J. Pethick and H.~Smith,
\newblock \emph{Bose--{{Einstein Condensation}} in {{Dilute Gases}}},
\newblock Cambridge University Press, 2 edn.,
\newblock ISBN 978-0-521-84651-6 978-0-511-80285-0,
\newblock \doi{10.1017/cbo9780511802850} (2008).

\bibitem{Gross1963}
E.~P. Gross,
\newblock \emph{Hydrodynamics of a superfluid condensate},
\newblock Journal of Mathematical Physics \textbf{4}(2), 195 (1963),
\newblock \doi{10.1063/1.1703944}.

\bibitem{pitaevskii1961vortex}
L.~P. Pitaevskii,
\newblock \emph{Vortex lines in an imperfect {{Bose}} gas},
\newblock Soviet Physics--JETP [translation of Zhurnal Eksperimentalnoi i
  Teoreticheskoi Fiziki] \textbf{13}(2), 451 (1961).

\bibitem{Pichery2023d}
A.~Pichery, M.~Meister, B.~Piest, J.~B{\"o}hm, E.~M. Rasel, E.~Charron and
  N.~Gaaloul,
\newblock \emph{Efficient numerical description of the dynamics of interacting
  multispecies quantum gases},
\newblock AVS Quantum Science \textbf{5}(4), 044401 (2023),
\newblock \doi{10.1116/5.0163850}.

\bibitem{Fitzek2020}
F.~Fitzek, J.-N. Siem{\ss}, S.~Seckmeyer, H.~Ahlers, E.~M. Rasel, K.~Hammerer
  and N.~Gaaloul,
\newblock \emph{Universal atom interferometer simulation of elastic scattering
  processes},
\newblock Scientific Reports \textbf{10}(1), 22120 (2020),
\newblock \doi{10.1038/s41598-020-78859-1}.

\bibitem{Stein2003}
E.~M. Stein and R.~Shakarchi,
\newblock \emph{Fourier analysis: an introduction}, vol.~1,
\newblock Princeton University Press (2003).

\bibitem{Hormander2003}
L.~H{\"o}rmander,
\newblock \emph{The analysis of linear partial differential operators. I.
  Classics in Mathematics},
\newblock Springer-Verlag, Berlin (2003).

\bibitem{Stein2005}
E.~M. Stein and R.~Shakarchi,
\newblock \emph{Real analysis: measure theory, integration, and hilbert
  spaces}, vol.~1,
\newblock Princeton University Press (2005).

\bibitem{Trefethen2000}
L.~N. Trefethen,
\newblock \emph{Spectral methods in MATLAB},
\newblock SIAM (2000).

\bibitem{Pharr2023}
M.~Pharr, W.~Jakob and G.~Humphreys,
\newblock \emph{Physically Based Rendering: From Theory to Implementation},
\newblock The MIT Press, Cambridge London, fourth edition edn.,
\newblock ISBN 978-0-262-04802-6 (2023).

\bibitem{Oppenheim2013}
A.~Oppenheim and R.~Schafer,
\newblock \emph{Discrete-Time Signal Processing},
\newblock Pearson Deutschland,
\newblock ISBN 9781292025728 (2013).

\bibitem{Unser2000}
M.~Unser,
\newblock \emph{Sampling - 50 years after {{Shannon}}},
\newblock Proceedings of the IEEE \textbf{88}(4), 569 (2000),
\newblock \doi{10.1109/5.843002}.

\bibitem{Jerri1977}
A.~Jerri,
\newblock \emph{The {{Shannon}} sampling theorem - {{Its}} various extensions
  and applications: {{A}} tutorial review},
\newblock Proceedings of the IEEE \textbf{65}(11), 1565 (1977),
\newblock \doi{10.1109/PROC.1977.10771}.

\bibitem{Muga2004}
J.~Muga, J.~Palao, B.~Navarro and I.~Egusquiza,
\newblock \emph{Complex absorbing potentials},
\newblock Physics Reports \textbf{395}(6), 357 (2004),
\newblock \doi{https://doi.org/10.1016/j.physrep.2004.03.002}.

\bibitem{NumpyFFTModule}
\emph{{{NumPy FFT Module}}},
\newblock
  \url{https://numpy.org/doc/2.2/reference/generated/numpy.fft.fft.html},
\newblock (accessed on 2025-04-06).

\bibitem{Press2007}
W.~H. Press, S.~A. Teukolsky, W.~T. Vetterling and B.~P. Flannery,
\newblock \emph{Numerical Recipes. {{The}} Art of Scientific Computing.},
\newblock Cambridge: Cambridge University Press, 3rd ed. edn.,
\newblock ISBN 978-0-521-88068-8 (2007).

\bibitem{Proakis2007}
J.~G. Proakis and D.~G. Manolakis,
\newblock \emph{Digital Signal Processing},
\newblock Pearson Prentice Hall, Upper Saddle River, N.J, 4th ed edn.,
\newblock ISBN 978-0-13-187374-2 (2007).

\bibitem{Meurer2023}
A.~Meurer, A.~Reines, R.~Gommers, Y.-L.~L. Fang, J.~Kirkham, M.~Barber,
  S.~Hoyer, A.~M{\"u}ller, S.~Zha, S.~Shanabrook \emph{et~al.},
\newblock \emph{Python array {{API}} standard: {{Toward}} array
  interoperability in the scientific python ecosystem},
\newblock In \emph{Proceedings of the 22nd Python in Science Conference}, pp.
  8--17,
\newblock \doi{10.25080/gerudo-f2bc6f59-001} (2023).

\bibitem{deMoura2008}
L.~{de Moura} and N.~Bj{\o}rner,
\newblock \emph{Z3: {{An Efficient SMT Solver}}},
\newblock In C.~R. Ramakrishnan and J.~Rehof, eds., \emph{Tools and
  {{Algorithms}} for the {{Construction}} and {{Analysis}} of {{Systems}}}, pp.
  337--340. Springer, Berlin, Heidelberg,
\newblock ISBN 978-3-540-78800-3,
\newblock \doi{10.1007/978-3-540-78800-3_24} (2008).

\bibitem{Brown1993}
S.~A. Brown, M.~Folk, G.~Goucher, R.~Rew and P.~F. Dubois,
\newblock \emph{Software for {{Portable Scientific Data Management}}},
\newblock Computers in Physics \textbf{7}(3), 304 (1993),
\newblock \doi{10.1063/1.4823180}.

\bibitem{Rew1990}
R.~Rew and G.~Davis,
\newblock \emph{{{NetCDF}}: An interface for scientific data access},
\newblock IEEE Computer Graphics and Applications \textbf{10}(4), 76 (1990),
\newblock \doi{10.1109/38.56302}.

\bibitem{Hoyer2017}
S.~Hoyer and J.~Hamman,
\newblock \emph{Xarray: {{N-D}} labeled {{Arrays}} and {{Datasets}} in
  {{Python}}},
\newblock Journal of Open Research Software \textbf{5}(1), 10 (2017),
\newblock \doi{10.5334/jors.148}.

\bibitem{fftarrayDocsCreationFunctions}
\emph{{FFTArray} creation functions},
\newblock
  \url{https://qstheory.github.io/fftarray/v0.5.1/api/creation_functions_array.html},
\newblock (accessed on 2025-09-03) (2025).

\bibitem{ArrayAPICompat}
\emph{{Array API compatibility library}},
\newblock \url{https://data-apis.org/array-api-compat/index.html},
\newblock (accessed on 2025-06-26), Version 1.12 (2025).

\bibitem{Moin2010}
P.~Moin,
\newblock \emph{Fundamentals of Engineering Numerical Analysis},
\newblock Cambridge University Press, Cambridge, 2 edn.,
\newblock \doi{10.1017/CBO9780511781438} (2010).

\bibitem{Sunaina2018}
{Sunaina}, M.~Butola and K.~Khare,
\newblock \emph{Calculating numerical derivatives using {{Fourier}} transform:
  Some pitfalls and how to avoid them},
\newblock European Journal of Physics \textbf{39}(6), 065806 (2018),
\newblock \doi{10.1088/1361-6404/aadda6}.

\bibitem{Jahnke2000}
T.~Jahnke and C.~Lubich,
\newblock \emph{Error {{Bounds}} for {{Exponential Operator Splittings}}},
\newblock BIT Numerical Mathematics \textbf{40}(4), 735 (2000),
\newblock \doi{10.1023/A:1022396519656}.

\bibitem{Thalhammer2008}
M.~Thalhammer,
\newblock \emph{High-{{Order Exponential Operator Splitting Methods}} for
  {{Time-Dependent Schr{\"o}dinger Equations}}},
\newblock SIAM Journal on Numerical Analysis \textbf{46}(4), 2022 (2008),
\newblock \doi{10.1137/060674636}.

\bibitem{Neuhauser2009}
C.~Neuhauser and M.~Thalhammer,
\newblock \emph{On the convergence of splitting methods for linear evolutionary
  {{Schr{\"o}dinger}} equations involving an unbounded potential},
\newblock BIT Numerical Mathematics \textbf{49}(1), 199 (2009),
\newblock \doi{10.1007/s10543-009-0215-2}.

\bibitem{Hansen2009}
E.~Hansen and A.~Ostermann,
\newblock \emph{Exponential splitting for unbounded operators},
\newblock Mathematics of Computation \textbf{78}(267), 1485 (2009),
\newblock \doi{10.1090/S0025-5718-09-02213-3}.

\bibitem{Kieri2015}
E.~Kieri,
\newblock \emph{Stiff convergence of force-gradient operator splitting
  methods},
\newblock Applied Numerical Mathematics \textbf{94}, 33 (2015),
\newblock \doi{10.1016/j.apnum.2015.03.005}.

\bibitem{An2021}
D.~An, D.~Fang and L.~Lin,
\newblock \emph{Time-dependent unbounded {{Hamiltonian}} simulation with vector
  norm scaling},
\newblock Quantum \textbf{5}, 459 (2021),
\newblock \doi{10.22331/q-2021-05-26-459}.

\bibitem{Burgarth2024}
D.~Burgarth, P.~Facchi, A.~Hahn, M.~Johnsson and K.~Yuasa,
\newblock \emph{Strong error bounds for {{Trotter}} and strang-splittings and
  their implications for quantum chemistry},
\newblock Physical Review Research \textbf{6}(4), 043155 (2024),
\newblock \doi{10.1103/PhysRevResearch.6.043155}.

\bibitem{matterwave}
\emph{Matterwave library},
\newblock \url{https://github.com/QSTheory/matterwave},
\newblock (accessed on 2025-09-03) (2025).

\bibitem{Berman1997}
P.~R. Berman,
\newblock \emph{Atom {{Interferometry}}},
\newblock Elsevier,
\newblock ISBN 978-0-12-092460-8,
\newblock \doi{10.1016/B978-0-12-092460-8.X5000-0} (1997).

\bibitem{Tino2014}
G.~M. Tino and M.~A. Kasevich, eds.,
\newblock \emph{Atom Interferometry: Proceedings of the {{International
  School}} of {{Physics}} "{{Enrico Fermi}}", Course 188, {{Varenna}} on {{Lake
  Como}}, {{Villa Monastero}}, 15 - 20 {{July}} 2013},
\newblock IOS Press, Amsterdam,
\newblock ISBN 978-1-61499-448-0 978-1-61499-447-3 978-88-7438-087-9 (2014).

\bibitem{Muller2008}
H.~M{\"u}ller, S.-w. Chiow, Q.~Long, S.~Herrmann and S.~Chu,
\newblock \emph{Atom {{Interferometry}} with up to
  24-{{Photon-Momentum-Transfer Beam Splitters}}},
\newblock Physical Review Letters \textbf{100}(18), 180405 (2008),
\newblock \doi{10.1103/PhysRevLett.100.180405}.

\bibitem{Chiow2011}
S.-w. Chiow, T.~Kovachy, H.-C. Chien and M.~A. Kasevich,
\newblock \emph{$102 \hbar k$ {{Large Area Atom Interferometers}}},
\newblock Physical Review Letters \textbf{107}(13), 130403 (2011),
\newblock \doi{10.1103/PhysRevLett.107.130403}.

\bibitem{Ahlers2016}
H.~Ahlers, H.~M{\"u}ntinga, A.~Wenzlawski, M.~Krutzik, G.~Tackmann, S.~Abend,
  N.~Gaaloul, E.~Giese, A.~Roura, R.~Kuhl, C.~L{\"a}mmerzahl, A.~Peters
  \emph{et~al.},
\newblock \emph{Double {{Bragg Interferometry}}},
\newblock Physical Review Letters \textbf{116}(17), 173601 (2016),
\newblock \doi{10.1103/PhysRevLett.116.173601}.

\bibitem{Gebbe2021}
M.~Gebbe, J.-N. Siem{\ss}, M.~Gersemann, H.~M{\"u}ntinga, S.~Herrmann,
  C.~L{\"a}mmerzahl, H.~Ahlers, N.~Gaaloul, C.~Schubert, K.~Hammerer, S.~Abend
  and E.~M. Rasel,
\newblock \emph{Twin-lattice atom interferometry},
\newblock Nature Communications \textbf{12}(1), 2544 (2021),
\newblock \doi{10.1038/s41467-021-22823-8}.

\bibitem{Meystre2001}
P.~Meystre,
\newblock \emph{Atom Optics}, vol.~33,
\newblock Springer Science \& Business Media,
\newblock ISBN 978-1-4419-2930-3 978-1-4757-3526-0,
\newblock \doi{10.1007/978-1-4757-3526-0} (2001).

\bibitem{Grynberg2010}
G.~Grynberg, A.~Aspect, C.~Fabre and C.~{Cohen-Tannoudji},
\newblock \emph{Introduction to {{Quantum Optics}}: {{From}} the
  {{Semi-classical Approach}} to {{Quantized Light}}},
\newblock Cambridge University Press, 1 edn.,
\newblock ISBN 978-0-521-55112-0 978-0-521-55914-0 978-0-511-77826-1,
\newblock \doi{10.1017/CBO9780511778261} (2010).

\bibitem{Steck}
D.~A. Steck,
\newblock \emph{Rubidium 87 {{D Line Data}}},
\newblock http://steck.us/alkalidata,
\newblock (revision 2.3.3, 28 May 2024).

\bibitem{AdaWhitepaper}
\emph{{{NVIDIA ADA GPU ARCHITECTURE}}},
\newblock
  \url{https://images.nvidia.com/aem-dam/Solutions/geforce/ada/nvidia-ada-gpu-architecture.pdf},
\newblock (accessed on 2024-05-16).

\bibitem{Ahlers2022}
H.~Ahlers, L.~Badurina, A.~Bassi, B.~Battelier, Q.~Beaufils, K.~Bongs,
  P.~Bouyer, C.~Braxmaier, O.~Buchmueller, M.~Carlesso, E.~Charron, M.~L.
  Chiofalo \emph{et~al.},
\newblock \emph{{{STE-QUEST}}: {{Space Time Explorer}} and {{QUantum
  Equivalence}} principle {{Space Test}}},
\newblock \doi{10.48550/arXiv.2211.15412} (2022), \eprint{2211.15412}.

\bibitem{Struckmann2024}
C.~Struckmann, R.~Corgier, S.~Loriani, G.~Kleinsteinberg, N.~Gox, E.~Giese,
  G.~M{\'e}tris, N.~Gaaloul and P.~Wolf,
\newblock \emph{Platform and environment requirements of a satellite quantum
  test of the weak equivalence principle at the $10^{-17}$ level},
\newblock Physical Review D \textbf{109}(6), 064010 (2024),
\newblock \doi{10.1103/PhysRevD.109.064010}.

\bibitem{Corgier2020}
R.~Corgier, S.~Loriani, H.~Ahlers, K.~{Posso-Trujillo}, C.~Schubert, E.~M.
  Rasel, E.~Charron and N.~Gaaloul,
\newblock \emph{Interacting quantum mixtures for precision atom
  interferometry},
\newblock New Journal of Physics \textbf{22}(12), 123008 (2020),
\newblock \doi{10.1088/1367-2630/abcbc8}.

\bibitem{fftarrayDualSpeciesExample}
\emph{{FFTArray} dual-species ground state example},
\newblock
  \url{https://github.com/QSTheory/fftarray/blob/v0.5.1/examples/two_species_groundstate.ipynb},
\newblock (accessed on 2025-09-03) (2025).

\bibitem{JaxStructuredControlFlow}
\emph{{{JAX}}: {{Structured}} control flow primitives},
\newblock
  \url{https://docs.jax.dev/en/latest/control-flow.html\#structured-control-flow-primitives},
\newblock (accessed on 2025-05-19).

\bibitem{torchgpeGasClassDoc}
\emph{{{TorchGPE Documentation}} - {{The Gas Class}}},
\newblock
  \url{https://qo-eth.github.io/TorchGPE/user_guide/fundamentals.gas_class.html},
\newblock (accessed on 2025-06-26).

\bibitem{Li2024c}
R.~Li, V.~J. {Mart{\'i}nez-Lahuerta}, S.~Seckmeyer, K.~Hammerer and N.~Gaaloul,
\newblock \emph{Robust double {{Bragg}} diffraction via detuning control},
\newblock Physical Review Research \textbf{6}(4), 043236 (2024),
\newblock \doi{10.1103/PhysRevResearch.6.043236}.

\bibitem{Lecoffre2025a}
J.~Lecoffre, A.~Hadi, M.~Bruneau, C.~Garcion, N.~Fabre, {\'E}.~Charron,
  N.~Gaaloul, G.~Dutier and Q.~Bouton,
\newblock \emph{Measurement of {{Casimir-Polder}} interaction for slow atoms
  through a material grating},
\newblock Physical Review Research \textbf{7}(1), 013232 (2025),
\newblock \doi{10.1103/PhysRevResearch.7.013232}.

\bibitem{Frye-Arndt2025a}
K.~{Frye-Arndt}, M.~Glaysher, M.~Glaeser, M.~Koch, S.~Seckmeyer, H.~Ahlers,
  W.~Herr, N.~Gaaloul, C.~Schubert and E.~Rasel,
\newblock \emph{Large, ultra-flat optical traps for uniform quantum gases},
\newblock \doi{10.48550/ARXIV.2505.14155} (2025).

\bibitem{Bruneau2025}
M.~Bruneau, J.~Lecoffre, G.~Routier, N.~Gaaloul, G.~Dutier, Q.~Bouton and
  T.~Emig,
\newblock \emph{Probing the geometry dependence of the {{Casimir-Polder}}
  interaction by matter-wave diffraction at a nano-grating},
\newblock \doi{10.48550/ARXIV.2505.10056} (2025).

\bibitem{A100ProductBrief}
\emph{{{NVIDIA A100}} {\textbar} {{Tensor Core GPU}}},
\newblock
  \url{https://www.nvidia.com/content/dam/en-zz/Solutions/Data-Center/a100/pdf/nvidia-a100-datasheet-nvidia-us-2188504-web.pdf},
\newblock (accessed on 2024-05-16).

\end{thebibliography}

\end{document}